\documentclass[12pt]{article}
\pdfoutput=1
\usepackage{latexsym}
\usepackage{amsmath}
\usepackage{amssymb}
\usepackage[usenames]{color}
\usepackage{cite}
\usepackage{amssymb}
\usepackage{fancybox}
\usepackage{underbracket}
\usepackage{bbding}
\def\al{\alpha}

\def\ga{\gamma} 
\def\ep{\epsilon}

\def\sig{\sigma}

\def\calA{{\cal A}}  \def\calC{{\cal C}}

  \def\calO{{\cal O}}

\def\atil{{\tilde{a}}}
\def\btil{{\tilde{b}}}

\def\Ybar{\bar{Y}}
\def\zbar{{\bar{z}}}
\def\Zbar{\bar{Z}}
\def\Xbar{\bar{X}}

\def\delbar{\bar{\del}}

\def\etabar{\bar{\eta}}

\def\del        {  \partial }
\def\half       {  {1\over 2}  }
\def\rootof#1   {  \left( #1 \right)^{1/2}  }

\def\abs#1      {  \vert #1 \vert  }
\def\ie         {{\it i.e.}\,\,}
\def\evalat#1   {  \left\vert_{#1} \right. }

\def\comma          {\, ,}
\def\period         {\, .}
\def\lsim      {\lower .65ex \hbox{\ $\stackrel{<}{\sim}$\ } }
\def\gsim      {\lower .65ex \hbox{\ $\stackrel{>}{\sim}$\ } }
\def\det       {{\rm det}\, }

\def\diag{{\rm diag}\,}
\def\bra#1{{\langle #1 | } }
\def\ket#1{{| #1 \rangle } }

\def\matel#1#2#3  {{\langle #1 | #2 | #3 \rangle } }

\def\lrvec#1    {\hbox{$\stackrel{\leftrightarrow}{#1}$}}
\def\lvec#1     {\hbox{$\stackrel{\leftarrow}{#1}$}}

\def\vecii#1#2      {  \left(\begin{array}{c}#1\\#2\end{array}\right)  }
\def\veciii#1#2#3   {  \left(\begin{array}{c}#1\\#2\\#3\end{array}
                     \right)  }
\def\veciv#1#2#3#4  {  \left(\begin{array}{c}#1\\#2\\#3\\#4
                                 \end{array}\right)  }
\def\vecfv#1#2#3#4#5 {  \left(\begin{array}{c}#1\\#2\\#3\\#4\\#5
                                 \end{array}\right)  }

\def\matrixii#1#2#3#4            {  \left(\begin{array}{cc}#1&#2\\#3&#4
                                       \end{array}\right) }
\def\matrixiii#1#2#3#4#5#6#7#8#9 {  \left(\begin{array}{ccc}#1&#2&#3\\
                                     #4&#5&#6\\#7&#8&#9\end{array}
                               \right)  }
\def\mativ#1#2#3#4               {  \left(\begin{array}{cccc}
                                       #1\\#2\\#3\\#4\end{array}\right) }
\def\matv#1#2#3#4#5              {  \left(\begin{array}{ccccc}
                                     #1\\#2\\#3\\#4\\#5\end{array}
                              \right)  }
\def\eqabegin         {  \begin{eqnarray}  }
\def\eqaend           {  \end{eqnarray}  }
\def\nn               {  \nonumber  }
\def\bracetwo#1#2     {  \left\{ \begin{array}{l} #1 \\ #2 \end{array}
                         \right.  }
\def\bracetwocases#1#2#3#4  {   \left\{ \begin{array}{ll} #1 &
                                 \qquad #2 \\
                                 #3 & \qquad #4 \end{array} \right.  }
\def\bracebegin#1     {  \left\{ \begin{array}{#1}   }
\def\braceend         {  \end{array}\right.   }
\def\boxit#1#2      {  \vbox{\hrule\hbox{ \hskip -4.1pt \vrule\kern3pt 
                     \vbox
                    {  \hsize #1 \strut\kern3pt #2 \kern3pt\strut  }
                       \kern3pt  \vrule} \hrule  } }
\def\centerbox#1#2  {  \mbox{  }\par\bigskip  \hfil \boxit{#1}{#2} \hfil
                       \par\bigskip\noindent }
\def\rightbox#1#2   {  \hfill\boxit{#1}{#2}  }
\def\leftbox#1#2    {  \boxit{#1}{#2}  }
\def\fullbox#1      {  \boxit{\textwidth}{#1}  }

\newcommand{\nullify}[1]{}

\def\mpg#1#2{\begin{minipage}[t]{#1} #2  \end{minipage} }

\def\epsfig#1#2#3{
{\lower #3 \hbox{
 \mpg{#1}{\begin{center} \includegraphics[width=#1,clip]{#2.eps} \\
 Fig. #2\end{center} }}}}

\definecolor{darkgreen}{rgb}{0,0.5,0}
\definecolor{darkblue}{cmyk}{0.9,0.9,0,0}
\definecolor{darkred}{rgb}{0.6,0,0.3}
\definecolor{MyRed}{cmyk}{0,1,1,0.15}
\definecolor{MyBlue}{cmyk}{1,1,0,0.25}

\newcommand{\beqa}{\begin{eqnarray}}
\newcommand{\eeqa}{\end{eqnarray}}
\newcommand{\bea}{\begin{array}}
\newcommand{\eea}{\end{array}}
\newcommand{\beqn}{\begin{equation}}
\newcommand{\eeqn}{\end{equation}}

\def\barz{\bar{z}}
\def\fn#1{\footnote{#1}}
\def\beq#1{\begin{align}#1\end{align}}
\def\pmatrix#1#2{\left( 
\begin{array}{#1}
#2\end{array} 
\right)}
\def\sl#1{\langle #1  \rangle}
\def\tr{ {\rm tr}  }
\def\det       {{\rm det}\, }

\def\eqref#1{(\ref{#1})}
\definecolor{darkgreen}{rgb}{0,0.5,0}
\definecolor{darkblue}{cmyk}{0.9,0.9,0,0}
\definecolor{darkred}{rgb}{0.6,0,0.3}

\def\upket{\ket{\!\uparrow}}
\def\downket{\ket{\!\downarrow}}

\def\ewick#1#2{\cunderbracket{$#1$}{ }{$#2$}}
\def\n{n}

\def\bracket#1#2{\langle #1 | #2 \rangle}

\makeatletter
\def\section{\@startsection {section}{1}{\z@}{-3.5ex plus -1ex minus 
-.2ex}{2.3ex plus .2ex}{\large\bf}}
\makeatother
\makeatletter
\def\subsection{\@startsection {subsection}{1}{\z@}{-3.5ex plus -1ex minus 
-.2ex}{2.3ex plus .2ex}{\normalsize\bf}}
\makeatother
\usepackage{graphicx} 
\usepackage{epsfig}
\usepackage[setpagesize=false,pagebackref=false, linktocpage, bookmarksopen=true, colorlinks=true, linkcolor=darkblue,citecolor=darkblue,urlcolor=darkblue]{hyperref}
\usepackage{hyperref}
\newcommand{\arXiv}[2]{\href{http://arxiv.org/abs/#1}{{\tt arXiv:#2}}}
\newcommand{\hep}[2]{\href{http://arxiv.org/abs/#1}{{\tt #2}}}
\def\picture#1#2{\includegraphics[#1]{#2.pdf}}
\def\Komabanumber#1#2#3{\hfill \begin{minipage}{6cm} {#1}
              \par\noindent {#2} 
              \par\noindent #3 \end{minipage}}
\def\Authors#1{\begin{center} {\it #1} \end{center}}
\begin{document}

\parskip 5pt plus 1pt   \jot = 1.5ex
\renewcommand{\thefootnote}{\fnsymbol{footnote}}
\setcounter{page}{1}
\setcounter{footnote}{0}
\setcounter{figure}{0}
\Komabanumber{RUP-16-6,  RIKEN-QHP-217}{UT-Komaba 16-02}{}
\vspace{-0.5cm}
\begin{center}
$$$$
{\Large\textbf{\mathversion{bold}
Classical Integrability for Three-point Functions: Cognate Structure at Weak and Strong Couplings
}\par}
\vskip 5ex

\Authors{\baselineskip 3ex
{\sc Yoichi Kazama$^{a,b,c}$\footnote[2]{{\tt yoichi.kazama@gmail.com}}, 
 Shota Komatsu$^{d}$\footnote[3]{{\tt skomatsu@perimeterinstitute.ca}} and Takuya Nishimura$^c$\footnote[4]{{\tt tnishimura@hep1.c.u-tokyo.ac.jp}} 
}
\vskip 3ex
${}^a$  Research Center for Mathematical Physics,  
Rikkyo University, \\  Toshima-ku, Tokyo  171-8501  
Japan 
  \vskip 0.3ex\noindent
${}^b$ Quantum Hadron Physics Laboratory, RIKEN  Nishina Center, \\
Wako 351-0198, Japan \vskip 0.3ex\noindent
${}^c$     Institute of Physics, University of Tokyo, \\
 Komaba, Meguro-ku, Tokyo 153-8902 Japan\vskip 0.3ex\noindent
 ${}^d$ Perimeter Institute for Theoretical Physics, \\31 Caroline Street North,
Waterloo, Ontario N2L 2Y5, Canada 
  }

\vspace{-0.5cm}
\renewcommand{\thefootnote}{\arabic{footnote}}
\numberwithin{equation}{section}
\parskip=0.9ex
\baselineskip 3.5ex
\par\vspace{1.4cm}
\thispagestyle{empty}
\textbf{{Abstract}}\vspace{1mm}
\end{center}
In this paper, we develop a new method of computing three-point functions in the SU(2) sector of the $\mathcal{N}=4$ super Yang-Mills theory in the semi-classical regime at weak coupling, which closely parallels the strong coupling analysis. The structure threading two disparate regimes is the so-called monodromy relation, an identity connecting the three-point functions with and without the insertion of the monodromy matrix. We shall show that this relation can be put to use directly for the semi-classical regime, where the dynamics is governed by the classical Landau-Lifshitz sigma model. Specifically, it reduces the problem to a set of functional equations, which can be solved once the analyticity in the spectral parameter space is specified. To determine the analyticity, we develop a new universal logic applicable at both weak and strong couplings. As a result, compact semi-classical formulas are obtained for a general class of three-point functions at weak coupling including the ones whose semi-classical behaviors were not known before. In addition, the new analyticity argument applied to the strong coupling analysis leads to a modification of the integration contour, producing the results consistent with the recent hexagon bootstrap approach. This modification also makes the Frolov-Tseytlin limit perfectly agree with the weak coupling form.

\noindent
\setcounter{page}{1}
\renewcommand{\thefootnote}{\arabic{footnote}}
\setcounter{footnote}{0}
\newpage
\tableofcontents
\baselineskip 3.2ex
\section{Introduction}
In the preceding few years, a number of substantial advancement have been made concerning the evaluation of the two- and three-point correlation functions in the $\mathcal{N}=4$ super Yang-Mills theory in the large $N$ limit, not only in the weak and the strong coupling regimes but also at finite couplings, with clever ideas and some assumptions.

As for the two-point functions, one can now compute them quite accurately at an arbitrary coupling using the so-called quantum spectral curve\cite{QSC}, which is a vastly evolved form of its precursor, the thermodynamic Bethe ansatz formalism\cite{TBA1,TBA2,TBA3}.

Equally important have been the developments on the computation of the three-point functions involving non-BPS operators, which are imperative in understanding the dynamical aspects of the AdS/CFT correspondence\cite{AdS/CFT1,AdS/CFT2,AdS/CFT3}. 

At weak coupling, one can systematically study these objects by mapping them to scalar products of the integrable spin chain as demonstrated in the pioneering works \cite{OT,RV,ADGN,Tailoring1}.
As for the strong coupling, due to the lack of the method of  quantization  for a string  in a relevant curved spacetime, only the semiclassical saddle-point computation appeared to be feasible. However, the initial attempts for some fully non-BPS three-point functions revealed that such a method is already  rather challenging and 
only some partial results were obtained\cite{JW,KK1}. It was only after 
some non-trivial efforts that these difficulties were overcome and finally rather general  class of  three-point  functions  in the SU(2) sector were evaluated\cite{KK3, KK2}. 

Very recently, in a different vein,  a nonperturbative framework capable of studying these objects at finite coupling  was put forward in \cite{BKV}. The basic idea of this approach is to decompose the three-point functions into more fundamental building blocks called the hexagon form factors and determine them using assumed all-loop  integrability\footnote{An attempt in a similar spirit using the assumed all-loop integrability to determine the string field theory vertex was made in \cite{BJ}.}. 
Although quite powerful, as this method  refers only to the magnon and its mirror excitations without referring to their specific origins,  it is difficult to see 
how the gauge theory and the string theory are related. In this sense,  our present work connecting the weak and the strong coupling representations based on 
 the known integrability properties should be considered as  complementary to such a universal approach.

Now concerning both the weak-coupling  and the hexagon form factor
methods, the three-point functions are expressed  in terms of the sums over partitions of rapidities\fn{
One can sometimes further simplify the expression into a determinant form \cite{Foda}. However, such an expression is known at the moment only for certain rank 1 sectors at weak coupling.}, which become increasingly more complicated as the number of magnons increases. However, it turned out that, in the semiclassical limit, where both the number of magnons and the length of the spin chain become very large, the result at weak coupling can be written in a surprisingly concise form, namely a simple integral on  the spectral curve, whose integrand  is expressed  solely in terms of  the so-called pseudo-momenta\cite{Tailoring3,Kostov,Kostov2}. 
Now it is important to recall at this point that also at strong coupling 
 in the semiclassical approximation  the form of the three-point functions 
exhibits the same simple  structure. A   natural question then is whether there is an underlying physical mechanism by which one can  produce  such a simple expression more directly.  

In the case of the two-point functions, a similar question was addressed in \cite{Kruczenski,KMMZ}. In the semiclassical limit, the collective dynamics of magnons is described by the so-called Landau-Lifshitz model, a classically integrable nonrelativistic sigma model which can be obtained  as a continuum  limit of the Heisenberg spin chain.  This formulation allows one to compute the semi-classical two-point function directly using the classical integrability and moreover makes it possible to describe the weak and strong coupling computations in a similar manner.

So the main purpose of the present work is to develop a formulation for the 
computation of the three-point functions at weak coupling, which in the semiclassical limit produces in a direct way the compact integral expressions similar to those 
in the strong coupling and to understand its basic mechanism. This will not only be 
quite useful from the point of view of the computation of the semiclassical limit, 
which is often physically most interesting, but the understanding of its mechanism 
would also reveal  an aspect of integrability common to apparently disparate regimes.  We will indeed see that a formulation extremely similar to that of 
the strong coupling analysis performed in \cite{KK3} is possible and  it will not only 
 reproduce existing results in the literature but also make predictions for a class of three-point functions whose semi-classical limit have not yet been computed. 

Let us now describe the idea and the structure of our formulation more explicitly. The basic starting point is the result of our previous paper\cite{KKN1}  where the tree-level three-point function in the SU(2) sector can be  expressed as the overlap between the singlet state and the three spin-chain states. 
By preparing a coherent state basis, we can then express such an overlap 
 as a product of integrals over the coherent state variables.  Now for the semiclassical situation of our interest, each spin chain reduces to a Landau-Lifshitz string 
and, more importantly, the overlap can be evaluated by the saddle point method. 
The situation is quite similar  to the one at strong coupling, and just as in that case the determination of the saddle point configuration is quite difficult. 
However, the similarity to the strong coupling case goes further in the semiclassical situation. We also have the monodromy relation identical in form, derived in \cite{KKN1,JKPS} for the weak coupling, which was one of the crucial ingredients in the strong coupling case in determining  the three-point function without the knowledge of the saddle point configuration. This relation is natural and powerful as it is a direct consequence of the classical integrability of the string sigma model and encodes infinitely many conservation laws. 

Now, with such a monodromy relation at hand, most of the crucial ingredients for the strong coupling computation can be transplanted, with some modifications, to the present weak coupling case. More precisely, what this means is the following:
\begin{itemize}
\item The semi-classical three-point functions can  again be expressed in terms of the ``Wronskians'' between the eigenvectors of the monodromy matrices.
\item The monodromy relations, which are identical  in form to the strong coupling case, determine the product of the Wronskians in terms of the quasimomenta.
\item The individual Wronskian can be projected out by solving the Riemann-Hilbert problem using the analyticity property concerning the positions of the zeros and the poles.
\end{itemize}

It should be noted, however, that there is an important difference from the strong coupling case,  concerning  the determination of the analyticity property of the Wronskians. For the strong coupling case, the analyticity was determined by assuming the smoothness of the worldsheet for the saddle-point configuration connecting the three strings. In the present case,  however, the three spin chains are  glued together directly by the singlet projector, which is nothing but a convenient way of performing the Wick contractions dictated by 
the  super Yang-Mills dynamics. There is no concept of worldsheet and hence 
the smoothness argument above does not apply. 

Therefore, in this paper we developed  a new different argument, which 
 is more powerful and universal. The basic idea is to study the response 
 of certain fundamental  quantities to an addition of  a small number of Bethe 
roots.
In the semiclassical context, such an addition corresponds to the continuous variation of the filling fraction of the Bethe roots and when applied to the  (log of) the structure constant $\ln C_{123}$, it reveals that $\ln C_{123}$ plays the role of the generating function of the angle variables and provides the  key equation for obtaining $C_{123}$. 
On the other hand, as it will be elaborated fully in section \ref{sec:Wronskian}, we can also apply such a variation to the norm of the exact spin-chain eigenstate. When the original and the deformed states are both on-shell Bethe eigenstates, 
 they must be orthogonal and we demand that this exact quantum property  must be smoothly  connected to the semiclassical structure  for consistency. This requirement will turn out to be  powerful enough to determine the configuration of the zeros and the poles on the spectral curve. The Wronskians determined through  this logic  not only  leads to the known  semiclassical results for the three-point functions in the literature but also allow us to compute more general SU(2) correlators,   which have  not been computed before. 

It is then extremely interesting to apply this new orthogonality 
 argument to the strong coupling case and see what happens.
 It turned out that  this  more universal argument lead to the modification
 of the integration contours obtained in the previous investigation, and  
  the results with the modified contours are consistent with the 
hexagon form factor approach of \cite{BKV} and  exactly match the Frolov-Tseytlin  limit \cite{FrolovTseytlin} in the weak coupling regime. This indicates that, as already suspected and discussed in \cite{KK3}, the apparently natural requirement of smoothness of the saddle-point worldsheet configuration in the strong coupling case is not quite correct and 
our new logic for determining the analyticity in the semiclassical spectral curve 
is more reliable.

The rest of the paper is structured as follows. In section \ref{sec:semi-classical}, after reviewing the formulation of the tree-level structure constant in terms of the overlap with the singlet
projector, we derive a path-integral  representation  for such an overlap using the coherent state basis, which is subsequently evaluated by its saddle point in the semi-classical limit. We then show that the variation of the semi-classical structure constant with respect to a conserved charge of the spin chain states produces  the angle variable which is canonically conjugate to that  charge.  In section \ref{sec:LL}, we construct the angle variables for the Landau-Lifshitz model using its classical integrability.  Based on the results in the previous sections, we express, in section \ref{sec:Wronskian},  the semi-classical structure constant in terms of the Wronskians of the eigenvectors of the monodromy matrices. In section \ref{sec:ev}, we evaluate such Wronskians,  making use of the monodromy relation and the orthogonality of two on-shell states. Putting together 
all the results in the preceding sections, we finally derive the explicit expression for semi-classical structure constants at weak coupling in section \ref{sec:resultweak}. In section \ref{sec:strong}, we describe 
 how the argument developed in the present paper applied to the strong coupling computation modifies the results obtained previously.  We conclude in section \ref{sec:conclusion} and indicate several future directions. A few  appendices are provided for technical details.  
\section{Semi-classical structure constant 
 and the monodromy relation\label{sec:semi-classical}} 
\subsection{Wick contraction represented as the singlet projection\label{subsec:wick}}
We begin with a brief review of the two devices introduced in 
our previous work \cite{KKN2}, namely the 
double spin-chain representation for the 
SU(2) sector and the interpretation  of the Wick contraction as the 
group singlet projection, which greatly facilitate the construction of the correlation 
 functions. 

The four scalar fields $\phi_i$ ($i=1,2,3,4$) forming the so-called the SU(2) sector of the super Yang-Mills theory can be  assembled into a $2\times 2$ matrix $\Phi_{\atil a}$ given by 
\begin{align}
\Phi_{\atil a} &\equiv \matrixii{Z}{Y}{-\Ybar}{\Zbar} _{\atil a} \comma 
\end{align}
where $Z\equiv \phi_1+i\phi_2$, $Y\equiv \phi_3+i\phi_4$ and $\Zbar$ and $\Ybar$ are
 their hermitian conjugates respectively. Evidently, the symmetry of the SU(2) sector is actually SO(4)$=$SU(2)${}_L \times $ SU(2)${}_R$ and the matrix $\Phi$ transforms 
 under these two SU(2) factors as $\Phi \rightarrow U_L \Phi U_R$, where 
$U_L$ ($U_R$) belongs to SU(2)${}_L$ (SU(2)${}_R$).  This suggests that 
 it is natural  to consider the spin-chain consisting of these 
basic fields as forming a tensor product of two spin-chains, which we called 
 the double spin-chain. Consider first the individual spin states $\upket$ and 
$\downket$ and denote them by $\upket = \ket{1}$ and $\downket=\ket{2}$
 for convenience. Then, from the transformation property above, 
 the basic fields correspond to the tensor product states as
$Z \mapsto \upket_L \otimes \upket_R =\ket{1} \otimes \ket{1}$, etc.
It is easy to see that this mapping  is  succinctly summarized as 
\begin{align}
\Phi_{\atil a} \mapsto \ket{\atil} \otimes \ket{a} \comma 
\quad \atil ,a = 1,2 \period 
\end{align}

To construct the correlation functions at the tree level, we need to Wick contract 
these fields. For the Wick contraction of $\ewick{\Phi_{\tilde{a} a}}{\Phi_{\tilde{b} b}}$, 
 the only non-vanishing ones are (suppressing the coordinate dependence) $\ewick{Z}{\Zbar} =1$ and $\ewick{X}{\Xbar} =1$. This gives the simple formula
\begin{align}
\ewick{\Phi_{\atil a}}{\Phi_{\btil b}} &=\ep_{\atil\btil} \ep_{ab}  
\period
\end{align}
In terms of the corresponding spin states, this rule is equivalent to
\begin{align}
\ewick{\ket{\atil}}{\ket{\btil}} &= \ep_{\atil\btil} \comma \qquad 
\ewick{\ket{a}}{\ket{b}} = \ep_{ab} \period 
\end{align}
Now consider the  general linear combination of states  $F = \ket{\tilde{f}} \otimes \ket{f} $, with 
\begin{align}
 \ket{\tilde{f}} &= \tilde{f}^1 \upket + \tilde{f}^2 \downket = \tilde{f}^{\atil} \ket{\atil} \comma \label{defketn}\\
 \ket{f} &= f^1 \upket + f^2 \downket = f^a \ket{a} \period
\end{align}
From the above rules, the Wick contraction between $F_1$ and $F_2$, where $F_i=\ket{\tilde{f}_i}
\otimes \ket{f_i}$,  can be easily computed as 
$\ewick{F_1}{F_2} =  (\tilde{f}_1^\atil \ep_{\atil\btil} \tilde{f}_2^\btil)(f_1^a \ep_{ab} f_2^b)$. This form shows that one can perform the Wick contraction 
by taking the inner product  with the singlet projection operator 
\begin{align}
\bra{{\bf 1}} &= \ep_{ab} \bra{a} \otimes \bra{b}, \qquad \mbox{for both
 SU(2)$_L$ and SU(2)$_R$} \comma 
\end{align}
namely 
\beq{\label{wicktosinglet}
\ewick{F_1}{F_2} = \bra{{\bf 1}}\Big( \ket{\tilde{f}_1} \otimes \ket{\tilde{f}_2}
\Big)\bra{{\bf 1}}\Big( \ket{f_1} \otimes \ket{f_2}
\Big) \period
}
This representation allows us to perform the Wick contractions 
 for  any complicated operators easily and systematically\footnote{ For the full  PSU$(2,2|4)$ sector,  the singlet projection  operator has been constructed in 
\cite{KKN2,JKPS}.}. 
\begin{figure}[t]
 \begin{center}
 \begin{minipage}{0.3\hsize}
 \begin{center}
  \picture{clip,height=1.8cm}{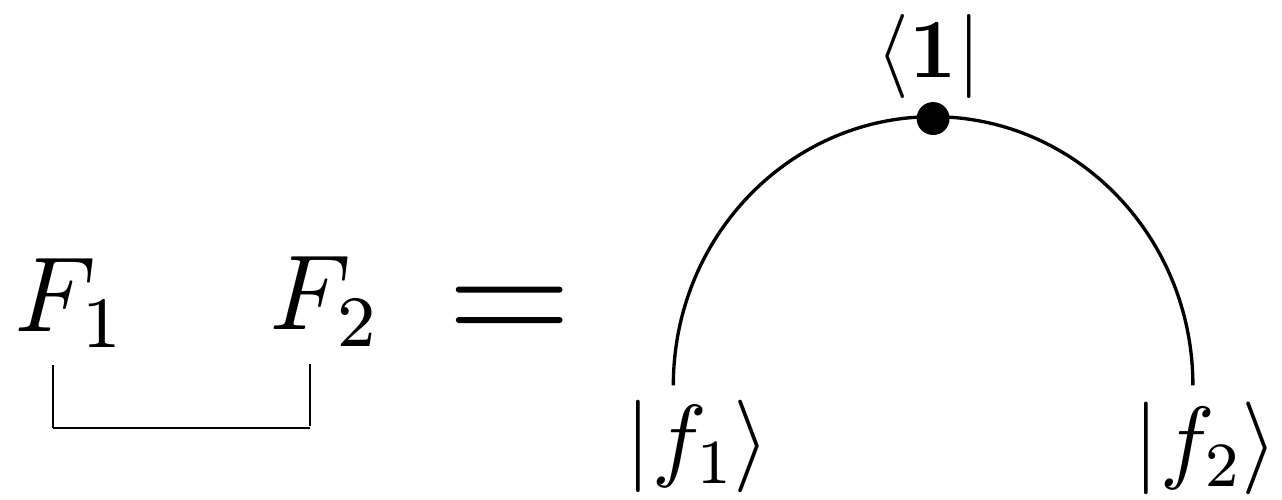}
  \end{center}
  \end{minipage}
 \begin{minipage}{0.3\hsize}
 \begin{center}
  \picture{clip,height=1.8cm}{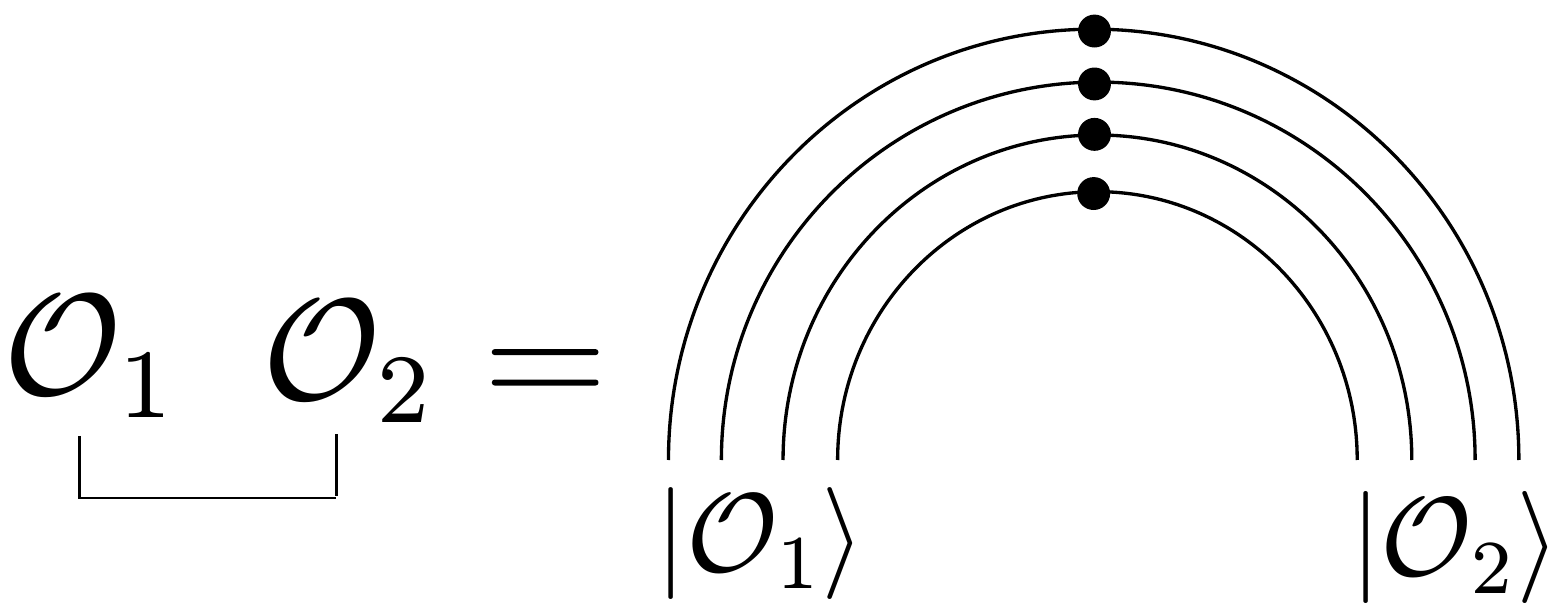}
  \end{center}
  \end{minipage}
  \begin{minipage}{0.34\hsize}
 \begin{center}
  \picture{clip,height=2.3cm}{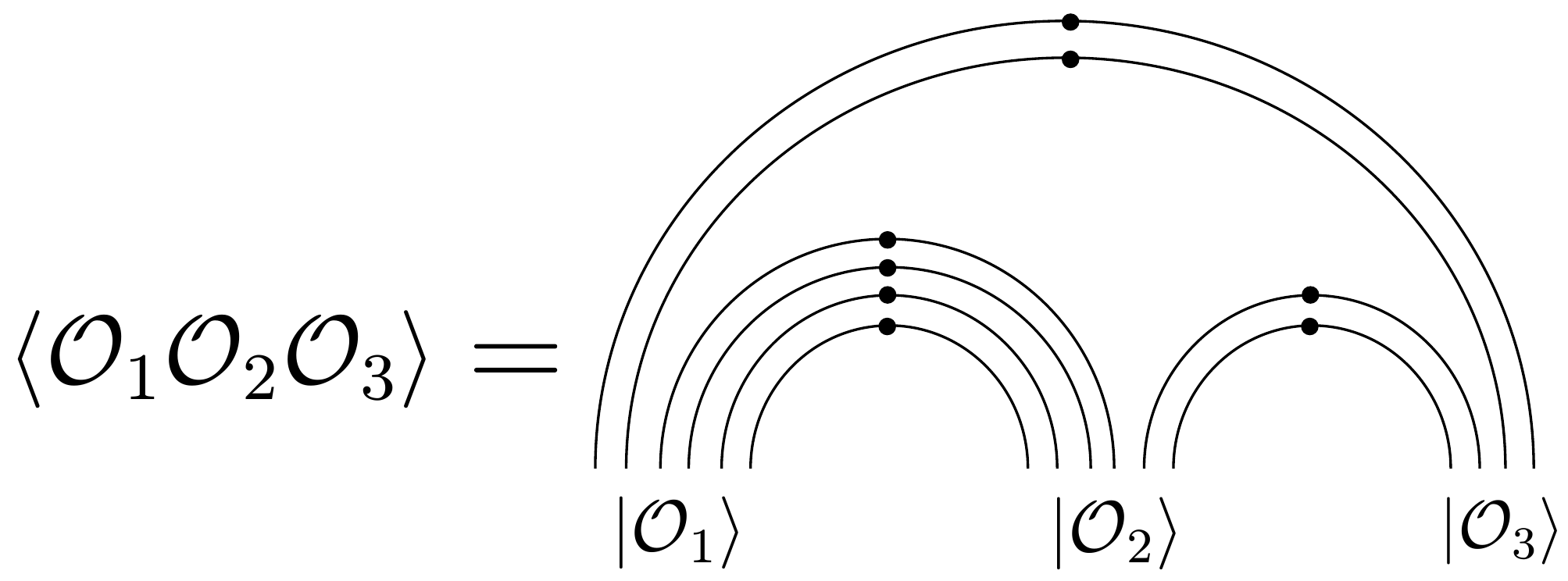}
  \end{center}
  \end{minipage}\\
  \begin{minipage}{0.3\hsize}
 \begin{center}
  $(a)$
  \end{center}
  \end{minipage}
 \begin{minipage}{0.3\hsize}
 \begin{center}
  $(b)$
  \end{center}
  \end{minipage}
  \begin{minipage}{0.34\hsize}
 \begin{center}
  $(c)$
  \end{center}
  \end{minipage}
\caption{The Wick contractions represented as the overlap with the singlet state. Here we only depict the ${\rm SU}(2)_R$ part. $(a)$ The Wick contraction between two fields can be expressed as the overlap between two spin states and the singlet, which is denoted by the black dot. $(b)$ The pictorial explanation of the formula for the two-point function \eqref{twoptaswick}. Here again each dot denotes the overlap with the singlet state. $(c)$ The three-point function expressed as the overlap with the singlet state \eqref{3ptaswick}.} 
\label{Wickpicture}
\end{center}
\vspace{-0.5cm}
\end{figure}

Now let us apply this scheme to the single-trace operators. The contractions which 
 survive in the large $N$ limit are the ones which connect the $(L+1-i)$-th field in the  operator $\calO_1$ with the $i$-th field in the operator $\calO_2$, 
where $L$ is the length common to both operators. Explicitly, an example of this structure looks like 
\beq{
\mathcal{O}_1:\tr\big(\cdots \setlength{\underbracketheight}{9pt}\cunderbracket{$X$}{\setlength{\underbracketheight}{3pt}\cUnderbracket{$Z$}{$\big)\qquad\mathcal{O}_2:\tr\big($}{$ \bar{Z}$}}{$\bar{X}$}\cdots \big) \period
}
This structure motivates us to consider the following tensor product of singlet states,
\beq{
\bra{{\bf 1}_{12}} = \prod_{i=1}^{L} \Big(\epsilon_{ab} \bra{a}_{L+1-i}^{(1)}\otimes \bra{b}_{i}^{(2)}\Big)\comma
}
where $\bra{\ast}^{(k)}_{i}$ denotes the single-spin state living on the $i$-the site of the spin chain corresponding to the operator $\mathcal{O}_k$. Then, the contractions between the operators can be reproduced by taking the inner product
\beq{\label{twoptaswick}
\ewick{\mathcal{O}_1}{\mathcal{O}_2}=\bra{{\bf 1}_{12}} \Big(\ket{\tilde{\mathcal{O}}_1}\otimes\ket{\tilde{\mathcal{O}}_2}  \Big)\bra{{\bf 1}_{12}} \Big(\ket{\mathcal{O}_1}\otimes\ket{\mathcal{O}_2}  \Big)\period
}
Here $ \ket{\tilde{\mathcal{O}}_k}\otimes \ket{\mathcal{O}_k}$ and $\ket{\tilde{\mathcal{O}}_k}\otimes \ket{\mathcal{O}_k} $ are the spin-chain states corresponding to the operators $\mathcal{O}_k$. For a pictorial explanation, see figure \ref{Wickpicture}-$(b)$.

Since the tree-level three-point function is essentially given by a product of Wick contractions, one can also map the computation of the three point function to that in the spin-chain Hilbert space:
\beq{\label{3ptaswick}
\langle \mathcal{O}_1\mathcal{O}_2\mathcal{O}_3\rangle=
\bra{{\bf 1}_{123}}\Big(\ket{\tilde{\mathcal{O}}_1}\otimes \ket{\tilde{\mathcal{O}}_2}\otimes \ket{\tilde{\mathcal{O}}_3} \Big)\bra{{\bf 1}_{123}}\Big(\ket{\mathcal{O}_1}\otimes \ket{\mathcal{O}_2}\otimes \ket{\mathcal{O}_3} \Big) \period
}
As in the previous case, the structure of the singlet state $\bra{{\bf 1}_{123}}$ is determined by the structure of the Wick contraction, which is depicted in figure \ref{Wickpicture}-$(c)$. Explicitly, it is given by
\beq{
\begin{aligned}
\bra{{\bf 1}_{123}}= \left(\prod_{i=1}^{L_{12}}\epsilon_{ab}\bra{a}_{L_1+1-i}^{(1)}\otimes \bra{b}^{(2)}_{i}\right)\otimes \left(\prod_{i=1}^{L_{23}}\epsilon_{ab}\bra{a}_{L_2+1-i}^{(2)}\otimes \bra{b}^{(3)}_{i}\right)\otimes \left(\prod_{i=1}^{L_{31}}\epsilon_{ab}\bra{a}_{L_3+1-i}^{(3)}\otimes \bra{b}^{(1)}_{i}\right)\period\nn
\end{aligned}
}
Here $L_k$ is the length of the operator $\mathcal{O}_k$ and $L_{ij}=(L_i+L_j-L_k)/2$ is the number of Wick contractions connecting $\mathcal{O}_i$ and $\mathcal{O}_j$.

Now taking into account the normalization factors correctly, we arrive at the following basic formula  for the structure constant\fn{See \cite{Tailoring1} for the origin of the prefactor in \eqref{eq-11}.}:
\beq{
C_{123}=\frac{\sqrt{L_1 L_2L_3}}{N_c} \bra{{\bf 1}_{123}}\Big(\ket{\tilde{\mathcal{O}}_1}\otimes \ket{\tilde{\mathcal{O}}_2}\otimes \ket{\tilde{\mathcal{O}}_3} \Big)\bra{{\bf 1}_{123}}\Big(\ket{\mathcal{O}_1}\otimes \ket{\mathcal{O}_2}\otimes \ket{\mathcal{O}_3} \Big)
\period\label{eq-11}
}
In the above,  $N_c$ denotes  the rank of the gauge group. 
\begin{figure}[t]
\begin{center}
\picture{clip, height=3.0cm}{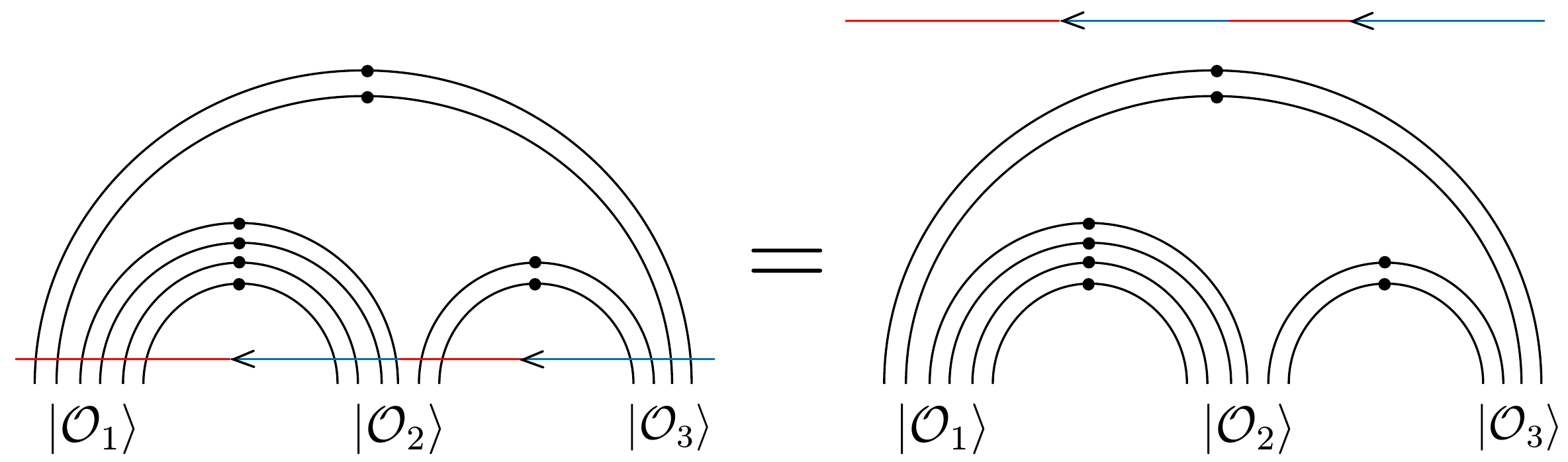}
\end{center}\vspace{-0.5cm}
\caption{The monodromy relation given in \eqref{monodexact}. The three-point function with and without the monodromy matrix are equal up to the prefactor, which we omit writing here. The red lines denote parts of the monodromy matrix with a $-i/2$ shift of the spectral parameter whereas the blue lines denote parts of the monodromy matrix with a $+i/2$ shift of the spectral parameter.}\label{monodpic}
\end{figure}

An important consequence of this formalism is the so-called monodromy relation, which is an identity connecting the structure constant with and without the insertion of the monodromy matrix. It was derived in \cite{KKN1,JKPS} and, for the ${\rm SU}(2)_R$ part, it reads\fn{Here we adopt the normalization of ${\rm L}_k(u)$ to be such that ${\rm L}_k(\infty) ={\bf 1}$, which is 
 slightly different from the one used in \cite{KKN1,JKPS}. The monodromy matrix in the present  normalization can be naturally identified with the monodromy matrix in the Landau-Lifshiz sigma model in the semi-classical limit.}
\beq{
\begin{aligned}\label{monodexact}
&\bra{{\bf 1}_{123}}\Big(\big(\Omega_1^{-} (u) \big)_{ij}\ket{\mathcal{O}_1}\otimes \big(\Omega_2^{+|-} (u) \big)_{jk}\ket{\mathcal{O}_2}\otimes \big(\Omega_3^{+} (u) \big)_{kl}\ket{\mathcal{O}_3}\Big)\\
 &= f_{123}(u)\delta_{il} \bra{{\bf 1}_{123}}\Big(\ket{\mathcal{O}_1}\otimes \ket{\mathcal{O}_2}\otimes \ket{\mathcal{O}_3}\Big)\period
\end{aligned}
}
Here $\Omega (u)$ is the monodromy matrix constructed from the Lax operator
\beq{
\begin{aligned}\label{defmonodhei}
\Omega (u) &\equiv {\rm L}_1 (u){\rm L}_2 (u)\cdots {\rm L}_{L} (u)\comma\\
 {\rm L}_k(u)&\equiv \pmatrix{cc}{1+iS_3^{k}/u&iS_{-}^{k}/u\\iS_{+}^{k}/u&1-iS_3^{k}/u}\comma
\end{aligned}
}
and the superscripts $\Omega^{\pm , +|-} (u)$ indicates the shift of the argument by $\pm i/2$ (for a precise definition, see figure \ref{monodpic}). The constant factor $f_{123}(u)$ is given by
\beq{
f_{123} (u) = \left(1+\frac{1}{u^2} \right)^{(L_1+L_2+L_3)/2}
}
The identity \eqref{monodexact} encodes infinitely many conservation laws for the structure constant. As we will see in section \ref{subsec:coherent}, the semi-classical limit of \eqref{monodexact} takes a form identical to the one at strong coupling and will play a key role in the subsequent analysis. 

 \subsection{On-shell Bethe states and polarization vectors\label{subsec:on-shellBethe}}
Before discussing the semi-classical limit, let us briefly explain how to characterize the general SU$(2)$ state in the double spin-chain representation.
 
 In the Bethe-ansatz approach, we construct the general eigenstates of the Hamiltonian by first considering the vacuum, which is typically taken to be $\tr \left(Z^{L} \right)$, and then introducing the magnons ($X$ or $\bar{X}$) with a set of rapidities satisfying the Bethe ansatz equation. An important property of such states, to be called the on-shell Bethe states, is that they are the highest weight state \cite{Faddeev} if the rapidities are all finite. In the double spin-chain representation, this translates to
 \beq{
 \tilde{S}_{+} \ket{\tilde{\mathcal{O}}}=0 \comma \quad S_{+} \ket{\mathcal{O}}=0\comma
  } 
  where $\tilde{S}_{+}$ ($S_{+}$) is the  raising operator for the total spin in ${\rm SU}(2)_{L}$ (${\rm SU}(2)_{R}$).

In the case of three-point functions, we cannot take all the states to be the ones constructed upon $\tr \left(Z^{L} \right)$ since such three-point functions vanish owing to the charge conservation. To study nonvanishing three-point functions, we have to consider the states constructed upon more general vacua, which can be obtained from  $\tr \left(Z^{L} \right)$ by the ${\rm SU}(2)_{L}\times {\rm SU}(2)_{R}$ rotations. As shown in \cite{KKN1}, such vacua can be characterized in terms of the {\it polarization vectors}\footnote{In the previous paper \cite{KK3}, they were called  polarization spinors.} $n$ and $\tilde{n}$, in the following way\fn{More explicitly, \eqref{rotatedvacuum} reads
\beq{
\tr \left((n_1\tilde{n}_1 Z + n_2 \tilde{n}_1 Y -n_1 \tilde{n}_2\bar{Y}+n_1\tilde{n}_2 \bar{Z})^{L} \right)\period
}
}:
\beq{\label{rotatedvacuum}
\tr\left( (\tilde{n}^{\tilde{a}} n^{a} \Phi_{\tilde{a}a})^{L}\right)\qquad  (\tilde{a},a=1,2)\period
}
The highest weight condition satisifed by the on-shell Bethe states constructed upon this rotated vacuum reads
\beq{\label{rotatedHW}
\tilde{S}^{\prime}_{+} \ket{\tilde{\mathcal{O}}}=0 \comma \quad S_{+}^{\prime} \ket{\mathcal{O}}=0\comma
 } 
 where $S_{+}^{\prime}$ and $\tilde{S}_{+}^{\prime}$ are rotated generators given by
 \beq{
 \begin{aligned}
 \pmatrix{cc}{\tilde{S}_3^{\prime} &\tilde{S}_{-}^{\prime}\\\tilde{S}_{+}^{\prime}&-\tilde{S}_3^{\prime}} &=\tilde{N}^{-1} \pmatrix{cc}{\tilde{S}_3 &\tilde{S}_{-}\\\tilde{S}_{+}&-\tilde{S}_3}\tilde{N}\comma\\
 \pmatrix{cc}{S_3^{\prime} &S_{-}^{\prime}\\S_{+}^{\prime}&-S_3^{\prime}} &=N^{-1} \pmatrix{cc}{S_3 &S_{-}\\S_{+}&-S_3}N\comma
 \end{aligned}\label{defSprime}
 }
 with
 \beq{
 N=\pmatrix{cc}{n^1 &-n^2\\n^2&n^1}\comma \quad \tilde{N}=\pmatrix{cc}{\tilde{n}^1 &-\tilde{n}^2\\\tilde{n}^2&\tilde{n}^1}\period
 }
The highest weight condition \eqref{rotatedHW} will play an important role when deriving the expression for the semi-classical structure constant in section \ref{subsec:anglewron}.

\subsection{Coherent-state representation and the semi-classical limit of $C_{123}$\label{subsec:coherent}}
We will now study the semi-classical limit of the expression \eqref{eq-11} for $C_{123}$. 
Unlike the previous methods \cite{Tailoring1,Foda, Kostov},  where one first evaluates this quantity exactly and then take the semi-classical limit, we shall take the semi-classical limit at the outset by deriving a path integral representation of  the structure constant and 
applying the saddle-point method.  This scheme will be seen to be  valuable  as a novel computational method universally applicable for a large class of SU(2) three-point functions, including the cases which previously could not be treated easily. 
Actually  the more important aspect of this method is that 
it  reveals  a cognate structure between 
 the weak coupling computation under consideration and the strong coupling 
 counterpart performed in \cite{KK3}, as we shall see. 

The semi-classical limit of our interest is a sort of the continuum limit of the Heisenberg spin chain. More precisely, it is the following scaling limit,
\beq{
L\to \infty \comma \quad M\to \infty \comma \quad L p , L/M :\text{ fixed}\period
}
Here $L, M$ and $p$ are, respectively,  the length of the spin chain,  the number of magnons and the momentum of each magnon. As such it is efficiently described by some continuous field along the chain, which should provide a representation of SU(2). The so-called coherent state representation is ideal for such a purpose. It is a representation realized on the coset space SU(2)$/$U(1), which is isomorphic to a unit sphere $S^2$.  As briefly reviewed in Appedix A,  a coherent state representation  for a single spin $1/2$ state can be taken to be 
\begin{align}
\ket{{\tt n}} &= 
\exp \left( i\theta {{\tt n}_0 \times {\tt n} \over  |{\tt n}_0 \times {\tt n}|} \cdot \vec{S}
\right) \upket = \cos{\theta \over 2} \upket -e^{i\phi} \sin{\theta\over 2} \ket{\downarrow} \comma 
\end{align}
where  ${\tt n}_0=(0,0,1)$ is a unit vector in the $z$ direction and ${\tt n} 
 = (\sin\theta \cos\phi, \sin\theta \sin\phi, \cos\theta)$ is  a unit vector in 
 a general direction. To express $C_{123}$ in this basis, we just need to insert the completeness relation
 \beq{
 1=\int \mathcal{D} {\tt n}\, \ket{\tt n}\bra{\tt n}\qquad \left( \mathcal{D} {\tt n} \equiv d^3 {\tt n}\, \delta({\tt n}^2-1)\right)\comma
 }
 to each inner product in \eqref{eq-11}. As a result, we obtain the following path-integral expression:
 \beq{
\begin{aligned}
C_{123}&=\frac{\sqrt{L_1L_2L_3}}{N_c}{\tt Left}\times {\tt Right} \comma \\
{\tt Left}&=\int \mathcal{D}\vec{\tilde{{\tt n}}}_1\mathcal{D}\vec{\tilde{{\tt n}}}_2\mathcal{D}\vec{\tilde{{\tt n}}}_3 e^{-S[\vec{\tilde{{\tt n}}}_1,\vec{\tilde{{\tt n}}}_2,\vec{\tilde{{\tt n}}}_3]}
\tilde{\Psi}_1 [\vec{\tilde{{\tt n}}}_1]\tilde{\Psi}_2 [\vec{\tilde{{\tt n}}}_2]\tilde{\Psi}_3 [\vec{\tilde{{\tt n}}}_3] \comma \\
{\tt Right}&=\int \mathcal{D}\vec{\tt n}_1\mathcal{D}\vec{\tt n}_2\mathcal{D}\vec{\tt n}_3 e^{-S[\vec{\tilde{{\tt n}}}_1,\vec{\tilde{{\tt n}}}_2,\vec{\tilde{{\tt n}}}_3]}\Psi_1 [\vec{\tt n}_1]\Psi_2 [\vec{\tt n}_2]\Psi_3 [\vec{\tt n}_3] \period
 \end{aligned}
 }
 Here $\vec{\tilde{{\tt n}}}$ and $\vec{{\tt n}}$ denote a chain of coherent states 
 \beq{
 \ket{\vec{\tt n}} \equiv \ket{{\tt n}}_1\otimes \ket{{\tt n}}_2 \otimes \cdots \otimes \ket{{\tt n}}_{L}\period
 }
 $e^{-S[\vec{{\tt n}},\vec{{\tt m}},\vec{{\tt l}}]}$ is the overlap between the singlet and the coherent states
 \beq{
 e^{-S[\vec{{\tt n}},\vec{{\tt m}},\vec{{\tt l}}]} \equiv \bra{{\bf 1}_{123}}\left(\ket{\vec{{\tt n}}}\otimes \ket{\vec{{\tt m}}}\otimes \ket{\vec{{\tt l}}}\right)\comma
 }
 while the wave functions  $\tilde{\Psi}$ and $\Psi$ are defined by
 \beq{
\tilde{\Psi}_k [\vec{\tilde{{\tt n}}}_k]=\langle \vec{\tilde{{\tt n}}}_k|\tilde{\mathcal{O}}_k\rangle \comma \qquad \Psi_k [\vec{\tt n}_k]=\langle \vec{{\tt n}}_k|\mathcal{O}_k\rangle\period
 }
 
Now in the semi-classical limit, this expression can be well-approximated by the saddle-point of the integrand, which gives
\beq{
\begin{aligned}\label{semiclassicalpsi}
&C_{123}=\frac{\sqrt{L_1L_2L_3}}{N_c}\Big( e^{-S[\vec{\tilde{\tt n}}_1^{\ast},\vec{\tilde{\tt n}}_2^{\ast},\vec{\tilde{\tt n}}_3^{\ast}]}\tilde{\Psi}_1 [\vec{\tilde{{\tt n}}}^{\ast}_1]\tilde{\Psi}_2 [\vec{\tilde{{\tt n}}}^{\ast}_2]\tilde{\Psi}_3 [\vec{\tilde{{\tt n}}}^{\ast}_3]\Big)\Big( e^{-S[\vec{{\tt n}}_1^{\ast},\vec{{\tt n}}_2^{\ast},\vec{{\tt n}}_3^{\ast}]}\Psi_1 [\vec{\tt n}^{\ast}_1]\Psi_2 [\vec{\tt n}^{\ast}_2]\Psi_3 [\vec{\tt n}^{\ast}_3]\Big)\comma
 \end{aligned}
}
where $\vec{{\tt n}}_k^{\ast}$ $(\vec{\tilde{{\tt n}}}_k^{\ast})$ represents the saddle point of the $\mathcal{D}\vec{{\tt n}}_k^{\ast}$ $(\mathcal{D}\vec{\tilde{{\tt n}}}_k^{\ast})$ integral. Evidently, the result \eqref{semiclassicalpsi} factorizes into the ${\rm SU}(2)_L$ part and the ${\rm SU}(2)_R$ part. In the discussions in the following sections, we mainly focus on the ${\rm SU}(2)_R$ part since the computation in the  ${\rm SU}(2)_L$ part is similar.

Let us now study the semi-classical limit of the monodromy relation. Since the monodromy matrix is an $O(1)$ quantity, the insertion of the monodromy matrix does not affect the saddle point. Thus, in the semi-classical limit, we can replace the monodromy matrix, which is originally the quantum operator acting on the spin chain, with the classical value evaluated on the saddle point given in \eqref{semiclassicalpsi}. Furthermore, since we scale the spectral parameter as $u \sim L$ in the semi-classical limit, the shifts of the arguments in $\Omega^{\pm}$ etc.~become negligible and the factor $f_{123}$ can be approximated by unity. Therefore we arrive at the relation
\beq{\label{semiclassicalmonod}
\left. \Omega_1 (u) \Omega_2(u) \Omega_3 (u) \right|_{\rm saddle} = {\bf 1}\period
 } 
 Importantly, \eqref{semiclassicalmonod} has exactly the same form as the monodromy relation in the string sigma model. This allows us to transplant most of the crucial ingredients for the strong coupling computation as we shall see in the next section.
\subsection{$\ln C_{123}$ as the ``generating function'' of the angle variable\label{subsec:generating}}
Once we choose the operators $\calO_i$ of definite conformal dimensions 
for which to compute the three-point function, each part of the expression in \eqref{semiclassicalpsi} can be explicitly computed in principle with the judicious use of the integrability. This is indeed the approach taken in the previous study at strong coupling \cite{KK3}. However, in this brute-force method, we shall encoutner extremely complicated intermediate expressions, most of which should cancel in the final result. 
Therefore, below we shall devise an entirely different method,  which at the same time reveals the important meaning of  $\ln C_{123}$ as a whole. This approach also enables us to study the semi-classical states with arbitrary number of cuts in the spectral curve, unlike the method in \cite{KK3}, which was restricted to the so-called one-cut solutions.

The basic idea is to see how  $\ln C_{123}$ changes as we introduce 
 a small number of additional Bethe roots. In the semi-classical context, this 
 means to study the variation of the structure constant with respect to 
 the variation of the filling fraction\fn{The precise definition of the filling fraction will be given in section \ref{subsec:action-angle}.} $S_i^{(m)}$ given by 
 \beq{
\frac{\del \ln C_{123}}{\del S_i^{(m)}} \comma\label{before}
 }
 where the subscript $i$   labels  the  filling fraction for the different cut belonging to the same operator, while the superscript $(m)$ labels the  three different operators. 
 By ``integrating''  this  quantity, one can determine the ratio between the structure constant involving non-BPS operators and the one for three BPS operators, for which all the filling fractions vanish. 

Specifically, the change of the filling fraction produces the following two effects:
(i)\ A slight change of the  saddle point configuration $\vec{{\tt n}}^\ast$ and (ii)\ the direct small change of the wave functions $\Psi[S_i, \vec{{\tt n}}^\ast]$ due to $\delta S_i$. Actually, the contribution from (i) takes the form,
\beq{
\frac{\del \vec{{\tt n}}^{\ast}_m}{\del S_{i}^{(m)}} \frac{\delta \left. \ln C_{123}\right|_{\rm saddle}}{\delta \vec{{\tt n}}^{\ast}_m}\comma
}
and hence it vanishes owing to the saddle-point equation $\delta \left. C_{123}\right|_{\rm saddle}/\delta  \vec{{\tt n}}^{\ast}_m=0$.

Now from the general theory,   the wave function in the semi-classical limit  is given by the following WKB form
\beq{\label{WKBwave}
\ln \Psi \sim i \sum_{k}\int P_k dQ_k\comma
}
where in the present case $Q_k$'s  correspond to the coherent-state variables, $\vec{\tt n}$, and $P_k$'s   to their canonical conjugates. The right hand side of \eqref{WKBwave} can be regarded as the generating function of the canonical transformation. Therefore, by differentiating with respect to the filling fraction, which is known to be the conserved action variables, we obtain
\beq{
\frac{\del}{\del S_i^{(n)}}\ln \Psi  = i\frac{\del}{\del S_i^{(n)}}\sum_{k} \int P_k dQ_k 
=i\phi_i^{(n)}\comma \label{gencan} }
where $\phi_i^{(n)}$ are the angle variables conjugate to $S_i^{(n)}$. 
Putting altogether, we find that $\ln C_{123}$ plays the role of the ``generating 
 function'' giving the angle variable under the  variation of the  
 filling fraction and we get the simple formula 
\beq{
 \frac{\del \ln C_{123}}{\del S_i^{(n)}}= i\phi_i^{(n)}\period\label{eq-16}
  } 
Concerning this formula,  two comments are in order. 
First, as we have already indicated by the use of quotation marks, the quantity
 $\ln C_{123}$ is not  the generating function of the action variables in the  usual sense. The precise meaning is that at the saddle point it behavs as if 
 it were a generating function of the value of the angle variable under the variation of $S_i^{(m)}$. \\
The second comment concerns the normalization of the structure constant $C_{123}$ or rather the normalization of the operators making the 
 three-point function.  As it will be discussed  in the next section, in the general integral expressing  
  the angle variable $\phi_i$ in \eqref{formphii},  we will not specify the initial point of integration. Therefore  the expression \eqref{eq-16} is actually ambiguous as it stands. To fix this ambiguity, we require that the operators we use produce the  normalized two-point functions correctly. This can be achieved in practice by replacing the right hand side of \eqref{eq-16} by the difference between the angle variable for the three-point function and the one for the two-point function in the following way:
  \beq{\label{unambiguous}
  \frac{\del \ln C_{123}}{\del S_i^{(n)}} = i \left(\phi_i^{(n)}-\phi_{i,{\rm 2pt}}^{(n)} \right) \equiv i\varphi_i^{(n)}\period
  }
  Unlike \eqref{eq-16}, the expression \eqref{unambiguous} is entirely unambiguous and we will adopt thhis form in the rest of this article.
\section{Classical integrability of the Landau-Lifshitz model\label{sec:LL}} 
We shall now apply the general formalism developed in the previous section 
 more explicitly to the semi-classical limit of the Heisenberg spin chain. 
It is  well-known that such a limit gives rise to  so-called the Landau-Lifshitz 
model, a classically integrable 
field theory in $1+1$ dimensions.
\subsection{Landau-Lifshitz model,  its Lax pair and the monodromy matrix\label{subsec:LLbasic}}
\begin{figure}[tb]
\begin{center}
\picture{clip, height=5cm}{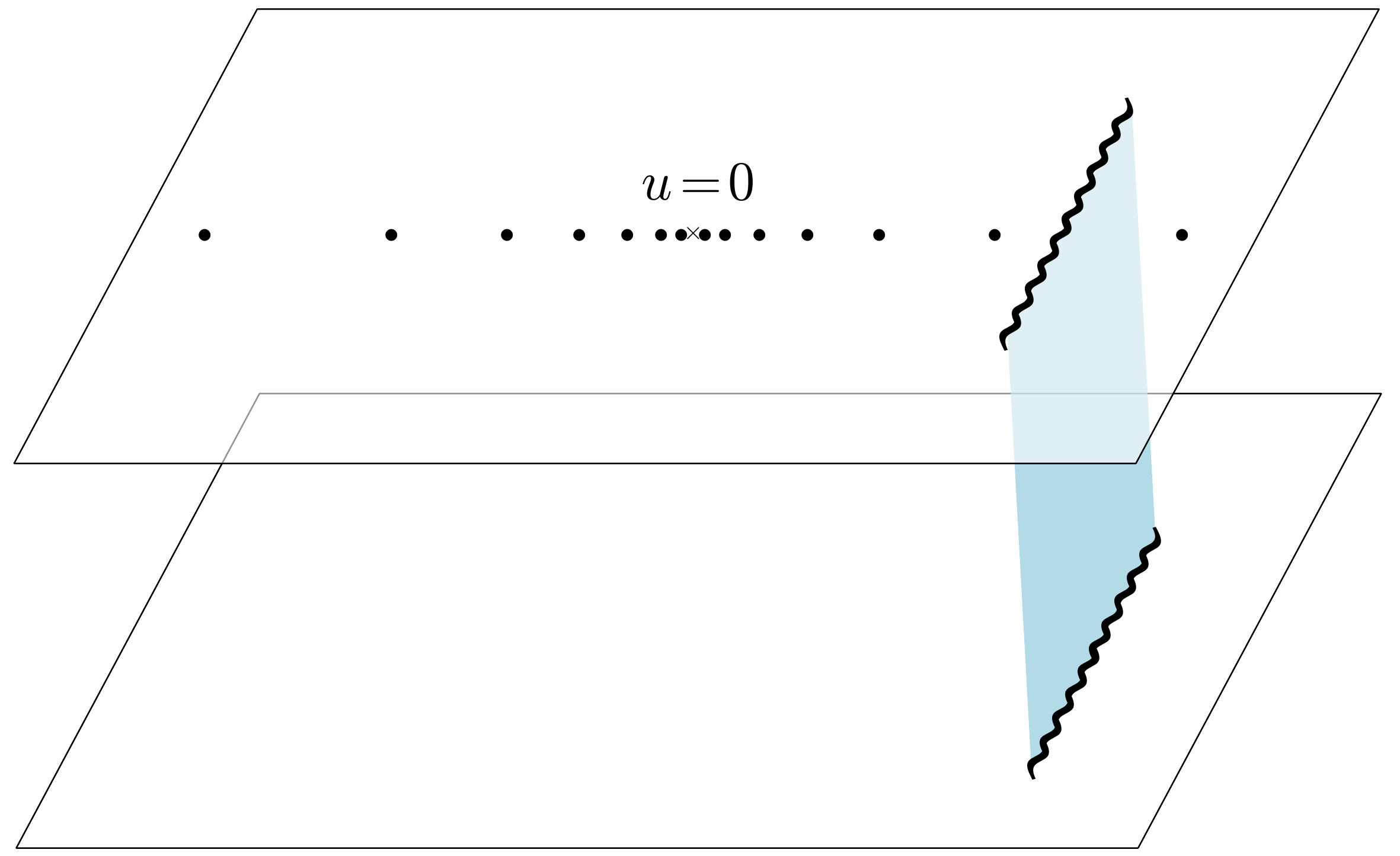}
\end{center}\vspace{-0.5cm}
\caption{The structure of the spectral curve at weak coupling. In general, it has several branch cuts and infinitely many singular points, denoted by black dots, which accumulate to $u=0$. The singular points can be regarded as degenerate branch points.}\label{curveweak}\vspace{-0.2cm}
\end{figure}

Let us briefly summarize the basic properties of the Landau-Lifshitz model and its 
 integrable nature. In the semi-classical limit, 
the coherent state variable $\vec{{\tt n}}(m, \tau)$,  where $m$ is  an  integer specifying the  position along the spin chain,  becomes 
 a continuous field $\vec{{\tt n}}(\sigma, \tau)$. It is convenient to take the range of 
 $\sigma $ to be  $0 \le \sigma \le L$, where $L$ is the length of the spin chain. 
The action is given by\footnote{A review of the derivation is provided in Appendix \ref{ap:a}.}
\beq{
S_{\rm LL}=\frac{1}{2}\int d\tau d\sigma \int_{0}^{1}ds\, \vec{{\tt n}}\cdot\left( \del_{\tau} \vec{{\tt n}}\times \del_{s}\vec{{\tt n}}\right) -\frac{g^2}{2}\int d\tau d\sigma \,\del_{\sigma}\vec{{\tt n}}\cdot \del_{\sigma}\vec{{\tt n}}\comma \label{ceq-1}
}
where $g=\sqrt{\lambda}/4\pi$ is the `t Hooft coupling constant. The first term in \eqref{ceq-1} is the Wess-Zumino term and 
 the $s$-dependence of $\vec{{\tt n}}$  is defined such that $\vec{{\tt n}}(s=1)=(0,0,1)$ and $\vec{{\tt n}}(s=0)=\vec{{\tt n}}$.
The equation of motion  obtained by varying the above action reads 
\beq{
\del_{\tau} \vec{{\tt n}} = 2g^2\vec{{\tt n}}\times \del_{\sigma}^2 \vec{{\tt n}}\period
}
One of the important features of this model is its classically integrability, whose clearest manifestation is the existence of the following Lax pair structure
\beq{
\begin{aligned}\label{lax-2}
&\left[\del_{\sigma}-J_{\sigma}\comma\del_{\tau}-J_{\tau}\right] =0\comma\\
&J_{\sigma}=\frac{i}{2 u}\vec{{\tt n}}\vec{\sigma} =\frac{i}{2 u}\pmatrix{cc}{{\tt n}_{3}&{\tt n}_{1}-i{\tt n}_{2}\\{\tt n}_{1}+i{\tt n}_{2}&-{\tt n}_{3}}\comma\quad
&J_{\tau}=\frac{2ig^2}{ u^2}\vec{{\tt n}}\vec{\sigma}+\frac{2ig^2}{ u}\left( \vec{{\tt n}}\times \del_{\sigma}\vec{{\tt n}}\right) \vec{\sigma}\comma 
\end{aligned}
}
 where  $u$ is   the spectral parameter. 
From the above Lax pair, one can construct the monodromy matrix in the 
 usual way\fn{The monodromy matrix defined here can be identified with the semi-classical limit of the monodromy matrix the Heisenberg spin chain \eqref{defmonodhei}.}:
\beq{
\Omega (u) \equiv  {\rm P}\exp \left( \int_{0}^{L} d\sigma J_{\sigma} \right)\period\label{asympt0} 
}
As in the case of the integrable string sigma model, one defines  the quasi-momentum $p(x)$ as the logarithm of the eigenvalue of the monodromy matrix:
\beq{
\Omega (u)\sim \pmatrix{cc}{e^{ip(u)}&0 \\0&e^{-ip(u)}}\period\label{asymptinf}
}
The asymptotic properties of the quasi-momentum at $u=0$ and $u=\infty$ 
can be easily obtained from the above definitions and contain useful information:
Its residue at $u=0$ is related to the length of the spin chain \cite{KMMZ} as 
\beq{\label{zerobehavior}
p(u)= -\frac{L}{2u}+ O(1)\comma 
}
while the leading behavior at infinity provides the information of the number $M$ of  magnon excitations of the system:
\beq{
p(u)=\frac{2M-L}{2u}+O(u^{-2})\period \label{pinfx}
}
The spectral curve is defined from the monodromy matrix as
\beq{
\det \left(y-\Omega (u) \right)= (y-e^{ip (u)})(y-e^{-ip (u)})=0\period
}
Owing to the singular behavior of the quasi-momentum \eqref{zerobehavior}, the spectral curve contains an infinite number of points satisfying $e^{2i p(u^{\ast})}=1$. Such points are called the singular points and can be regarded as the infinitesimal branch cut \cite{Vicedo1,Vicedo2} (see also figure \ref{curveweak}).

Now, it is well-known that the information contained  in  the non-linear Lax equation 
can be recovered by the simultaneous solution  of the auxiliary linear problem given by 
\begin{align}
(\del_\sig -J_\sig) \psi &= 0 \comma \qquad (\del_\tau -J_\tau) \psi =0 \period
\end{align}
In particular, for the $i$-th spin chain, the solutions which are at the same time 
the eigenfunctions of the monodromy matrix $\Omega_i$ with eigenvalues 
 $e^{\pm i p_i(u)}$ will be denoted by $i_\pm$, \ie 
\begin{align}
\Omega_i  i_\pm &= e^{\pm ip_i} i_\pm \comma 
\end{align}
 and they will be of great importance in what follows. 
\subsection{Action-angle variables\label{subsec:action-angle}}
As was already indicated in section \ref{subsec:generating}, the concept of action-angle variables 
plays an essential role in the computation of the structure constant. 
Therefore in this subsection, with the use of  the method of Sklyanin\cite{Sklyanin} we shall 
construct the action-angle variables for the Landau-Lifshitz model. 

First, we must compute the Poisson bracket between the elements of the 
 monodromy matrix, which is characterized by the classical r-matrix in the form 
\beq{
\{\Omega (u)\,\overset{\otimes}{\comma} \,\Omega (v) \} = \left[ \Omega (u)\otimes \Omega (v)\comma \mathbf{r}(u-v)\right]\period
\label{defrmatrix}
}
In the case of the Landau-Lifshitz sigma model, since the quantum R-matrix $R(x)$  for the XXX spin chain is well-known, a quick way\fn{For the first-principle derivation of the classical r-matrix, see Appendix \ref{ap:b}.} to obtain the classical r-matrix 
 is to take the classical limit of $R(x)$. Explicitly, we obtain 
\beq{
&R(u)=\mathbb{I}+i\frac{\mathbb{P}}{u}\longmapsto \mathbb{I}+ i \mathbf{r}(u)\comma \nn\\
&\Rightarrow\quad \mathbf{r}(u)=\frac{\mathbb{P}}{ u}\comma \label{rmatrix}
}
where $\mathbb{I}$ is the identity operator and $\mathbb{P}$ is the permutation operator. Using the form \eqref{rmatrix} in \eqref{defrmatrix}, one can 
 obtain the explicit form of the Poisson bracket between the individual 
 components of the monodromy matrix, which are denoted as usual in the form 
\beq{
\Omega (u)\equiv \pmatrix{cc}{\mathcal{A}(u)&\mathcal{B}(u)\\\mathcal{C}(u)&\mathcal{D}(u)}\period
}
The resulting Poisson brackets obtained in  this way are displayed in Appedix C. 

We now describe the Sklyanin's approach\cite{Sklyanin} for the construction 
 of the action-angle variables. 
Consider the  eigenfunctions  $\psi_\pm(u)$ of the monodromy matrix $\Omega(u)$, with eigenvalues $e^{\pm i  p(u)}$, \ie 
 defined by 
\beq{
\Omega (u) \psi_\pm (u;\tau)=e^{\pm i p(u)}\psi_{\pm} (u;\tau)\period 
\label{psipm}
}
Such eigenstates  are called Baker-Akhiezer vectors. Now an important information
 is encoded in the {\it normalized} Baker-Akhiezer vector $h(u;\tau)$ defined 
to be proportional to $\psi_+(u;\tau)$ and satisfying  the normalization condition
\begin{align}
\langle n\comma  h\rangle  \equiv \epsilon_{ab} n^{a} h^{b} = 1\comma \qquad h = \frac{1}{ \langle n\comma \psi_+ \rangle}
\psi_+ \comma \label{normvec}
\end{align}
Here $n=(n^1,n^2)^t$ is a constant vector with unit norm. In the original formalism by Sklyanin, $n$ can be arbitrary as long as it is independent of the spectral parameter. However, in the present context, we must choose it to be equal to the polarization vector diuscussed in section 2.2 in order to guarantee the highest weight property of the semi-classical wave function (see Appendix \ref{ap:c} for a detailed explanation).
  
For general solutions, there are infinitely many poles in $h(u,\tau)$, the position of which are denoted by $\gamma_i$, $i=1,2,\ldots$. Sklyanin observed  that 
 to each such pole $\ga_i$, which becomes a dynamical  variable through 
 the relations \eqref{psipm} and \eqref{normvec}, a canonical pair of variables are associated. 
Relegating the details of the derivation to Appendix \ref{ap:d}, 
the result is the following set of commutation relations 
\beq{
\{\gamma_i \comma \gamma_j \}=\{p(\gamma_i)\comma p(\gamma_j)\}=0\comma\quad -i\{\gamma_i \comma p(\gamma_j)\}=\delta_{ij}\period \label{eq-5}
}
where $p(\ga_i)$ is the quasi-momentum $p(u)$ with the substitution $u=\ga_i$. This shows that $\left(\gamma_i\comma - i p(\gamma_i) \right)$'s are canonical pairs of variables. 

Once the canonical pairs are obtained, one can easily construct the 
action variables, which should be identified with the conserved filling fractions
 $S_i$, as 
\beq{
S_i\equiv \frac{1}{2\pi i}\oint _{\mathcal{C}_i} p(u) du\period \label{eq-6} 
}
Here  $\mathcal{C}_i$ denotes  the $i$-th 
 branch cut.  

Now to construct the angle variables $\phi_i$ conjugate to $S_i$, we need to 
 to find the generating function of the type $F(\gamma_i\comma S_i )$,  which 
effects  the canonical transformation from $\left(\gamma_i\comma -ip(\gamma_i) \right)$ to the action-angle variables. Such a function is defined by 
the properties 
\beq{
&\frac{\del F}{\del \gamma_i}=- ip(\gamma_i) \comma
\qquad 
\frac{\del F}{\del S_i}=\phi_i\period
}
In the present context, the first equation should be viewed as defining the function $F$,  while the second equation should  be regarded as the definition of $\phi_i$.
Therefore, to determine $F$, we need to integrate the first equation with $S_i$ fixed. As the filling fractions are given by the integral of $p(u)$ on the spectral curve, fixing all $S_i$'s is equivalent to fixing the functional form of $p(u)$. Therefore, $F$ can be determined as
\beq{\label{generatingfunction}
F=-i\sum_i \int^{\gamma_i} p(u)du \period
}

Next we compute $\phi_i = \del F /\del S_i$. This requires changing $S_i$ with all the other filling fractions fixed. In the Heisenberg spin chain we started with, this corresponds to adding a small number of Bethe roots to the branch cut $\mathcal{C}_i$. As is clear from \eqref{asymptinf}, this addition of magnon inevitably changes the asymptotic behavior of $p(u)$ at $u=\infty$. Therefore, changing $S_i$ is precisely equivalent to adding to $p(u)du$ a one-form whose period integral is non-vanishing only for the cycle around $\mathcal{C}_i$ and the cycle at infinity. Such a one-form should be proportional to a holomorphic differential $\omega_i$ satisfying   the following properties:
\beq{
\oint_{\mathcal{C}_j} \omega_i =\delta_{ij}\comma \quad \int_{\infty^{+}}\omega_i =-1 = \oint_{\infty^-}(- \omega_i )\period
}
Here $\infty^{+}$ ($\infty^{-}$) denotes the infinity on the first (second) sheet.
Using such $\omega_i$, the partial derivative $\del F/\del S_i$ is expressed as\fn{This expression is a generalization of the so-called Abel map known in the theory of Riemann surfaces.}
\beq{
\phi_i = 2\pi \sum_j \int^{\gamma_j} \omega_i\period \label{formphii}
} 
\section{Angle variables and the Wronskians\label{sec:Wronskian}} 
In this section, we shall show that the angle variables constructed  in the 
 previous section can be expressed in terms of the skew-symmetric product, to 
 be called the Wronskians, of the solutions of the auxiliary linear problem corresponding to the Lax pair and of the polarization vectors. In what follows 
 the Wronskian of any  two-component vectors $\chi^a$ and $\phi^a$ is defined as
\begin{align}
\langle \chi, \phi \rangle &\equiv \chi^{a }\ep_{ab} \phi^{b} 
\period 
\end{align}
\subsection{Normalization of the solutions to the auxiliary linear problems\label{subsec:normalization}}
By using the Wronskian, we shall conveniently normalize the solutions $k_{\pm}$ of the auxiliary linear problem for the $k$-th spin chain as  
\begin{align}\label{normalizationcond121}
\langle k_+, k_- \rangle  =1 \period
\end{align}
In addition to this condition, it is consistent to require that the two solutions $k_{\pm}$ are related across the cut by
\beq{\label{normalizationcond12}
\left. k_{+}\right|_{\text{2nd-sheet}} = -i\left.k_{-}\right|_{\text{1st-sheet}}\comma \quad  \left. k_{-}\right|_{\text{2nd-sheet}} = -i\left.k_{+}\right|_{\text{1st-sheet}}\period
}
Then from \eqref{normalizationcond121} and \eqref{normalizationcond12}, one can show that $k_{\pm}$ develop  the following singularity at the branch points of the spectral curve
\beq{\label{branchsingular}
k_{\pm} \propto \frac{1}{\sqrt{u-u_b}}\quad (u\to u_b)\period
}
Let us briefly explain how this comes about. As the eigenvectors of the monodromy matrix are determined only up to an overall factor, we may first choose an eigenvector $k^{0}_{+}$ which remains non-singular even at the position of the branch points.   Then the other eigenvector $k^{0}_{-}$ can be obtained by the smooth analytic continuation of $k_{+}^{0}$ to the second sheet, since upon this operation the quasimomentum $p_k(u)$ flips sign and hence the eigenvalue changes
 from $e^{ip_k(u)}$ to $e^{-ip_k(u)}$.  By this definition, $k_{\pm}^0$ clearly satisfy the following relations:
\beq{
\left. k^0_{+}\right|_{\text{2nd-sheet}}=\left. k^0_{-}\right|_{\text{1st-sheet}} \comma \quad \left. k^0_{-}\right|_{\text{2nd-sheet}}=\left. k^0_{+}\right|_{\text{1st-sheet}}\period
}
Now let us normalize these two eigenvectors so that they satisfy the normalization condition \eqref{normalizationcond121}. This can be achieved by the rescaling
\beq{\label{rescalingk}
k_{+} \equiv \frac{1}{\sqrt{\langle k_{+}^{0}\comma k_{-}^0\rangle}}k_{+}^0\comma \quad k_{-} \equiv \frac{1}{\sqrt{\langle k_{+}^{0}\comma k_{-}^0\rangle }}k_{-}^0\period
}
Since two eigenvectors $k_{\pm}^{0}$ become degenerate at the branch points, $\langle k_{+}^{0}\comma k_{-}^0\rangle$ has a simple zero at such points. This yields the singularity structure given in \eqref{branchsingular}. 

Note that the aforementioned conditions do not completely fix the normalization, since we can always ``renormalize'' the eigenvectors as
 \beq{\label{renormalizenorm}
 k_{+}\to c(u) k_{+}\comma  \qquad k_{-}\to  k_{-}/c(u)\comma
 } 
 without violating the conditions \eqref{normalizationcond121} and \eqref{normalizationcond12}, if the function $c(u)$ satisfies
\beq{
\left. c(u)\right|_{\text{1st-sheet}}=\left.\frac{1}{c(u)}\right|_{\text{2nd-sheet}}\period
}
In section \ref{subsec:anglewron}, we will utilize this freedom to express the angle variable in terms of the Wronskians.
\subsection{Separated variables for two-point functions and orthogonality\label{subsec:orthogonality}}
In order to obtain the formula for the difference of the angle variables 
 appearing in \eqref{unambiguous} in terms of  appropriate Wronskians, 
we must first clarify  the structure of the separated variables 
 on the two-sheeted spectral curve. Similar information was crucial also 
 in the case of the strong coupling, treated by the string theory representation. 
In that case,  certain assumptions on the  analyticity as a function  on the string worldsheet helped determine some important structure. However, in the present case
 there is no worldsheet and we must devise a different logic  to get a handle on 
 the structure of the separated variables. 

Before delving  into the discussion of the case of the three-point function, 
 it is necessary  to understand in detail the separated variables for the two-point 
 functions. It will turn out that the logic that we shall employ is of such a  general 
 validity  that it can also be applied  for strong coupling, as well as for the 
weak coupling that we are analyzing. 

Let us consider the norm of a physical spin-chain state $\bracket{\Psi}{\Psi}$ 
(or equivalently a two-point function) and perturb one of the states by adding a 
 small number of Bethe roots to produce  the inner product  $\bracket{\Psi}{\Psi +\delta\Psi}$. Clearly $\langle \Psi | \Psi+\delta \Psi\rangle$ should be  non-vanishing for a general perturbation.  However, when the perturbed state is such 
  that it becomes on-shell again,  $\langle \Psi | \Psi+\delta \Psi\rangle$ must vanish  because of the orthogonality of  different eigenstates of the 
spin Hamiltonian. Therefore we have 
\beq{
\langle \Psi | \Psi+\delta \Psi\rangle =0 \comma \quad \text{if $|\Psi\rangle$ and $|\Psi +\delta \Psi\rangle$ are on-shell}\period\label{eq-18}
}
It should be emphasized  that this is an exact quantum statement. 

Now we perform the same type of perturbation in the semi-classical regime. 
Specifically, consider the norm $\bracket{\Psi}{\Psi}$ of a semiclassical on-shell state and perturb only the ket state $\ket{\Psi}$ by adding a small cut at the position of one of the singular points\fn{As discussed in \cite{GV,GNV}, the on-shell perturbation of the classical solution corresponds to the insertion of an infinitesimal cut at singular points.}, $u_{\ast}$, which corresponds to adding a small number of Bethe roots.  
When the added cut is small enough, the log of this quantity (normalized by the 
original norm) can be expressed as 
\begin{align}
\ln \left( {\bracket{\Psi}{\Psi+\delta \Psi} \over \bracket{\Psi}{\Psi}}
\right) \simeq {\del \ln  \bracket{\Psi}{\Psi'} \over \del S_{u_{\ast}}}
\Bigg|_{\Psi'=\Psi} \delta S_{u_{s}}\comma \label{addcut}
\end{align}
where the derivative with respect to $S_{u_{\ast}}$ acts only on $\ket{\Psi'}$. 
We have denoted the action variable associated with the degenerate cut at the singular point $u_{\ast}$ by 
$S_{u_{\ast}}$ and $\delta S_{u_{\ast}}$ denotes the filling fraction corresponding to the small cut  added. Since the state $\ket{\Psi}$ is semiclassical, we can evaluate the quantity $\del \bracket{\Psi}{\Psi'} /\del S_{u_{\ast}}$ using the saddle 
 point approximation. This operation is exactly the same as the one performed 
 on $\ln C_{123}$ previously, and taking into account  the saddle point equation itself the contribution that remains is 
\begin{align}
{\del \ln  \bracket{\Psi}{\Psi'} \over \del S_{u_{\ast}}}
\Bigg|_{\Psi'=\Psi} 
= i\phi_{u_{\ast}} \comma \label{anglevaratxast}
\end{align}
where $\phi_{u_{\ast}}$ is the angle variable evaluated on the unperturbed 
state. As the small cut added in this regime is actually made  of some number 
$m$ of on-shell Bethe roots, with the positive integer $m$ being of $\calO(1)$, 
we can identify $\delta S_{u_{\ast}}$ as $m$ and hence  \eqref{addcut}
together with \eqref{anglevaratxast} can be written as\fn{Note that in the semi-classical limit, anything which does not scale as the length of the chain $L$ 
can be regarded as small numbers.}
\begin{align}
\ln \left( {\bracket{\Psi}{\Psi+\delta \Psi} \over \bracket{\Psi}{\Psi}}
\right) \simeq i m \phi_{u_{\ast}} \period 
\end{align}
This means that when the perturbed state $\ket{\Psi+\delta \Psi}$ is again on-shell, 
according to  the exact quantum property \eqref{eq-18}, which must hold 
 in the semi-classical regime as well, we must have 
\begin{align}
{\bracket{\Psi}{\Psi+\delta \Psi} \over \bracket{\Psi}{\Psi}}
\simeq e^{im \phi_{u_{\ast}}} \rightarrow 0 \period  \label{vanishexp}
\end{align}
To examine this,  let us compute $\phi_{u_{\ast}}$ using the formula \eqref{formphii} applied to this case. We have 
\begin{align}
\phi_{u_{\ast}} = 2\pi \sum_j \int^{\ga_j} \omega_{u_{\ast}} \comma 
\end{align}
where $\ga_j$ are the separated variables and $ \omega_{u_{\ast}}$ is the holomorphic 
 differential which satisfies the following properties on the first and the second 
sheet. 
\begin{align}
\begin{aligned}
\text{1st sheet:}\qquad 
&\oint_{u_{\ast}} \omega_{u_{\ast}} =1 \comma \quad 
 \oint_{\infty} \omega_{u_{\ast}} =-1 \comma \quad  \oint_{\calC_i} \omega_{u_{\ast}} =0 \comma \\
& \omega_{u_{\ast}} \sim {1\over 2\pi i} {1\over u-u_{\ast} }\quad (u\to u_{\ast}) \comma  \\
\text{2nd sheet:}\qquad 
&\oint_{u_{\ast}} \omega_{u_{\ast}} =-1 \comma \quad 
 \oint_{\infty} \omega_{u_{\ast}} =1 \comma \quad  \oint_{\calC_i} \omega_{u_{\ast}} =0 \comma \\
& \omega_{u_{\ast}} \sim -{1\over 2\pi i} {1\over u-u_{\ast} }\quad (u\to u_{\ast})\period
\end{aligned}
\end{align}
This means that when one of the $\ga_j$'s  is at $u=u_{\ast}$ on the first sheet, 
$\phi_{u_{\ast}}$ behaves like 
\begin{align}
\phi_{u_{\ast}} \sim {1\over i } \ln (u-u_{\ast}) \quad (u\to u_{\ast})\comma
\end{align}
while if such a situation occurs on the second sheet, we have 
\begin{align}
\phi_{u_{\ast}} \sim -{1\over i } \ln (u-u_{\ast}) 
\quad (u\to u_{\ast})\comma
\end{align}
Thus in order for $e^{im \phi_{u_{\ast}}}$ to vanish as dictated by \eqref{vanishexp}, there must be a separated variable at each singular point on the first sheet. This information will be of prime importance in section 4.3 and 5.2: 
In section 4.3, it will provide the information of the zeros and poles  of the important quantity $\langle n, \psi_+^{{\rm 3pt}}\rangle$. Such analyticity  properties  will in turn be  imperative in determining those of the Wronskians, in terms of which the correlation functions will be expressed. 

We once again stress that  the preceding argument only uses the exact 
quantum property and its validity for the semi-classical regime as a special case, 
 it is applicable regardless of the strength of the coupling constant. 
\subsection{Angle variables expressed  in terms of  the Wronskians\label{subsec:anglewron}}
Using the properties discussed above, we now rewrite the angle variables in terms of the Wronskians. 
 
As described in section \ref{subsec:action-angle}, to construct the angle variables, we need to know the separated variables, which are associated to  the poles of the normalized eigenvector of the monodromy matrix given in \eqref{normvec}.  Clearly some of the zeros of $\langle n\comma \psi_{+}\rangle$, where $\psi_{+}$ is the {\it unnormalized} eigenvector, correspond to such poles. However,  $\langle n \comma \psi_{+}\rangle$ may contain additional zeros, which do not appear in the normalized eigenvector $\psi_{+}/\langle n \comma \psi_{+}\rangle$ since $\psi_{+}$ itself  becomes a zero-vector at such points and the ratio becomes finite. In addition to zeros, $\langle n \comma \psi_{+}\rangle$ in general has poles where $\psi_{+}$ itself diverges. Likewise,  these poles do not appear in the normalized eigenvector as they cancel between the numerator and the denominator. In what follows, we call these zeros and poles {\it spurious zeros and poles}. It is important to note that  the positions of the spurious zeros and poles may move if we change the normalization of the eigenvector as \eqref{renormalizenorm} whereas those of the separated variables do not.

With these properties in mind, let us study the analytic structure of the factor $\langle n\comma \psi_{+}^{\rm 2pt}\rangle$ for the two-point function. When the spectral curve contains $m$-cuts, there are $m$ ``dynamical'' separated variables, which evolve in the worldsheet time \cite{Vicedo1}. In addition to those, there are infinitely many non-dynamical separated variables which are trapped at the singular points on the first sheet of the spectral curve as discussed in the previous subsection. Both of them must correspond to zeros of $\langle n\comma \psi_{+}^{\rm 2pt}\rangle$. However, if we construct the solution explicitly using the finite gap method \cite{Vicedo1}, we do not find infinitely many zeros corresponding to the nondynamical separated variables. This is because $\langle n\comma \psi_{+}^{\rm 2pt}\rangle$ has spurious poles, which cancel the zeros associated with those separated variables. Furthermore, it has a square-root singularity at the branch points as shown in \eqref{branchsingular}. Thus the divisor of $\langle n\comma \psi_{+}^{\rm 2pt}\rangle$ is given by\fn{The solution for the two-point function has moduli, and for generic values of the moduli $  \langle n\comma \psi_{+}^{\rm 2pt}\rangle$ can have other spurious zeros and poles. However, on physical grounds, we expect that it is possible to choose a solution for which such poles and zeros are absent (although the argument is not completely rigorous). For a more detailed discussion on this point, see Appendix \ref{ap:e}.}:
\beq{\label{logdertwo}
\left(  \langle n\comma \psi_{+}^{\rm 2pt}\rangle\right) = \sum_k \gamma_k^{\rm 2pt} -\sum_l s_l-\frac{1}{2}\sum_mb_{m}\period
 } 
Here $\gamma_k^{\rm 2pt}$'s  correspond to the separated variables (both dynamical and nondynamical), $s_l$'s denote the singular points on the first sheet, and $b_m$'s denote the branch points. 

 We now turn to the corresponding quantity for the three-point function $\langle n\comma \psi_{+}^{\rm 3pt}\rangle$. To compute the normalized three-point functions, it is convenient to use the same normalization for  the eigenvectors $\psi^{\rm 2pt}_+$  and  $\psi^{\rm 3pt}_{+}$. More precisely, we require $\psi_{+}^{\rm 3pt}$ (and therefore $\langle n\comma \psi_{+}^{\rm 3pt}\rangle$) to have the same spurious zeros and poles as $\psi_{+}^{\rm 2pt}$.  This can always be achieved by multiplying by an appropriate function of the spectral parameter as \eqref{renormalizenorm}. However, in that process, we often need to introduce extra spurious zeros and poles to $\langle n\comma \psi_{+}^{\rm 3pt}\rangle$ in order to make $c(u)$ in \eqref{renormalizenorm} to be single-valued on the spectral curve. Therefore, the general structure of  the divisor takes the following form:
 \beq{
 \begin{aligned}\label{ourfavorite}
 &\left( \langle n\comma \psi_{+}^{\rm 3pt}\rangle\right)=\sum_k \gamma_k^{\rm 3pt} -\sum_l s_l-\frac{1}{2}\sum_mb_{m}+\sum_{n} (\eta_{n}-\delta_{n})\period
 \end{aligned}
 }
Here $\gamma_k^{\rm 3pt}$ are the separated variables for the three-point functions whereas the last term ($\eta_n$ and $\delta_n$) correspond to the extra spurious zeros and poles alluded to above.

 Let us make two important remarks regarding \eqref{ourfavorite}. First, owing to the normalization condition $\langle \psi_{+}^{\rm 3pt}\comma \psi_{-}^{\rm 3pt}\rangle=1$, $\psi_{-}^{\rm 3pt}$ has zeros at $\delta_n$ and poles at $\eta_n$. Since $\psi_{\pm}^{\rm 3pt}$ are related to each other by \eqref{normalizationcond12}, $\eta_n $ and $\delta_n$ must satisfy
\beq{\label{holoetadel}
\eta_n =\hat{\sigma}\delta_n \comma
   }   
   where $\hat{\sigma}$ is the holomorphic involution, which exchanges two sheets of the Riemann surface.
Second, as \eqref{ourfavorite} shows, $\psi_+^{\rm 3pt}$ becomes singular at the singular points on the first sheet. This property plays a key role in the determination of the analyticity structure in section \ref{subsec:analyticity}.

Taking into account the analyticity structure discussed above, 
we now compute the right hand side of \eqref{unambiguous}, which is the shift of the angle variables for the three-point function 
relative to those of the two-point function. (In what follows, we suppress the indices distinguishing operators until the very end when we write down the final expression \eqref{eq-34}.) For this purpose, it is 
 useful  to introduce a one-form $df$ defined by 
\beq{
df=d\ln  \frac{\langle  n\comma \psi_{+}^{\rm 3pt}\rangle}{ \langle n\comma  \psi^{\rm 2pt}_{+}\rangle}
\comma }
the divisor of which  is given by 
\beq{
\left( df\right)=\sum_{k}\gamma_k^{\rm 3pt}-\gamma_k^{\rm 2pt}+\sum_n \eta_{n}-\delta_{n}\period
}
 Now, using the formula \eqref{formphii}, we can express the shift $\varphi_k$ as
 \begin{align}
\varphi_k &= 2\pi \sum_{j=1}^\infty \int_{\ga_j^{\rm 2pt}}^{\ga_j^{\rm 3pt}} \omega_k \period \label{deltaphik}
\end{align}
where $\omega_k$ satisfies
\beq{
\oint_{\mathcal{C}_j} \omega_k =\delta_{jk}\comma \quad \int_{\infty^{+}}\omega_k =-1 = \oint_{\infty^-}(- \omega_k )\period
}
This expression can be simplified further using the generalized Riemann bilinear identity\footnote{This 
 form is given in \cite{Vicedo3} and was used in a similar manner as below in \cite{KK2,KK3}.}, which reads 
\begin{align}
\int_Q^P \tilde{\omega}_{RS;k} &= \int_S^R \tilde{\omega}_{PQ;k} \period
\end{align}
Here  $\tilde{\omega}_{PQ;k}$ and $\tilde{\omega}_{RS;k}$ are normalized Abelian differentials  satisfying\fn{To make connection with the formulas in \cite{Vicedo3}, we need to choose the basis of the $a$-cycle as the cycles around the cuts except $\mathcal{C}_k$. Then \eqref{defabelian} coincides with the definition of the normalized Abelian differential of the third kind.}
\begin{align}\label{defabelian}
\oint_P \tilde{\omega}_{PQ;k} = 1 \comma \quad \oint_Q\tilde{\omega}_{PQ;k} = -1
\comma \quad \oint_{\mathcal{C}_j}\tilde{\omega}_{PQ;k} =-\delta_{jk} \period 
\end{align}
Since $\omega_k $ can be identified with $-\tilde{\omega}_{\infty^+\infty^-;k}$,  we can use the Riemann bilinear identity 
 to rewrite $ \varphi_k$ as 
\begin{align}
\varphi_k &= 2\pi \sum_{j=1}^\infty \int_{\ga_j^{\rm 2pt}}^{\ga_j^{\rm 3pt}} \omega_k = -2\pi \int_{\infty^-}^{\infty^+}
\sum_{j=1}^\infty \tilde{\omega}_{\ga_j^{\rm 3pt}\ga_j^{\rm 2pt};k} 
 = i \int_{\infty^-}^{\infty^+} ( df -e_k) \label{eq-30}
\end{align}
where the integration contour starts from $\infty^{-}$, goes through the cut $\mathcal{C}_k$ and ends at $\infty^{+}$, and $e_k$ is the one-form uniquely characterized by the following conditions:
\beq{
\begin{aligned}
&\left( e_k\right) =\sum_n (\eta_{n}-\delta_{n})\comma\qquad e_k (\hat{\sigma}u)=-e_k(u)\comma \qquad \oint_{\calC_{j}} e_k =0 \quad \text{for $j\neq k$}\period\label{eq-32}
\end{aligned}
}
Here the second property follows from \eqref{holoetadel}. 
Finally, substituting the definition of $df$ into \eqref{eq-30}, we obtain\fn{Here we used the relation that $\psi_{+}(\infty^{-})=\psi_{-}(\infty^{+})/i$ which follows from \eqref{normalizationcond12}.}
\beq{
\varphi_{k}
&=\left.i\ln \frac{\langle n\comma \psi^{\rm 3pt}_{+}\rangle\langle n\comma \psi^{\rm 2pt}_{-}\rangle}{\langle n\comma \psi^{\rm 3pt}_{-}\rangle\langle n\comma \psi^{\rm 2pt}_{+}\rangle}\right|_{x=\infty^{+}} -i\int_{\infty^{-}}^{\infty^{+}}e_{k}\period\label{eq-33}
}

This expression can be further rewritten in terms of the Wronskians by judicious use of the highest weight condition. To see this, recall that the on-shell states constructed upon the rotated vacuum with the polarization vector $n$ is annihilated by the operator $S^{\prime}_{+}$, given in \eqref{defSprime}. 
Such global charges can be read off from the asymptotic behavior of the monodromy matrix as
\beq{
\begin{aligned}
\Omega (u)&= {\bf 1}+\frac{i}{u}\pmatrix{cc}{S_{3}&S_{-}\\S_{+}&-S_{3}}+O(u^{-2})\\
&= {\bf 1}+\frac{i}{u}\,N\pmatrix{cc}{S^{\prime}_{3}&S^{\prime}_{-}\\S^{\prime}_{+}&-S^{\prime}_{3}}N^{-1}+O(u^{-2}) \period
\end{aligned}
}
where the second equality follows from \eqref{defSprime}.
This leads to the following asymptotic form of the monodromy matrix in the semi-classical limit:
\beq{
\left.\Omega (u)\right|_{\rm saddle}= {\bf 1}+\frac{i}{2u}\, N\pmatrix{cc}{L-2M &\ast\\0&-(L-2M)}N^{-1}+O(u^{-2}) \qquad (u\to \infty^{+})\period\label{eq-25}
}
Note that this is true both for two-point functions and multi-point functions. From \eqref{eq-25} and the asymptotic form of the quasi-momentum \eqref{pinfx}, we can determine the asymptotic behavior of the eigenvectors $\psi_{\pm}$ to be of the form
\beq{
\psi_{-}(\infty^{+})=a \,n \comma \qquad \psi_{+}(\infty^{+}) 
=-a^{-1} (i \sigma_2 n) + b \, n \comma \label{psi-psi+}
}
where $a,b$ are arbitrary and we have imposed the normalization condition $\langle \psi_+ \comma \psi_- \rangle=1$. In the special case of two-point functions, by the explicit construction based on the finite-gap method, we can check\fn{See also Appendix \ref{ap:e}.} that $a_{\rm 2pt}=1$. Thus the ratio appearing in \eqref{eq-33} can be evaluated as
\beq{
\left.\frac{\langle n\comma \psi^{\rm 3pt}_{+}\rangle\langle n\comma \psi^{\rm 2pt}_{-}\rangle}{\langle n\comma \psi^{\rm 3pt}_{-}\rangle\langle n\comma \psi^{\rm 2pt}_{+}\rangle}\right|_{x=\infty^{+}}=a_{\rm 3pt}^{-2}\period
}
For three-point functions, the same quantity can be extracted from different combinations of the Wronskians. For instance, it is easy to verify, using \eqref{psi-psi+}, that the following combination gives the $a_{\rm 3pt}^{-2}$ for the operator $\mathcal{O}_i$:
\beq{
\left. a_{\rm 3pt}^{-2}\right|_{\mathcal{O}_i}=\left.\frac{\langle n_i\comma n_j\rangle\langle n_k\comma n_i\rangle}{\langle n_j\comma n_k\rangle}\frac{\langle j_{-}\comma k_{-}\rangle}{\langle i_{-}\comma j_{-}\rangle\langle k_{-}\comma i_{-}\rangle}\right|_{x=\infty^{+}}\period
}

Thus, restoring the indices for the operators, we arrive at the final expression,
\beq{
\varphi_{k_i}^{(i)}=i\ln \left(\left.\frac{\langle n_i\comma n_j\rangle\langle n_k\comma n_i\rangle}{\langle n_j\comma n_k\rangle}\frac{\langle j_{-}\comma k_{-}\rangle}{\langle i_{-}\comma j_{-}\rangle\langle k_{-}\comma i_{-}\rangle}\right|_{x=\infty^{+}} \right)-i\int_{\infty^{-}}^{\infty^{+}}e_{k_i}^{(i)}\period\label{eq-34}
}
Here $e_{k_i}^{(i)}$ is a one form defined on the spectral curve of the $i$-th operator $\mathcal{O}_i$, which satisfies
\beq{
\begin{aligned}
\left(e_{k_i}^{(i)}\right)=\sum_n \eta^{(i)}-\delta_n^{((i)} \comma \qquad e_{k_i}^{(i)}(\hat{\sigma}_iu)=e_{k_i}^{(i)}(u)\comma\qquad \oint_{\calC_{s}^{(i)}} e^{(i)}_{k_i} =0 \quad \text{for $s\neq k_i$}\comma\label{defofnormekii}
\end{aligned}
}
where $\hat{\sigma}_i$ and $\mathcal{C}_s^{(i)}$ denote the holomorphic involution and the branch cuts respectively, for the spectral curve of $\mathcal{O}_i$.

Let us make a remark on the nature of the angle variables for the general  
 multi-cut solutions  that we are capable of dealing with in this paper, in comparison to the previous work\cite{KK3}, where only the one-cut solution could be 
studied. In that work, the only allowed perturbation 
is to vary the filling fraction associated with the unique cut and at the same time 
the one at infinity, \ie $S_\infty$,  in the opposite direction for consistency.
 This is why the angle variable conjugate 
 to the global charge $S_\infty$ showed up in the previous work. However, 
 in the more general case of multi-cut solutions, 
we have to specify the cut to be perturbed among many and the corresponding angle variable is not necessarily the one conjugate to the global charge  but the one  associated with the more general filling fraction. 
\section{Evaluation of the Wronskians}\label{sec:ev}
Now that we have expressed the structure constant in terms of Wronskians, our final task  is to evaluate these Wronskians. 
\subsection{Products of Wronskians from monodromy relation\label{subsec:WfromM}}
Let us recall that at strong coupling, the monodromy relation was of crucial importance and it allowed us to express certain products of Wronskians in terms of quasi-momenta \cite{JW,KK1,KK3}. Since the relation derived in \eqref{semiclassicalmonod} is identical in form to the one in that analysis, one can apply the same argument also to the present case.

First, take the basis in which  $\Omega_1$ is diagonal of the form  ${\rm diag}(e^{ip_1},e^{-ip_1})$. Since the eigenvectors of different monodromy matrices are related with each other as
\beq{
i_{\pm}=\langle i_{\pm},j_{-}\rangle j_{+}-\langle i_{\pm},j_{+}\rangle j_-\comma
}
$\Omega_2$ in this basis is given by
\beq{
\Omega_2 = {\rm M}_{12} \, {\rm diag} (e^{ip_2},e^{-ip_2})\, {\rm M}_{21}\comma 
}
where ${\rm M}_{ij}$ is the basis-transformation matrix defined by
\beq{
{\rm M}_{ij}=\pmatrix{cc}{-\langle i_{-},j_{+}\rangle&\langle i_{-},j_{-}\rangle\\\langle i_{+},j_{+}\rangle&\langle i_{+},j_{-}\rangle}\period
}
Now, using the relation  $\Omega_1 \Omega_2 \Omega_3=1$ \eqref{semiclassicalmonod},  we can express the trace of the monodromy matrix $\Omega_3$ as
\beq{
\tr\, \Omega_3 \left(=2 \cos p_3\right)=\tr \Omega_2^{-1}\Omega_1^{-1}\period
\label{use-of-monod} 
}
Substituting the explicit expressions of $\Omega_1$ and $\Omega_2$  above to the right hand side of \eqref{use-of-monod}, we get
\beq{
\cos p_3 = \cos (p_1-p_2) \langle 1_{+}, 2_{+}\rangle\langle 1_{-}, 2_{-}\rangle -\cos (p_1 +p_2) \langle 1_{+}, 2_{-}\rangle \langle 1_{-}, 2_{+}\rangle\period 
}
Combining this equation with the Schouten identity,
\beq{
1=\langle 1_{+}, 2_{+}\rangle\langle 1_{-}, 2_{-}\rangle -\langle 1_{+}, 2_{-}\rangle \langle 1_{-}, 2_{+}\rangle\comma
}
we can determine the products $\langle 1_{+}, 2_{+}\rangle\langle 1_{-}, 2_{-}\rangle$ and $\langle 1_{+}, 2_{-}\rangle \langle 1_{-}, 2_{+}\rangle$. Products of other Wronskians can be computed in a similar manner and the results 
can be summarized as 
\beq{\label{products}
\begin{aligned}
\sl{i_+,j_+}\sl{i_-,j_-}& =\frac{\displaystyle \sin \frac{p_i+p_j+p_k}{2} \sin \frac{p_i+p_j-p_k}{2}}{\sin p_i \sin p_j}\comma\\
\sl{i_+,j_-}\sl{i_-,j_+}& =\frac{\displaystyle \sin \frac{p_i-p_j+p_k}{2} \sin \frac{p_i-p_j-p_k}{2}}{\sin p_i \sin p_j}\comma
\end{aligned}
 } 
where the  cyclic permutation of $(1,2,3)$ should be applied to $(i,j,k)$.

\subsection{Analytic properties of the Wronskians\label{subsec:analyticity}}
Now to compute three-point functions, we need to know the individual Wronskians, not just their products \eqref{products}. For this purpose, below we need to determine the analytic properties of the Wronskians as a function of the spectral parameter. This knowledge will later be indispensable in  decomposing  the products into individual Wronskians.

Before proceeding, let us make one general remark: Since each set of eigenvectors $i_{\pm}$ live on a two-sheeted Riemann surface, Wronskians generally live on a $2^{3}$-sheeted Riemann surface. Each of these eight-fold  sheets will be 
denoted as $[\ast, \ast,\ast]$-sheet, where the $n$-th entry $\ast$ is written as ``$u$'' for the upper sheet of $p_n (u)$ and ``$l$'' when it refers to the lower sheet of $p_n (u)$. The determination of the analyticity on this eight-sheeted Riemann surface may at first sight seem a formidable  task. However, as the eigenvectors on different sheets are related with each other by \eqref{normalizationcond12}, once we know the analyticity of all the Wronskians on the $[u,u,u]$-sheet, the analyticity on the other sheets can be automatically deduced. For instance, the analyticity of $\sl{1_+, 2_+}$ on the $[l,u,u]$-sheet are the same as the analyticity properties of  $\sl{1_-, 2_+}$ on the $[u,u,u]$-sheet. Thus, in what follows, it suffices to  determine   the analyticity on the $[u,u,u]$-sheet.
 \subsubsection*{BPS correlators}
We first study the simplest possible case, namely  the three-point function of BPS operators. A distinct feature of such correlators is that the quasi-momenta do not contain any branch cuts. This simplifies the determination of the analyticity of Wronskians drastically, as we see below.
 \begin{figure}[t]
 \begin{center}
 \begin{minipage}{0.4\hsize}
 \begin{center}
  \picture{clip,height=3.3cm}{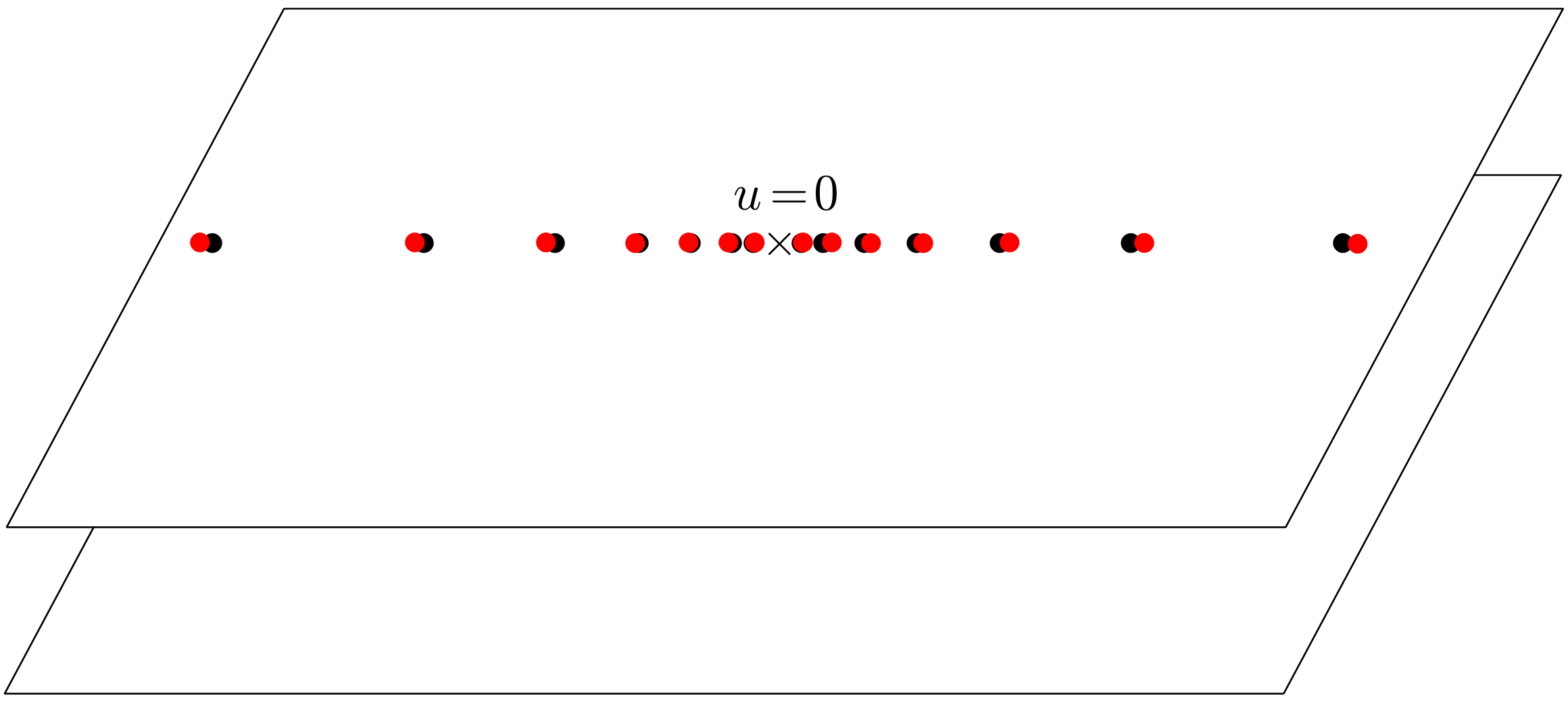}
  \end{center}
  \end{minipage}
  \hspace{20pt}
 \begin{minipage}{0.4\hsize}
 \begin{center}
  \picture{clip,height=3.3cm}{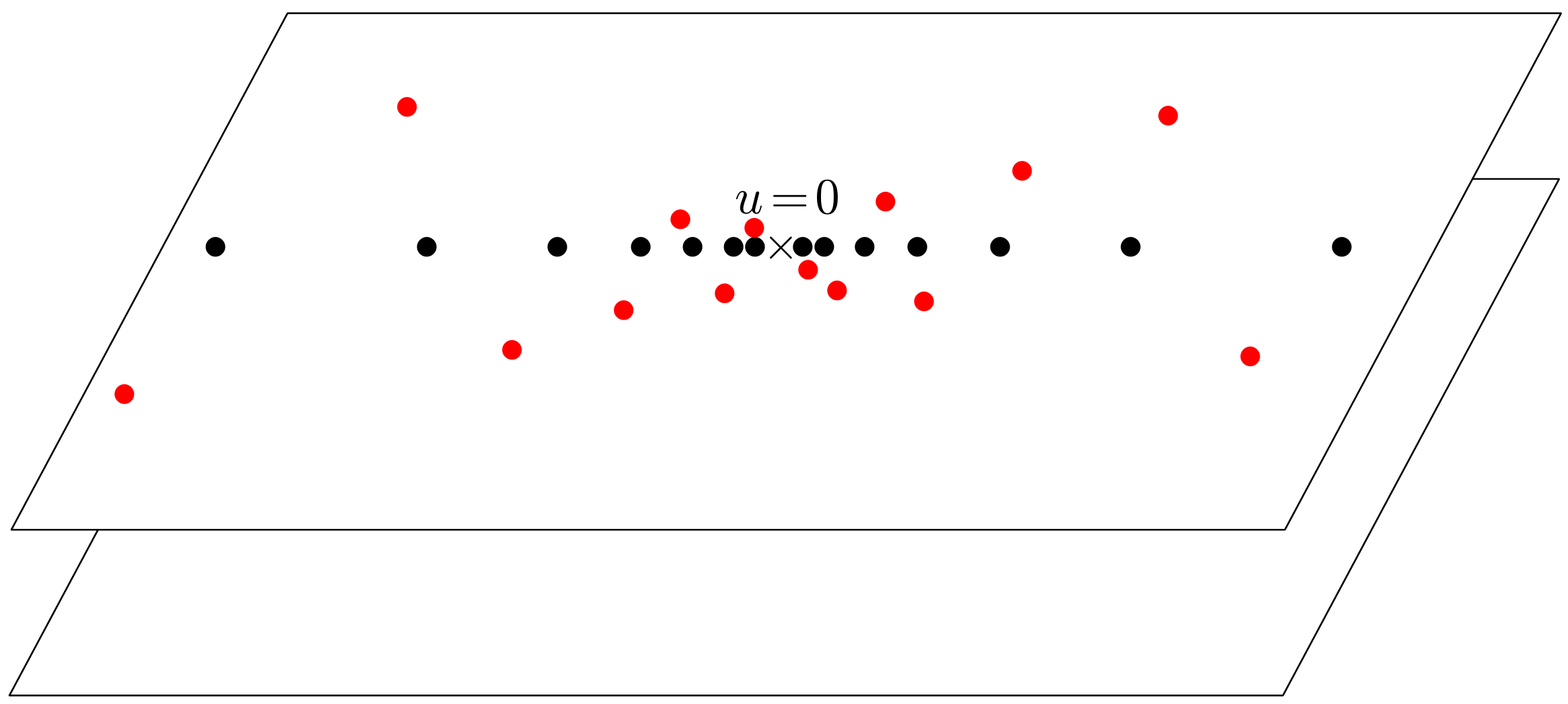}
  \end{center}
  \end{minipage}
\caption{The analytic structure of $\sl{i_+,j_+}$ on the $[u,u,u]$-sheet when all the operators are BPS. Left: In the limit $p_i\to p_j$ and $p_k\to 0$, the poles (denoted by black dots) and the zeros (denoted by red dots) are on top of each other, and the Wronskians are free of singularities. Right: Since there are no branch cuts, even after $p_k$ becomes nonzero and $p_i$ starts to differ from $p_j$, the zeros and the poles cannot move away from the $[u,u,u]$-sheet. } 
\label{BPSzero}
\end{center}
\vspace{-0.9cm}
\end{figure}

Let us first consider the Wronskians with the same signs, $\sl{i_{+}, j_{+}}$ and $\sl{i_{-}, j_{-}}$. As \eqref{products} shows, their products contain poles at $\sin p_i =0$ and $\sin p_j=0$, which are at the singular points of the spectral curve. In our normalization of the eigenvectors, the plus solutions $i_{+}$ and $j_{+}$ become singular at the singular points on the first sheet of the spectral curve (see the discussion below \eqref{ourfavorite}). This means that the poles on the $[u,u,u]$-sheet belong to $\sl{i_{+},j_{+}}$ (and not to  $\sl{i_{-},j_{-}}$). In addition to poles, the right hand side of \eqref{products} has zeros at $\sin (p_i+p_j+p_k)/2 =0$ and $\sin (p_i+p_j-p_k)/2 =0$. To determine which Wronskian contains these zeros, we first consider the limit where $p_k\to 0$ and $p_i\to p_j$. In this limit, the classical solution for the three-point function approaches to the one for  the two-point function. As mentioned  in section \ref{subsec:anglewron}, for the two-point function, the eigenvectors are nonsingular even at $\sin p_i =0$ and so are the Wronskians. This means that in this limit the zeros of the Wronskians must cancel the pole singularities. In order for such  cancellations to occur,  all the zeros on the $[u,u,u]$-sheet must belong to $\sl{i_{+}, j_{+}}$ when $p_k$ is very small. Now, since all the operators are BPS and there are no branch cuts connecting different sheets, those zeros cannot leave the  $[u,u,u]$-sheet even if we increase the value of $p_k$ (see figure \ref{BPSzero}). We therefore conclude that all the zeros on the $[u,u,u]$-sheet must always belong to $\sl{i_{+}, j_{+}}$, not to $\sl{i_{-}, j_{-}}$,  when the three operators are BPS. 
 
Next we consider the Wronskians with opposite signs $\sl{i_{+}, j_{-}}$ and $\sl{i_{-}, j_{+}}$. Also in this case, the determination of the poles are straightforward since they  are associated with the eigenvectors themselves. By the same reasoning as above, we conclude that the poles at $\sin p_i=0$ belong to $\sl{i_{+}, j_{-}}$ whereas the poles at $\sin p_j$ belong to $\sl{i_{-},j_{+}}$.  On the other hand, the determination of zeros is more complicated and we need to resort to the argument given in \cite{KK3}. As reviewed in Appendix \ref{ap:f}, it leads to the following general rules:
{\it 
\begin{enumerate}
\item When a factor $\sin \left(\sum_i \epsilon_i\, p_i /2\right)$ vanishes, the Wronskians which vanish are the ones among $\{1_{\epsilon_1},2_{\epsilon_2},3_{\epsilon_3}\}$ or the ones among $\{1_{-\epsilon_1},2_{-\epsilon_2},3_{-\epsilon_3}\}$. (Here $\epsilon_i$ takes $+$ or $-$.)
\item On the $[u,u,u]$-sheet, the Wronskians from the group which contains more than one $+$ eigenvectors all vanish whereas the Wronskians from the other group do not.
\end{enumerate}
}
\noindent Applying these rules, we can determine the analyticity on the $[u,u,u]$-sheet as given in tables \ref{tab1} and \ref{tab2}. As already mentioned,  the analyticity on other sheets can be deduced from the relations \eqref{normalizationcond12}.
\begin{table}[htb]
\begin{center}
\begin{tabular}{c||c|c||c|c}
&$1/\sin p_i$&$1/\sin p_j$&${\displaystyle \sin \frac{p_i+p_j+p_k}{2}}$&${\displaystyle \sin \frac{p_i+p_j-p_k}{2}}$\\\hline
$\sl{i_+,j_+}$&\checkmark&\checkmark&\checkmark&\checkmark\\
$\sl{i_-,j_-}$&&&&
\end{tabular}
\caption{The analytic properties of $\sl{i_+,j_+}$ and $\sl{i_-,j_-}$ on the $[u,u,u]$-sheet. The checked entries indicate existence of  poles/zeros. \label{tab1}}
\end{center}
\end{table}
\begin{table}[htb]
\begin{center}
\begin{tabular}{c||c|c||c|c}
&$1/\sin p_i$&$1/\sin p_j$&${\displaystyle \sin \frac{p_i-p_j+p_k}{2}}$&${\displaystyle \sin \frac{-p_i+p_j+p_k}{2}}$\\\hline
$\sl{i_+,j_-}$&\checkmark&&\checkmark&\\
$\sl{i_-,j_+}$&&\checkmark&&\checkmark
\end{tabular}
\caption{The analytic properties of $\sl{i_+,j_-}$ and $\sl{i_-,j_+}$ on the $[u,u,u]$-sheet. \label{tab2}}
\end{center}
\end{table}
 \subsection*{Extension to non-BPS}
\begin{figure}[tb]
\begin{center}
\picture{clip, height=3cm}{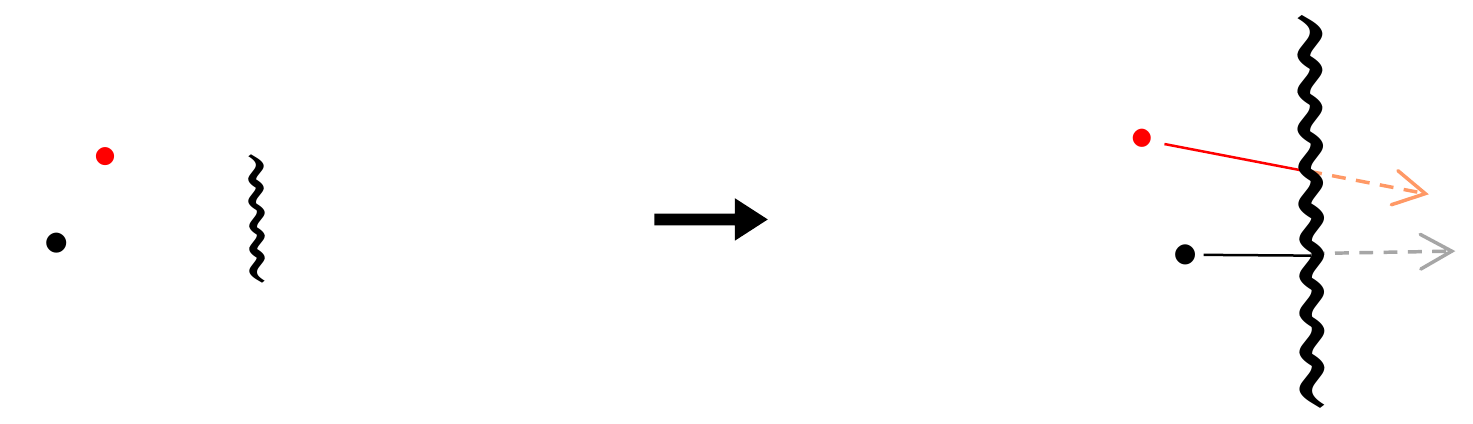}
\end{center}
\caption{The analyticity of the Wronskians for the non-BPS correlators. Under the continuity assumption, the analyticity remains the same as the one for the BPS correlators as long as the cuts are sufficiently small (left figure). As we increase the size of the cut, the zeros and the poles start to move and, at some point, they cross the cut and cause the change in the analyticity (right figure). Such effects affect the integration contours of the final result as we shall see in section \ref{subsec:finalweak}.}
\label{nonBPSzero}
\end{figure}

Now we turn to non-BPS correlators. Their analytic properties can be inferred from those  of the BPS correlators if we make the following physically reasonable assumption:
\begin{itemize}
\item[] {\small \bf Continuity Assumption}:\\
When all the branch cuts of the quasi-momenta $p_i (x)$ shrink to zero  size, the classical solution for the non-BPS correlator smoothly goes over to those for 
the BPS correlators.
\end{itemize}
This assumption implies in particular that the Wronskians for the BPS and the non-BPS cases are also smoothly connected. Now, let us consider the three BPS correlator discussed above and insert a very small cut to make it non-BPS. Because of the continuity assumption, the zeros and the poles of the Wronskians cannot move  to a different sheet as long as the cut is sufficiently small and therefore the analyticity of Wronskians for such non-BPS correlators must be the same as the one for the BPS correlators. If  we gradually increase the sizes of the cuts, at some point the zeros and the poles start crossing the branch cuts and move over  to a different sheet, leading to a change in the analytic property of the Wronskians. A simple way to take such effects into account is to first compute the correlators with small cuts and then analytically continue the final results with respect to the sizes of the cuts (see figure \ref{nonBPSzero}). This analytic continuation to larger cuts will be  discussed in section \ref{subsec:finalweak}, and  we will comment on how it affects the integration contours\fn{A similar phenomenon was observed in the context of one-loop corrections to the classical energy both at weak \cite{BBG} and strong coupling \cite{GV,GNV}. In both cases, as the sizes of the cuts become bigger, some of the physical excitations cross the cuts. At weak coupling, this leads to the change in the distribution of the Bethe roots. Nevertheless, the final result turns out to be a smooth function of the sizes of the cuts, and we expect that this is also the case for three-point functions.}. Thus, until then, we will restrict ourselves to the spectral curves with small cuts. 
\subsubsection*{Spurious zeros and poles}
In addition to the structures we discussed so far, there are extra spurious zeros and poles of the eigenvectors ($\eta_n$ and $\delta_n$ in \eqref{ourfavorite}). Owing to the normalization condition $\sl{i_+,i_-}=1$, whenever $i_{+}$ has such a zero $i_{-}$ has a pole, and vice versa. Thus, these extra zeros and poles cancel out if we consider the products of the Wronskians appearing on the left hand side of \eqref{products}. Nevertheless we should keep in mind that such poles and zeros are present as we solve for the individual Wronskians below. In the next section, we first subtract such zeros and poles from the Wronskians and study the Riemann-Hilbert problem for these subtracted quantities.
 
\subsection{Solving the Riemann-Hilbert problem\label{subsec:RH}}
Let us now use the analytic properties to set up and solve the Riemann-Hilbert problem to determine the Wronskians. Here we will only discuss $\sl{i_{+}, j_{+}}$ and $\sl{i_{-},j_{-}}$ since these are the Wronskians relevant for the computation of the structure constant. (The argument below can be straightforwardly generalized to other Wronskians but we will not elaborate on  it here.)

In what follows, we analyze the logarithmic derivative of the relation \eqref{products}, namely
\beq{\label{logder}
\begin{aligned}
&\del_u \ln \sl{i_{+}\comma j_{+}} + \del_u \ln \sl{i_{-}\comma j_{-}}\\
& = \del_u \ln \sin \frac{p_i+p_j-p_k}{2} + \del_u \ln \sin \frac{p_i+p_j+p_k}{2} -\del_u \ln\sin p_i -\del_x \ln \sin p_j\period
\end{aligned}
}
Since the Wronskians contain extra zeros and poles which are absent on the right hand side of \eqref{logder} as mentioned above, it is convenient to consider the following quantities from which extra zeros or poles are subtracted:
\beq{
\begin{aligned}
W_{++}^{ij}&=\del_{u} \ln \sl{i_{+},j_{+}}-e^{(i)}_{k_i}-e^{(j)}_{k_j}\comma\\
W_{--}^{ij}&=\del_{u} \ln \sl{i_{-},j_{-}}+e^{(i)}_{k_i}+e^{(j)}_{k_j}\period
\end{aligned}
 } 
Here $e^{(i)}_{k_i}$ is a one-form given by \eqref{defofnormekii} in section \ref{subsec:anglewron}. As explained there, it depends on the choice of the cut $\mathcal{C}^{(i)}_{k_i}$, which we are perturbing. In terms of $W_{\pm\pm}^{ij}$, \eqref{logder} can be written as
\beq{
\begin{aligned}\label{RH}
&W_{++}^{ij}+W_{--}^{ij}\\
& = \del_u \ln \sin \frac{p_i+p_j-p_k}{2} + \del_u \ln \sin \frac{p_i+p_j+p_k}{2} -\del_u \ln\sin p_i -\del_x \ln \sin p_j\period
\end{aligned}
}
This is the Riemann-Hilbert problem we need to solve.

To uniquely characterize the solution to the Riemann-Hilbert problem, we have to specify its period integrals in addition to its analyticity. In the case at hand,  the period integrals of $W_{\pm\pm}^{ij}$ are given by
\beq{
\begin{aligned}\label{propertyofW}
&\oint_{\mathcal{C}_s^{(i)}}W_{\pm\pm}^{ij}=0 \quad (s\neq k_{i})\comma\qquad \oint_{\mathcal{C}_s^{(j)}} W_{\pm\pm}^{ij}=0 \quad (s\neq k_{j})\period
\end{aligned}
}
These properties can be shown as follows: Since  $\sl{i_{\pm},j_{\pm}}$ is a single-valued function on the Riemann surface, the integral of $\del_u\ln\sl{i_{\pm},j_{\pm}}$ along any closed curve gives  $n \pi$ where $n$ is an integer. As we are considering  the spectral curves  with small cuts, which are continuously connected to the curves without cuts, $n$ must be zero in such a  case. Together with the property of $e_{k_i}^{(i)}$ given in \eqref{defofnormekii}, this leads to \eqref{propertyofW}. This  property will  be later utilized in checking  the correctness of the expressions  of $W^{ij}_{\pm\pm}$  we shall construct. 
\subsubsection*{Wiener-Hopf method}
\begin{figure}[tb]
\begin{center}
\picture{clip, height=2.5cm}{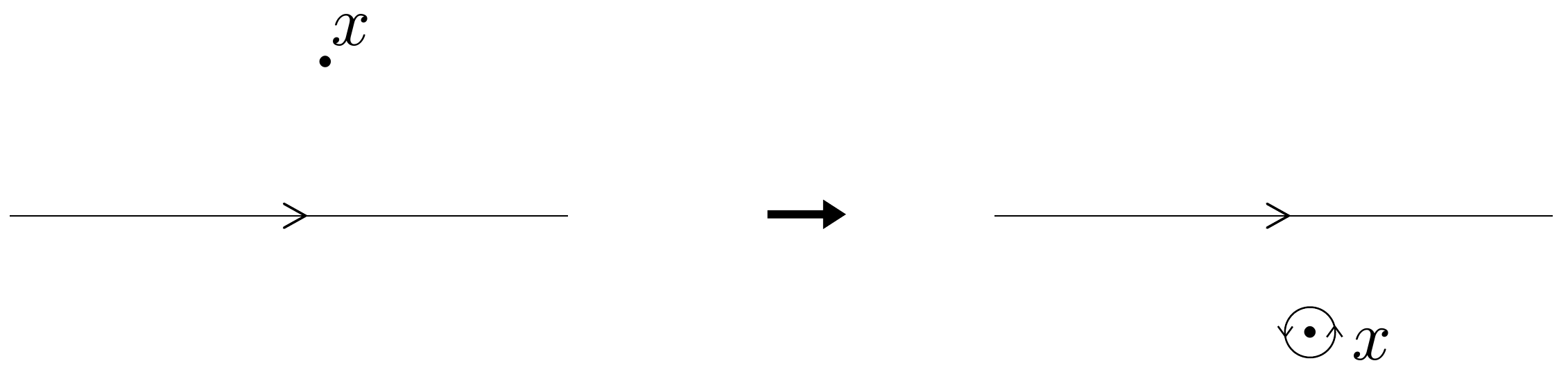}
\end{center}\vspace{-0.5cm}
\caption{Analytic continuation of the Wiener-Hopf integral. As a result of the analytic continuation, the integral picks up a pole at $x^{\prime}=x$. This yields the first term in \eqref{analyticcontiWiener}.}
\label{changecontour}\vspace{-0.3cm}
\end{figure}

Before solving \eqref{RH}, let us briefly review the standard Wiener-Hopf method, which decomposes a function into the part regular on the upper-half plane and the part regular on the lower-half plane. Suppose $f(x)$ is a function which decreases sufficiently fast at infinity and does not  have a pole on the real axis. Then $f(x)$ can be decomposed as $f(x)=f_+ (x)+f_- (x)$ where $f_+$ and $f_-$ are defined on the half planes by
\beq{
\begin{aligned}\label{defful}
f_+ (x)&= \int_{-\infty}^{\infty} \frac{dx^{\prime}}{2\pi i} \frac{1}{x^{\prime}-x}f(x^{\prime})&& ({\rm Im}\,x>0)\comma\\
f_- (x)&= -\int_{-\infty}^{\infty} \frac{dx^{\prime}}{2\pi i} \frac{1}{x^{\prime}-x}f(x^{\prime})&& ({\rm Im}\,x<0)\period
\end{aligned}
}
Using \eqref{defful}, it is easy to verify that $f_+$ is regular on the upper-half plane and $f_-$ is regular on the lower-half plane. When $x$ is not in the region specified in \eqref{defful}, we need to analytically-continue these formulas. This leads, for instance, to the following expression for $f_+(x)$ on the lower-half plane $({\rm Im}\,x<0)$
\beq{\label{analyticcontiWiener}
f_+(x)=f(x)+ \int^{\infty}_{-\infty}\frac{dx^{\prime}}{2\pi i} \frac{1}{x^{\prime}-x}f(x^{\prime})=f(x)-f_{-}(x)\period
}
Here the first term $f(x)$ is produced by  the integral along a small circle around $x^{\prime}=x$ as shown in figure \ref{changecontour}. An important feature of this method is that the contour of integration separates domains with different analyticity structures. This is true also in a more complicated situation where functions are defined on a multi-sheeted Riemann surface as in \eqref{RH}.

For later convenience, let us also present the version of \eqref{defFul} 
obtained by integration by parts:
\beq{
\begin{aligned}\label{defFul}
f_+ (x)&= -\int_{-\infty}^{\infty} \frac{dx^{\prime}}{2\pi i} \frac{1}{(x^{\prime}-x)^2} F(x^{\prime})&& ({\rm Im}\,x>0)\comma\\
f_- (x)&= \int_{-\infty}^{\infty} \frac{dx^{\prime}}{2\pi i} \frac{1}{(x^{\prime}-x)^2}F(x^{\prime})&& ({\rm Im}\,x<0)\period
\end{aligned}
}
Here $F(x)$ is the integral of $f(x)$, \ie,  $F(x)=\int^{x} f(x^{\prime})$. It is this form of the Wiener-Hopf decomposition that will  be generalized below 
in order  to deal with the multi-sheeted Riemann surface on which $p_i(x)$'s  
 are defined. 
\subsubsection*{Decomposition of the poles}
We first study the factors $\del_u \ln\sin p_i$ and $\del_u \ln\sin p_j$, which give rise to the poles of the Wronskians. Below we focus on $\del_u \ln\sin p_i$ since the case for  $\del_u\ln \sin p_j$ is completely similar. 

As in the standard Wiener-Hopf decomposition, we should be able to decompose it by considering a convolution integral whose contour separates the domains with different analyticity. As summarized in table \ref{tab1}, the poles of $1/\sin p_i$ belong to $\sl{i_+,j_+}$ when the rapidity is on the first sheet of $p_i$ while they belong to  $\sl{i_-,j_-}$ when it is on the second sheet of $p_i$. Obviously, these two regions are separated by the branch cuts of $p_i$ and so the contour should be taken to go around the cuts.  Now what we must properly deal with is the choice of the convolution kernel,  as we have a two-sheeted Riemann surface instead of a simple complex plane. The natural generalization of the kernel \eqref{defFul} in the present case is given by the  bidifferential characterized by the properties listed below,  which is often called the {\it Bergman kernel} in  physics literature\footnote{
This quantity is introduced by J. Fay\cite{Fay} as the bidifferential made from the prime form  and is called  ``the normalized bidifferential of the second kind" (see also 
 \cite{Mumford}). Although in mathematics the  Bergman kernel, strictly speaking,  refers to slightly broader notion,  we shall follow the physics nomenclature. We thank M. Jimbo and A. Nakayashiki for useful information on these matters.} . To define the Bergman kernel, we must first pick a  basis of cycles. Then, the Bergman kernel $B(p,q)$ is a differential in both $p$ and $q$ and is uniquely specified by the following properties;
{\it 
\begin{enumerate}
\item Symmetry: 
\beq{B(p,q)=B(q,p)\period\label{symBerg}}
\item Normalization:
\beq{\oint_{p\in a_j}B(p,q)=0\comma\label{normBerg}} where $\{a_j\}$ is the basis of $a$-cycles.
\item Analyticity: $B (p,q)$ is meromorphic in $p$ with only a double pole at $p=q$ with the following structure:
\beq{
B(p,q) \sim \frac{1}{2\pi i(\zeta(p)-\zeta(q))^2} d\zeta(p) d\zeta(q)\period\label{analyBerg}
}
Here $\zeta$ is a local coordinate around $p \simeq q$.
\end{enumerate}
}
\noindent
In addition to these properties, when the curve is hyperelliptic, it satisfies
{\it
\begin{enumerate}
\item[4.] Involution property:
\beq{B(\hat{\sigma}p,\hat{\sigma}q)=B(p,q)\label{Bholosym} \period} 
\end{enumerate}
}\noindent
This is because the kernel $B(\hat{\sigma}p,\hat{\sigma}q)$ satisfies all three properties listed above, which specify the Bergman kernel uniquely.
In the present case,  we can define the Bergman kernel for each of the three spectral curves and we denote them by 
\beq{
B_{k_i}^{(i)}(p,q)\qquad i=1,2,3\period
}
Here the subscript $k_i$ designates our choice of the basis of the cycles; namely we choose the $a$-cycles as the cycles that surround each cut except $\mathcal{C}_{k_i}^{(i)}$\fn{This means that the integration of the Bergman kernel around $\mathcal{C}_{k_i}^{(i)}$ does not vanish, $\oint_{p \in \mathcal{C}_{k_i}^{(i)}}B_{k_i}^{(i)}\neq 0$.}.

\begin{figure}[tb]
 \begin{minipage}{0.32\hsize}
 \begin{center}
\picture{clip, height=3.2cm}{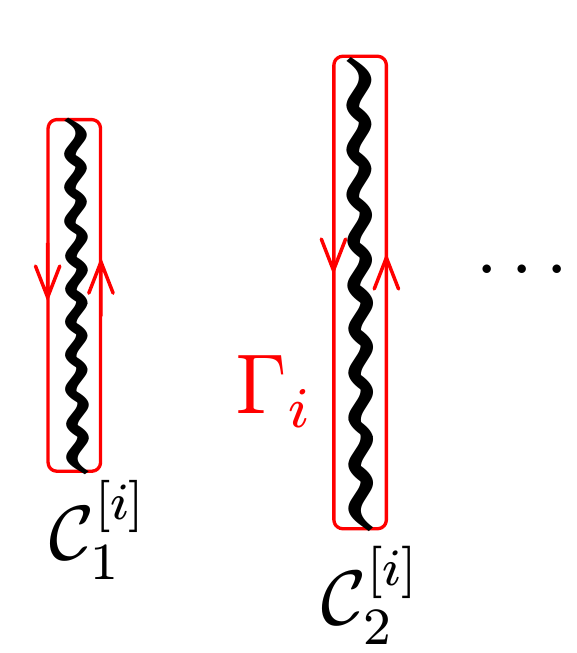}\\
$(a)$
\end{center}
 \end{minipage}
 \begin{minipage}{0.32\hsize}
 \begin{center}
\picture{clip, height=3.2cm}{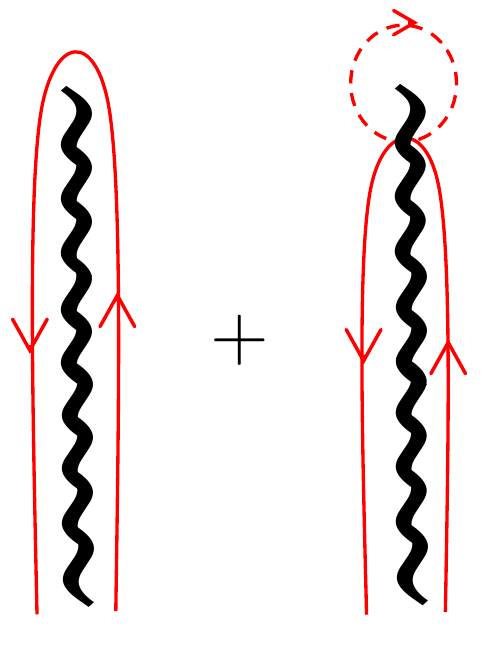}\\
$(b)$
\end{center}
 \end{minipage}
 \begin{minipage}{0.32\hsize}
 \begin{center}
\picture{clip, height=3.2cm}{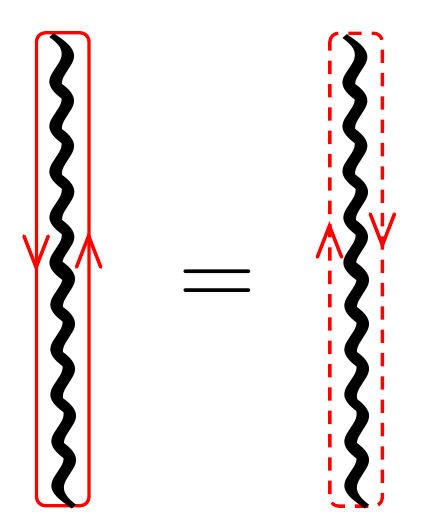}\\
$(c)$
\end{center}
 \end{minipage}
\caption{The integration contours relevant to the decomposition of poles. $(a)$ The contour $\Gamma_i$ goes along the branch cuts $\mathcal{C}_{s}^{(i)}$ on the first sheet of $p_i$ counterclockwise. $(b)$ Near the branch point, one must average over all possible ways to avoid the branch point as shown in the figure. The dashed curve denotes the contour on the second sheet. $(c)$ By the continuous deformation, one can move the contour to the second sheet of $p_i$. The contour on the second sheet has the opposite orientation from the one on the first sheet. This leads to the minus sign on the right hand side of \eqref{swapminus}}
\label{contourmanipulation}
\vspace{-0.2cm}
\end{figure}

Now, using these kernels, one can decompose $\del_u \ln \sin p_i$ as follows:
\beq{
\begin{aligned}\label{decomposepole}
W_{++}^{ij}du&= \oint_{u^{\prime}\in\Gamma_i}\!\! B_{k_i}^{(i)}(u,u^{\prime}) \ln \sin p_i(u^{\prime})+{\tt rest}&&(u\in \text{2nd sheet of $p_i$}) \comma\\
W_{--}^{ij}du&= -\oint_{u^{\prime}\in\Gamma_i}\!\!B_{k_i}^{(i)}(u,u^{\prime}) \ln \sin p_i(u^{\prime})+{\tt rest}&&(u\in \text{1st sheet of $p_i$})\period
\end{aligned}
}
Here ${\tt rest}$ represents the terms coming from decomposition of the rest of terms on the right hand side of \eqref{RH}. The integration contour $\Gamma_i$ is on the 1st sheet of $p_i$ and goes along the cuts $\mathcal{C}_{s}^{(i)}$ as depicted in figure \ref{contourmanipulation}-$(a)$. Let us now make a remark on the contour: Unlike other poles, the poles at the branch points are shared equally by  $\del_u \ln\sl{i_+,j_+}$ and $\del_u \ln\sl{i_-,j_-}$, since each eigenvector has a square-root singularity as shown in \eqref{branchsingular}. To realize this structure, one must average over different ways of avoiding the branch points as shown in figure \ref{contourmanipulation}-$(b)$. Apart from this small subtlety, these formulas are natural generalization of \eqref{defFul} and more importantly they are consistent with the property of $W_{\pm\pm}^{ij}$ \eqref{propertyofW}, thanks to the normalization of the Bergman kernel \eqref{normBerg}. As in the standard Wiener-Hopf method, the expressions in the other regions can be obtained by  analytic continuation.
 
 Before proceeding, let us rewrite \eqref{decomposepole} in a form where the action of the holomorphic involution is more clearly seen. For this purpose, we first make a change of the integration variable from $u^{\prime}$ to $\hat{\sigma}_i u^{\prime}$, with $\hat{\sigma}_i$ being the holomorphic involution for the spectral curve of $p_i$. This, of course, leaves the value of the integral intact, but its form gets slightly modified. For instance, the integrand is transformed as
 \beq{
&\ln \sin p_i (\hat{\sigma}_i u^{\prime}) = \ln \left( -\sin p_i(u^{\prime})\right) \comma\\
&B_{k_i}^{(i)}(u,\hat{\sigma}_iu^{\prime})= \check{B}_{k_i}^{(i)} (u,u^{\prime})\period\label{checkB}
  } 
Here, the new kernel $\check{B}^{(i)}_{k_i}(p,q)$ is defined by \eqref{checkB} and has a pole when $p = \hat{\sigma}_i q$,
\beq{
\check{B}_{k_i}^{(i)}(p,q)\sim \frac{1}{2\pi i (\zeta(p)-\zeta(\hat{\sigma}_i q))}d\zeta(p) d\zeta(\hat{\sigma}_i q)\comma
}
where $\zeta$ is a local coordinate around $p$ and $\hat{\sigma}_i q$. Similarly, the integration contour is modified as follows (see figure \ref{contourmanipulation}-$(c)$ for more explanation):
\beq{\label{swapminus}
\oint_{\Gamma_i} d\left(\hat{\sigma}_i u^{\prime}\right) = -\oint_{\Gamma_i} d u^{\prime}
}
From these transformation rules, we can rewrite the integral appearing in \eqref{decomposepole} as
\beq{\label{newrep}
\oint_{u^{\prime}\in\Gamma_i}\!\! B_{k_i}^{(i)}(u,u^{\prime}) \ln \sin p_i(u^{\prime}) = -\oint_{u^{\prime}\in\Gamma_i}\!\! \check{B}_{k_i}^{(i)}(u,u^{\prime}) \ln \sin p_i(u^{\prime})\period
}
Here and below we neglect the term coming from $\ln (-1)$ since it changes the structure constant only by an overall phase. Now, by averaging two sides of \eqref{newrep}, we arrive at the following expressions for $W_{\pm\pm}^{ij}$:
\beq{
\begin{aligned}\label{decomposepole2}
W_{++}^{ij}du&= \oint_{\Gamma_i} A_{k_i}^{(i)}\ast \ln \sin p_i+{\tt rest}&&(u\in \text{2nd sheet of $p_i$}) \comma\\
W_{--}^{ij}du&= -\oint_{\Gamma_i}A_{k_i}^{(i)}\ast \ln \sin p_i+{\tt rest}&&(u\in \text{1st sheet of $p_i$})\period
\end{aligned}
}
Here $A_{k_i}^{(i)}$ is the ``anti-symmetrized'' kernel defined by
\beq{\label{defofAki}
A_{k_i}^{(i)}(p,q)\equiv \frac{1}{2}\left( B_{k_i}^{(i)}(p,q) -\check{B}_{k_i}^{(i)}(p,q)\right)=\frac{1}{2}\left( B_{k_i}^{(i)}(p,q) -B_{k_i}^{(i)}(p,\hat{\sigma}_iq)\right)\comma
}
and the notation $\oint_{\mathcal{C}} F\ast f$ denotes the  convolution integral 
\beq{
\oint_{\mathcal{C}} F\ast f = \oint_{u^{\prime}\in \mathcal{C}} F(u,u^{\prime}) f(u^{\prime})\period
 } 
 
The kernel $A_{k_i}^{(i)}$ is odd under the holomorphic involution of each of the arguments:
 \beq{\label{antisymAki}
 A_{k_i}^{(i)}(p,\hat{\sigma}_i q)=-A_{k_i}^{(i)}(p, q)\comma \qquad A_{k_i}^{(i)}(\hat{\sigma}_i p, q)=-A_{k_i}^{(i)}(p, q)\period 
  } 
  The first equality follows immediately from the definition whereas the second equality can be derived using the property of the Bergman kernel \eqref{Bholosym}. This property is used in section \ref{sec:resultweak} when we write down the expression for the semi-classical structure constant.
\subsubsection*{Decomposition of the zeros}
We now decompose $\del_u \ln\sin (p_i + p_j +p_k)/2$ and $\del_u \ln\sin (p_i+p_j-p_k)/2$, which are responsible for the zeros of the Wronskians. Since these quantities are defined on the $2^3$-sheeted Riemann surface, both the integration contour and the convolution kernel must also be defined on the same eight-sheeted surface. 

\begin{figure}[tb]
\begin{center}
\picture{clip, height=5.6cm}{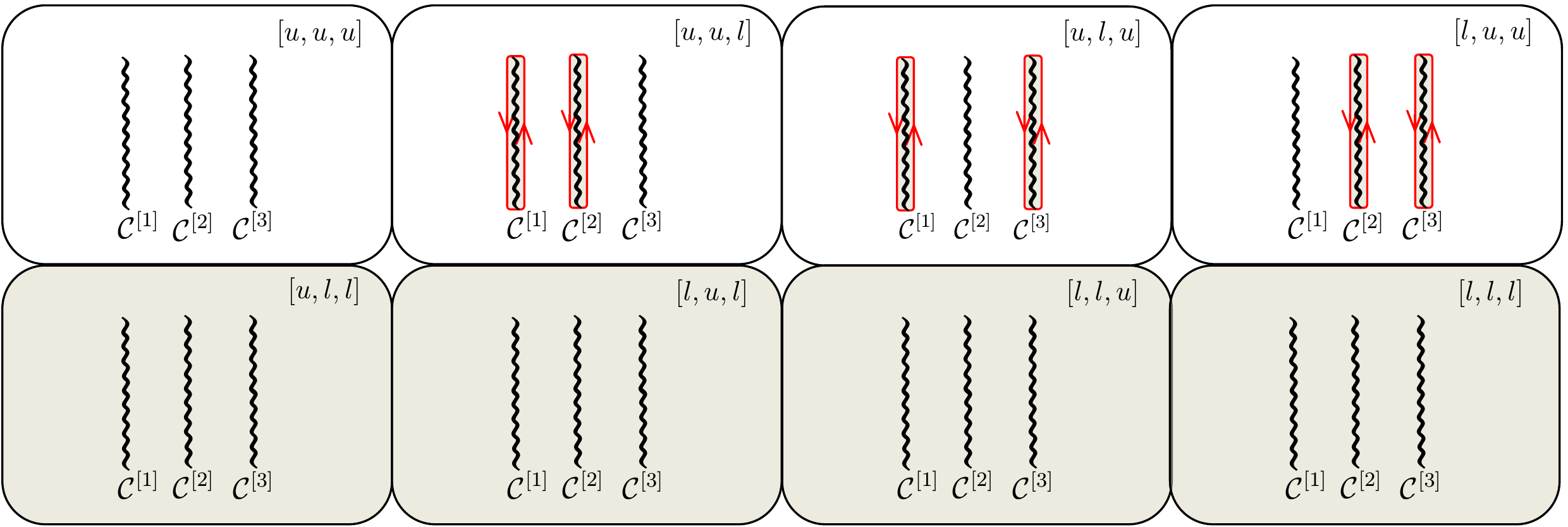}
\end{center}\vspace{-0.5cm}
\caption{The analyticity structure and the integration contour for the decomposition of $\sin (p_i+p_j+p_k)/2$. In the region depicted in white, the Wronskians among $\{1_{+},2_{+},3_{+}\}$ have zeros,  whereas in the region depicted in gray, the Wronskians among $\{1_-,2_-,3_-\}$ have zeros. The integration contour $\Gamma_{+++}$, denoted in red, separates the white region from the gray region.}
\label{gamma+++}
\vspace{-0.2cm}
\end{figure}

Let us first specify the integration contour. As in the previous case,  the contour 
should be taken such that it separates the domains with different analyticity. As stated in the rules in section \ref{subsec:analyticity}, when $\sin \left(\sum_i \epsilon_i p_i\right)$ vanishes, the Wronskians that vanish are the ones among $\{1_{\epsilon_1},2_{\epsilon_2},3_{\epsilon_3}\}$ or the ones among $\{1_{-\epsilon_1},2_{-\epsilon_2},3_{-\epsilon_3}\}$. Depending on which of the two groups contain vanishing Wronskians, the eight-sheeted Riemann surface is divided into two regions. Then the integration contour, denoted by $\Gamma_{\epsilon_1\epsilon_2\epsilon_3}$, will be placed at the boundary of the two regions. For instance, for the case of $\sin (p_1+p_2+p_3)/2$, the two regions and the contour are depicted in figure \ref{gamma+++}. To find  the analyticity structure and the contour of other factors, one just needs to exchange the sheets appropriately,  thanks to the property \eqref{normalizationcond12}. For example, the analyticity structure and the contour of $\sin (p_1+p_2 -p_3)/2$ are given by those in figure \ref{gamma+++} with $[\ast,\ast,u]$-sheets and $[\ast,\ast,l]$-sheets swapped.

We next determine the convolution kernel. To carry out the desired decomposition, we use the kernel $B_{\rm all}(p,q)$, which satisfies the first and the third 
 properties, \eqref{symBerg} and  \eqref{analyBerg} respectively, of the Bergman kernel and the following slightly different normalization condition:
{\it
\begin{itemize}
\item[] Normalization:
\beq{\label{newnorm}
\oint_{p\in \mathcal{C}_{s}^{(i)}} B_{\rm all}(p,q)=0 \comma \qquad s\neq k_i\comma \quad i=1,2,3\period 
}
\end{itemize}
}
\noindent
Using this kernel, we can, for instance, decompose $\del_u \ln (p_i+p_j+p_k)/2$ in the following way:
\beq{\label{zerodecomp}
\begin{aligned}
W_{++}^{ij}du=&-\oint_{u^{\prime}\in\Gamma_{+++}} \!\!\!\!\!\!\!\!B_{\rm all}(u,u^{\prime}) \ln \sin \frac{p_i+p_j+p_k}{2}(u^{\prime})+{\tt rest} \\
&(u\in \text{gray region in figure \ref{gamma+++}}) \comma\\
W_{--}^{ij}du=&\oint_{u^{\prime}\in\Gamma_{+++}}\!\! \!\!\!\!\!\!B_{\rm all}(u,u^{\prime}) \ln \sin \frac{p_i+p_j+p_k}{2}(u^{\prime})+{\tt rest}\\
& (u\in \text{white region in figure \ref{gamma+++}})\period
\end{aligned}
}
Again, in virtue of  the normalization condition \eqref{newnorm}, this is consistent with the property of $W_{\pm\pm}^{ij}$ \eqref{propertyofW}. The decomposition of $\del_u \ln\sin (p_i+p_j-p_k)$ can be performed in a similar manner.

Let us make a clarifying remark.  Although  the existence of the kernel $B_{\rm all}$ with the properties listed above has not  been explicitly proven,  the convolution integral \eqref{zerodecomp}  can be rewritten entirely  in terms of the standard Bergman kernel, the existence of which is firmly established. To show this, we again make use of the holomorphic involution. To illustrate the idea, let us consider the following terms that appear in the expression for $W_{++}^{12}$ and $W_{--}^{12}$:
\beq{\label{torewrite}
\begin{aligned}
\oint_{u^{\prime}\in\Gamma_{+++}} \!\!\!\!\!\!\!\!B_{\rm all}(u,u^{\prime}) \ln \sin \frac{p_1+p_2+p_3}{2}(u^{\prime})+\oint_{u^{\prime}\in\Gamma_{++-}} \!\!\!\!\!\!\!\!B_{\rm all}(u,u^{\prime}) \ln \sin \frac{p_1+p_2-p_3}{2}(u^{\prime}) \period
\end{aligned}
}
If we change the integration variables from $u^{\prime}$ to $\hat{\sigma}_3 u$, the integrand and the contour of \eqref{torewrite} transform as
\beq{\label{Balltransf}
\begin{aligned}
&\ln \sin \frac{p_1+p_2 \pm p_3}{2} (\hat{\sigma}_3u^{\prime})=\ln \sin \frac{p_1+p_2\mp p_3}{2}(u^{\prime})\comma\\
&B_{\rm all}(u,\hat{\sigma}_3u^{\prime}) = B_{\rm all}^{(3)} (u,u^{\prime})\comma\\
&\oint_{\Gamma_{++\pm}}d(\hat{\sigma}_3u^{\prime})=\oint_{\Gamma_{++\mp}}du^{\prime}\comma
\end{aligned}
 }
 where the new kernel, $B_{\rm all}^{(3)} (p,q)$, has a pole at $p=\hat{\sigma}_3 q$. Using this transformation rule, we can re-express the integral \eqref{torewrite} as
 \beq{
\begin{aligned}
\oint_{u^{\prime}\in\Gamma_{+++}} \!\!\!\!\!\!\!\!B^{(3)}_{\rm all}(u,u^{\prime}) \ln \sin \frac{p_1+p_2+p_3}{2}(u^{\prime})+\oint_{u^{\prime}\in\Gamma_{++-}} \!\!\!\!\!\!\!\!B^{(3)}_{\rm all}(u,u^{\prime}) \ln \sin \frac{p_1+p_2-p_3}{2}(u^{\prime})  \period
\end{aligned} 
 }
Considering all possible combinations of holomorphic involutions, we obtain $2^3$ different expressions for \eqref{torewrite}. Then averaging over these $2^3$ expressions, we get
\beq{\label{rewritten1}
\begin{aligned}
&\frac{1}{8}\oint_{u^{\prime}\in \Gamma_{+++}} \!\!\!\!\!\!\!\! K(u,u^{\prime}) \ln \sin \frac{p_1+p_2+p_3}{2}(u^{\prime})+(\text{4 other terms})\comma
 \end{aligned}
 } 
with
\beq{
\begin{aligned}
K(p,q)&\equiv \left(B_{\rm all}+B^{(3)}_{\rm all}-B^{(12)}_{\rm all}-B^{(123)}_{\rm all}\right)(p,q)\comma\\
B^{(12)}_{\rm all} (u,u^{\prime}) &\equiv  B_{\rm all} (u,\hat{\sigma}_1\hat{\sigma_2}u^{\prime})\comma\quad
B^{(123)}_{\rm all} (u,u^{\prime}) \equiv  B_{\rm all} (u,\hat{\sigma}_1\hat{\sigma_2}\hat{\sigma}_3u^{\prime})\period
\end{aligned}
}
Now the kernel $K(p,q)$ has four double poles as shown in table \ref{tab3}. 
\begin{table}[tb]
\begin{center}
\begin{tabular}{c|c|c|c|c|c|c|c|c}
&$q$&$\hat{\sigma}_1q$&$\hat{\sigma}_2q$&$\hat{\sigma}_3q$
&$\hat{\sigma}_1\hat{\sigma}_2q$&$\hat{\sigma}_1\hat{\sigma}_3q$
&$\hat{\sigma}_2\hat{\sigma}_3q$
&$\hat{\sigma}_1\hat{\sigma}_2\hat{\sigma}_3q$\\
\hline
$K(p,q)$&$+1$&&&$+1$&$-1$&&&$-1$\\
$A_{k_1}^{(1)}(p,q)$&$+1/2$&$-1/2$&$+1/2$&$+1/2$&$-1/2$&$-1/2$&$+1/2$&$-1/2$\\
$A_{k_2}^{(2)}(p,q)$&$+1/2$&$+1/2$&$-1/2$&$+1/2$&$-1/2$&$+1/2$&$-1/2$&$-1/2$
\end{tabular}
\end{center}\vspace{-0.5cm}
\caption{The structure of the poles of the kernels $K$, $A_{k_1}^{(1)}$ and $A_{k_2}^{(2)}$. The first row designates the position of the double pole as a function of $p$ and the numbers within the table are the coefficient of each pole. One can easily see that $K$ and $A_{k_1}^{(1)}+A_{k_2}^{(2)}$ have the same pole structure.}
\label{tab3}
\vspace{-0.2cm}
\end{table}
As is clear from table \ref{tab3}, the analytic properties of $K(p,q)$ are identical to those of $A_{k_1}^{(1)}+A_{k_2}^{(2)}$, which are expressed in terms 
 of the usual Bergman kernels. Thus we can replace $K(p,q)$ in \eqref{rewritten1} with  $A_{k_1}^{(1)}+A_{k_2}^{(2)}$. Performing similar analysis to other 4 terms, we obtain the following expression for $W_{++}^{12}$:
\beq{
\begin{aligned}\label{finaldecomp1}
&W_{++}^{12}du={\tt rest}\\
&-\frac{1}{8}\left(\oint_{\Gamma_{+++}}\!\!\!\!\!(A_{k_1}^{(1)}+A_{k_2}^{(2)})\ast\ln\sin \frac{p_1+p_2+p_3}{2} +\oint_{ \Gamma_{++-}}\!\!\!\!\!(A_{k_1}^{(1)}+A_{k_2}^{(2)})\ast\ln\sin \frac{p_1+p_2-p_3}{2}\right.\\
&\left.+\oint_{\Gamma_{+-+}}\!\!\!\!\!\!\!(A_{k_1}^{(1)}-A_{k_2}^{(2)})\ast\ln\sin \frac{p_1-p_2+p_3}{2}+\oint_{\Gamma_{-++}}\!\!\!\!\!\!\!(-A_{k_1}^{(1)}+A_{k_2}^{(2)})\ast\ln\sin \frac{-p_1+p_2+p_3}{2}\right)\period
\end{aligned}
}
As in the standard Wiener-Hopf decomposition, this integral expression is valid in the region where $W_{++}^{12}$ does not have any poles, which in this case correspond to the $[l,l,\ast]$-sheets\fn{As discussed in section \ref{subsec:analyticity}, when the spectral parameter is on these sheets, $\sl{1_+,2_+}$ does not have any poles or zeros except for extra poles and zeros which are now subtracted. See tables \ref{tab1} and \ref{tab2}.}. 

Finally, let us discuss the simplification of the integration contours. 
The contours of \eqref{finaldecomp1} are defined on the eight-sheeted Riemann surface. However, for comparison with the results in the literature, it is more convenient to convert them  into integrals defined purely on the $[u,u,u]$-sheet. This can be achieved again by making use of the holomorphic involution: For instance, take the integral along $\Gamma_{+++}$ in \eqref{finaldecomp1} and consider the portion of the integral on the $[u,l,u]$-sheet. If we perform the holomorphic involution $\hat{\sigma}_2$ to the integrated variable $u^{\prime}$, this contribution becomes identical to the third term in \eqref{finaldecomp1} evaluated on the $[u,u,u]$-sheet. Repeating the same analysis for the other relevant integrals, we arrive at
 the expression
\beq{
\begin{aligned}\label{finaldecomp2}
&W_{++}^{12}du={\tt rest}\\
&-\frac{1}{2}\left(\oint_{ \gamma_{+++}}\!\!\!\!\!(A_{k_1}^{(1)}+A_{k_2}^{(2)})\ast\ln\sin \frac{p_1+p_2+p_3}{2} +\oint_{ \gamma_{++-}}\!\!\!\!\!(A_{k_1}^{(1)}+A_{k_2}^{(2)})\ast\ln\sin \frac{p_1+p_2-p_3}{2}\right.\\
&\left.+\oint_{\gamma_{+-+}}\!\!\!\!\!\!\!(A_{k_1}^{(1)}-A_{k_2}^{(2)})\ast\ln\sin \frac{p_1-p_2+p_3}{2}+\oint_{\gamma_{-++}}\!\!\!\!\!\!\!(-A_{k_1}^{(1)}+A_{k_2}^{(2)})\ast\ln\sin \frac{-p_1+p_2+p_3}{2}\right)\comma
\end{aligned}
}
where $\gamma_{\epsilon_1\epsilon_2\epsilon_3}$ is  a  portion of $\Gamma_{\epsilon_1\epsilon_2\epsilon_3}$ on the $[u,u,u]$-sheet. It is clear from figure \ref{gamma+++} that $\Gamma_{+++}$ does not have any portion on the $[u,u,u]$-sheet, and thus $\gamma_{+++}=\varnothing$. The other contours are along the cuts of some of the quasi-momenta as shown below:
\beq{\label{explicitcontour}
\begin{aligned}
\gamma_{++-}= \Gamma_1 \cup\Gamma_2\comma\qquad\gamma_{+-+}=\Gamma_1\cup\Gamma_3\comma \qquad\gamma_{-++}=\Gamma_2\cup\Gamma_3 \period
\end{aligned}
}
Here $\Gamma_i$'s are the contours given in figure \ref{contourmanipulation}-$(a)$. Substituting \eqref{explicitcontour} into \eqref{finaldecomp2} and restoring the terms coming from the decomposition of poles, we finally obtain
\beq{
\begin{aligned}\label{finaldecomp3}
&W_{++}^{12}du=\\
&\oint_{\Gamma_1}A_{k_1}^{(1)}\ast\ln \sin p_2+\oint_{\Gamma_2}A_{k_2}^{(2)}\ast\ln \sin p_1-\frac{1}{2}\left(\oint_{ \Gamma_1 \cup\Gamma_2}\!\!\!\!\!(A_{k_1}^{(1)}+A_{k_2}^{(2)})\ast\ln\sin \frac{p_1+p_2-p_3}{2}\right.\\
&\left.+\oint_{\Gamma_1 \cup\Gamma_3}\!\!\!\!\!\!\!\!\!(A_{k_1}^{(1)}-A_{k_2}^{(2)})\ast\ln\sin \frac{p_1-p_2+p_3}{2}+\oint_{\Gamma_2 \cup\Gamma_3}\!\!\!\!\!\!\!\!\!(-A_{k_1}^{(1)}+A_{k_2}^{(2)})\ast\ln\sin \frac{-p_1+p_2+p_3}{2}\right)\period
\end{aligned}
}

Similarly, we can write down the expression for $W_{--}^{12}$, which is valid when the spectral parameter is on the $[u,u,\ast]$-sheets:
 \beq{
\begin{aligned}\label{finaldecomp4}
&W_{--}^{12}du=\\
&-\oint_{\Gamma_1}A_{k_1}^{(1)}\ast \ln \sin p_1 -\oint_{\Gamma_2}A_{k_2}^{(2)}\ast \ln \sin p_2 +\frac{1}{2}\left( \oint_{\Gamma_1 \cup\Gamma_2}\!\!\!\!\!(A_{k_1}^{(1)}+A_{k_2}^{(2)})\ast\ln\sin \frac{p_1+p_2-p_3}{2}\right.\\
&\left.+\oint_{\Gamma_1 \cup\Gamma_3}\!\!\!\!\!\!\!\!\!(A_{k_1}^{(1)}-A_{k_2}^{(2)})\ast\ln\sin \frac{p_1-p_2+p_3}{2}+\oint_{\Gamma_2 \cup\Gamma_3}\!\!\!\!\!\!\!\!\!(-A_{k_1}^{(1)}+A_{k_2}^{(2)})\ast\ln\sin \frac{-p_1+p_2+p_3}{2}\right)\period
\end{aligned}
   }
   The expressions for the other $W_{++}^{ij}$'s  and $W_{--}^{ij}$'s  can be obtained from \eqref{finaldecomp3} and \eqref{finaldecomp4} by the permutation of the indices.
\section{Results at weak coupling\label{sec:resultweak}}
Now we combine the results of sections \ref{sec:semi-classical}, \ref{sec:Wronskian} and \ref{sec:ev} and write down the explicit integral expression for the structure constant.
\subsection{Integral expression for the semi-classical structure constant\label{subsec:integralc123}}
\begin{figure}[tb]
\begin{center}
\picture{clip,height=4cm}{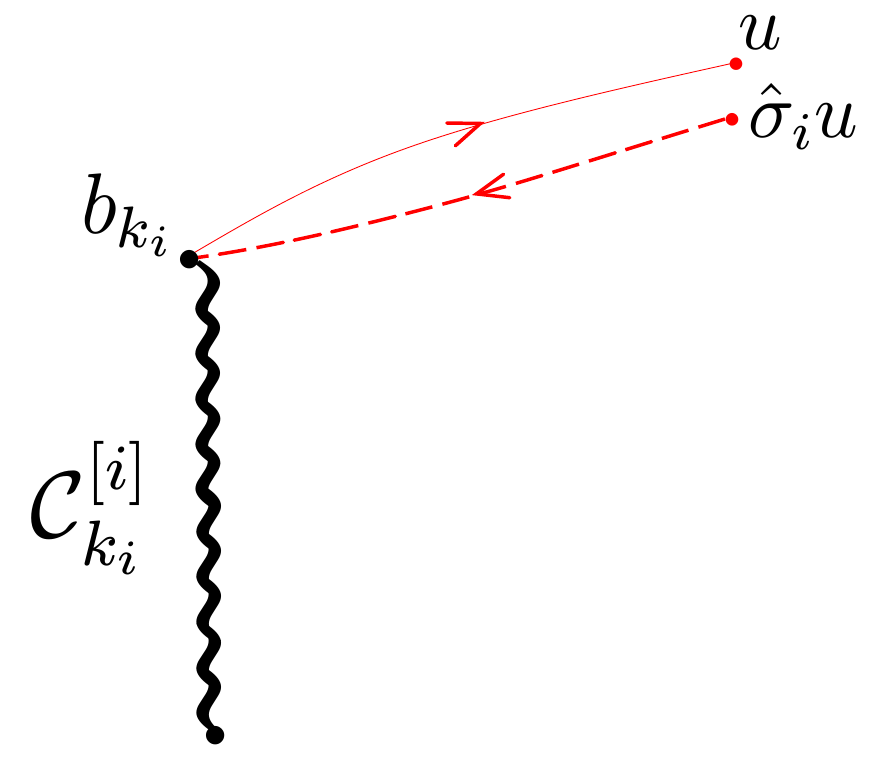}
\end{center}
\vspace{-0.5cm}
\caption{The branch point $b_{k_i}$ and the integration contour in \eqref{simpleEkiAki}.}
\label{uhatu}
\end{figure}

As shown in \eqref{eq-34}, the variation of the semi-classical structure constant is given in terms of the Wronskians. To compute those Wronskians, we integrate the results obtained in the previous section \eqref{finaldecomp3} and \eqref{finaldecomp4}. The net effect of integration is to replace the kernels $A_{k_i}^{(i)}(u,u^{\prime})$ by their integrals $\int^{v=u}_{v=v_0}A_{k_i}^{(i)}(v,u^{\prime})$. This is, however, still ambiguous since the initial point of the $u$-integration $v_0$ is not fixed. To determine $v_0$, we impose the following condition which comes from the normalization of the eigenvectors \eqref{normalizationcond12}:
\beq{\label{involreq}
\begin{aligned}
\sl{i_{-},j_{-}}(\hat{\sigma}_i\hat{\sigma}_ju)=-\sl{i_{+},j_{+}}(u) \period
 \end{aligned}
 } 
 In terms of the logarithm of the Wronskians, this reads
 \beq{\label{involreq2}
 \ln \sl{i_{-},j_{-}}(\hat{\sigma}_i\hat{\sigma}_ju)=\ln\sl{i_{+},j_{+}}(u) \period
 }
As in the previous analyses,  we have neglected the minus sign in \eqref{involreq}, which only affects the overall phase of the final result. We shall now show that  \eqref{involreq2} is satisified if we choose $v_0$ to be the branch point of $\mathcal{C}_{k_i}^{(i)}$, which we denote by $b_{k_i}$ (see figure \ref{uhatu}). Under this choice, the Wronskians are given by
\beq{
\ln\sl{i_+,j_+}&=E_{k_i}^{(i)}+E_{k_j}^{(j)}+\oint_{\Gamma_i}\alpha_{k_i}^{(i)}\ast \ln \sin p_i +\oint_{\Gamma_j}\alpha_{k_j}^{(j)}\ast \ln \sin p_j\nn\\
&-\left( \oint_{\Gamma_i\cup \Gamma_j}\!\!\!\!\!(\alpha_{k_i}^{(i)}+\alpha_{k_j}^{(j)})\ast \ln \sin \frac{p_i+p_j-p_k}{2}+\oint_{\Gamma_i\cup \Gamma_k}\!\!\!\!\!\!\!(\alpha_{k_i}^{(i)}-\alpha_{k_j}^{(j)})\ast \ln \sin \frac{p_i-p_j+p_k}{2}\right.\label{ipjp}\\
&\left. +\oint_{\Gamma_j\cup \Gamma_k}\!\!\!\!\!(-\alpha_{k_i}^{(i)}+\alpha_{k_j}^{(j)})\ast \ln \sin \frac{-p_i+p_j+p_k}{2}\right)\comma\nn\\
\ln\sl{i_-,j_-}&=-E_{k_i}^{(i)}-E_{k_j}^{(j)}
-\oint_{\Gamma_i}\alpha_{k_i}^{(i)}\ast \ln \sin p_i -\oint_{\Gamma_j}\alpha_{k_j}^{(j)}\ast \ln \sin p_j\nn\\
&+ \oint_{\Gamma_i\cup \Gamma_j}\!\!\!\!\!(\alpha_{k_i}^{(i)}+\alpha_{k_j}^{(j)})\ast \ln \sin \frac{p_i+p_j-p_k}{2}+\oint_{\Gamma_i\cup \Gamma_k}\!\!\!\!\!\!\!(\alpha_{k_i}^{(i)}-\alpha_{k_j}^{(j)})\ast \ln \sin \frac{p_i-p_j+p_k}{2}\label{imjm}\\
& +\oint_{\Gamma_j\cup \Gamma_k}\!\!\!\!\!(-\alpha_{k_i}^{(i)}+\alpha_{k_j}^{(j)})\ast \ln \sin \frac{-p_i+p_j+p_k}{2}\comma\nn
}
with $ E_{k_i}^{(i)}$ and $\alpha_{k_i}^{(i)}$ given by 
\beq{
\label{alki}E_{k_i}^{(i)}\equiv \int_{v=b_{k_i}}^{v=u}e_{k_i}^{(i)}(v)\comma\qquad 
\alpha_{k_i}^{(i)}(u,u^{\prime})\equiv \int_{v=b_{k_i}}^{v=u}A_{k_i}^{(i)}(v,u^{\prime})\period
}
As with the expressions in the previous section, \eqref{ipjp} and \eqref{imjm} are valid on the $[l,l,l]$-sheet and on the $[u,u,u]$-sheet respectively. To see that \eqref{ipjp} and \eqref{imjm} indeed satisfy the condition \eqref{involreq}, we just need to use the fact that $e_{k_i}^{(i)}$ and $A_{k_i}^{(i)}$ are odd while the branch point $b_{k_i}$ is invariant under the holomorphic involution (see \eqref{defofnormekii} and \eqref{antisymAki}). Using these properties, we can express $E_{k_i}^{(i)}$ and $\alpha_{k_i}^{(i)}$ in a more symmetric way as follows\fn{For instance, the expression for $E_{k_i}^{(i)}$ can be derived as follows:
\beq{
\begin{aligned}\label{firststepE}
E_{k_i}^{(i)} &= \int_{v=b_{k_i}}^{v=u} e_{k_i}^{(i)}(v) =\frac{1}{2}\left( \int_{v=b_{k_i}}^{v=u} e_{k_i}^{(i)}(v)+\int_{\hat{\sigma}_i v=b_{k_i}}^{\hat{\sigma}_iv= u} e_{k_i}^{(i)}(\hat{\sigma}_i v)\right)\\
&=\frac{1}{2}\left( \int_{v=b_{k_i}}^{v=u} e_{k_i}^{(i)}(v)+\int_{ v=b_{k_i}}^{v= \hat{\sigma}_i u} e_{k_i}^{(i)}(\hat{\sigma}_i v)\right)=\frac{1}{2}\int_{v=\hat{\sigma}_iu}^{v=u}e_{k_i}^{(i)}(v)\period
\end{aligned}
}
}:
\beq{
\label{simpleEkiAki}
E_{k_i}^{(i)}&=\frac{1}{2}\int_{v=\hat{\sigma}_iu}^{v=u}e_{k_i}^{(i)}(v)\comma\qquad \alpha_{k_i}^{(i)} (u,u^{\prime})= \frac{1}{2}\int^{v=u}_{v=\hat{\sigma}_i u} B_{k_i}^{(i)}(v,u^{\prime})\period
}
Here, the precise integration contour is the one given in figure \ref{uhatu}.

Having determined the Wronskians, we can now compute the angle variable by evaluating \eqref{imjm} at $u=\infty$ and substituting them into \eqref{eq-34}. It turns out that the terms $E_{k_i}^{(i)}(u=\infty)$ precisely cancel the last term in \eqref{eq-34}. Thus, as anticipated, the contribution from extra zeros and poles do not appear  in the final expression, which takes the form 
\beq{
\begin{aligned}\label{yoshida}
 \varphi_{k_i}^{(i)} &= i\left(\ln \frac{\sl{n_i,n_j}\sl{n_k,n_i}}{\sl{n_j,n_k}} +2\oint_{\Gamma_i}\bar{\alpha}_{k_i}^{(i)}\, \ln \sin p_i- \oint_{\Gamma_i\cup \Gamma_j}\!\!\!\!\!\bar{\alpha}_{k_i}^{(i)}\, \ln \sin \frac{p_i+p_j-p_k}{2}\right.\\
&\left.-\oint_{\Gamma_i\cup \Gamma_k}\!\!\!\!\!\bar{\alpha}_{k_i}^{(i)}\, \ln \sin \frac{p_i-p_j+p_k}{2}+\oint_{\Gamma_j\cup \Gamma_k}\!\!\!\!\!\bar{\alpha}_{k_i}^{(i)} \,\ln \sin \frac{-p_i+p_j+p_k}{2}\right)\period
 \end{aligned} 
  }
  Here the one form $\bar{\alpha}_{k_i}^{(i)}$ is defined by
  \beq{
  \bar{\alpha}_{k_i}^{(i)}(u^{\prime})\equiv \alpha_{k_i}^{(i)}(\infty, u^{\prime}) =\frac{1}{2}\int_{v=\hat{\sigma}_i\infty}^{v=\infty} B_{k_i}^{(i)}(v,u^{\prime})\period
  }

From the properties of the Bergman kernel, one can show that $\bar{\alpha}_{k_i}^{(i)}$ has the following analytic properties:
\beq{
\begin{aligned}\label{alphanalytic}
&\underset{u=\infty}{\rm Res}\,\bar{\alpha}_{k_i}^{(i)} =-\frac{1}{2}\comma\qquad \underset{u=\hat{\sigma}_i\infty}{\rm Res}\,\bar{\alpha}_{k_i}^{(i)} =+\frac{1}{2}\comma\\
&\oint_{\mathcal{C}_s^{(i)}} \bar{\alpha}_{k_i}^{(i)}=0 \quad (s\neq k_i)\comma\qquad\oint_{\mathcal{C}_{k_i}^{(i)}} \bar{\alpha}_{k_i}^{(i)}=+\frac{1}{2}\period
\end{aligned}
}
Now, it is easy to check that the one form\fn{One can show \eqref{delpdu} using the argument similar to the one given in section \ref{subsec:action-angle}: To perturb $S_{k_i}^{(i)}$, one needs to add to $p_i du$ a one form whose period integral does not vanish only along the cycle at infinity and the cycle around $\mathcal{C}_{k_i}^{(i)}$. By comparing the residues carefully, we arrive at \eqref{delpdu}.}
\beq{\label{delpdu}
\frac{\del \left( p_i du/4\pi i\right)}{\del S_{k_i}^{(i)}}
}
also satisfies the same analytic properties. Since \eqref{alphanalytic} specifies the one form uniquely, this means that $\bar{\alpha}_{k_i}^{(i)}$ is identical to \eqref{delpdu}. Using this fact and the identity,
\beq{
\int_0^{x}dx^{\prime}\ln \sin x^{\prime} = \frac{i}{2}\left( {\rm Li}_2 (e^{2ix})-\frac{\pi^2}{6}\right) +\ln (i/2) x-\frac{i}{2} x^2\comma
}
we can integrate the relation $\del \ln C_{123}/\del S_{k_i}^{(i)}=i\delta \phi_{k_i}^{(i)}$ to get the following integral expression:
\beq{
\left.\ln \left( \frac{C_{123}}{C_{123}^{\rm BPS}}\right)\right|_{\text{${\rm SU}(2)_R$}}\!\!\!\!\!=&\sum_{\{i,j,k\}\in{\rm cperm}\{1,2,3\}}\left[(M_k-M_i-M_j)\ln \sl{n_i,n_j}+\frac{1}{2}\oint_{\Gamma_i\cup \Gamma_j}\frac{du}{2\pi}{\rm Li}_2 \left(e^{ip_i+ip_j-ip_k}\right)\right]\nn\\
&-\frac{1}{2} \sum_{k=1}^{3}{\rm Li}_2 (e^{2ip_k})\period\label{forthemoment}
}
Here the summation in the first line denotes the sum over the cyclic permutation (abbreviated as ``cperm")  of $\{1,2,3\}$ and $M_i$ is the total number of magnons in $p_i$.
Note that \eqref{forthemoment} is the contribution from the ${\rm SU}(2)_R$ sector only.  For the complete result for the structure function 
 for the distinct types of three-point functions, to be analyzed in the next section,  
it must be combined  with the contribution from the ${\rm SU}(2)_L$ sector as well.%
\subsection{Results and comparison with the literature\label{subsec:finalweak}}
The operators  forming the three-point functions we are studying transform
 under a single group ${\rm SO}(4)={\rm SU}(2)_L\times {\rm SU}(2)_{R}$. 
For such correlators, there are two distinct classes, as discussed in \cite{KK3}. 
\subsubsection*{Type I-I-II three-point function}
\begin{figure}[tb]
 \begin{minipage}{0.45\hsize}
 \begin{center}
\picture{clip, height=5.5cm}{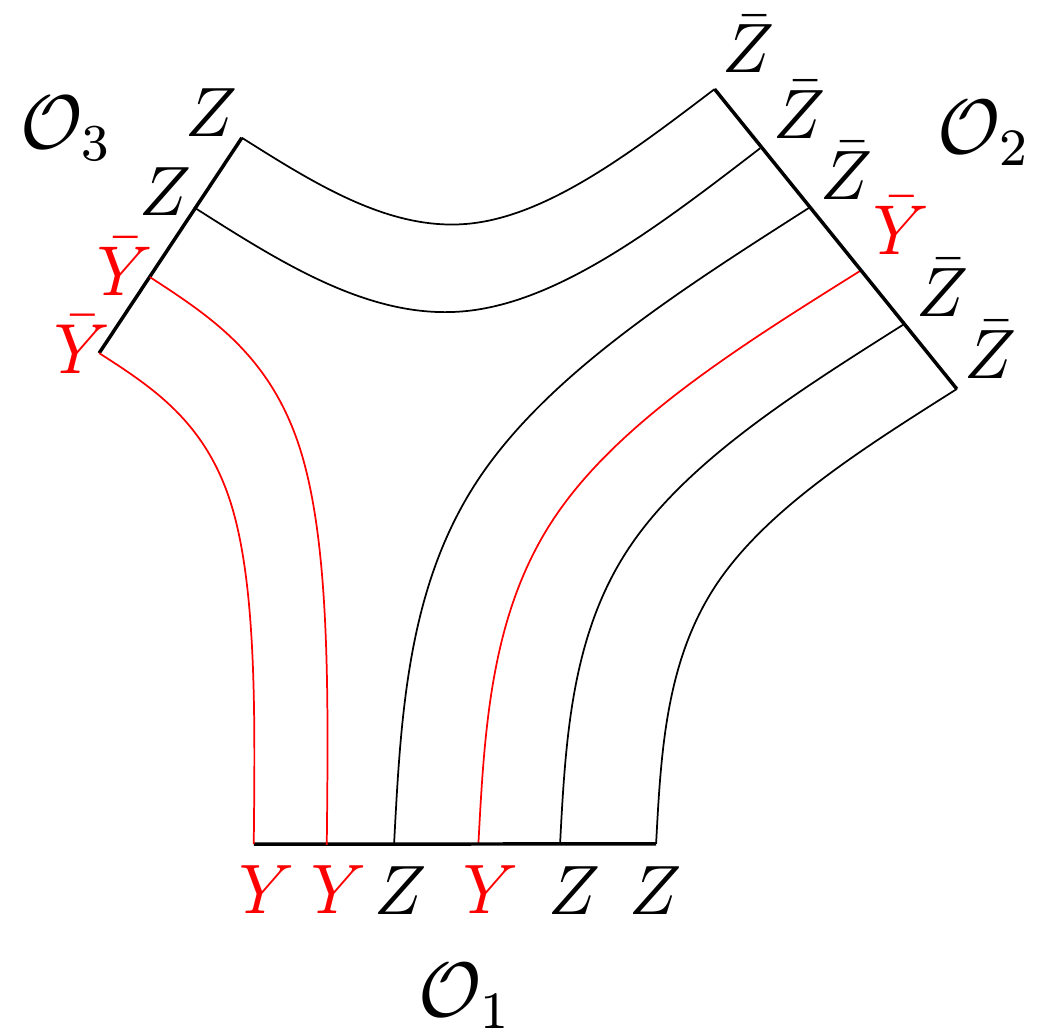}\\
\hspace{-20pt}$(a)$
\end{center}
 \end{minipage}
 \begin{minipage}{0.45\hsize}
 \begin{center}
\picture{clip, height=5.5cm}{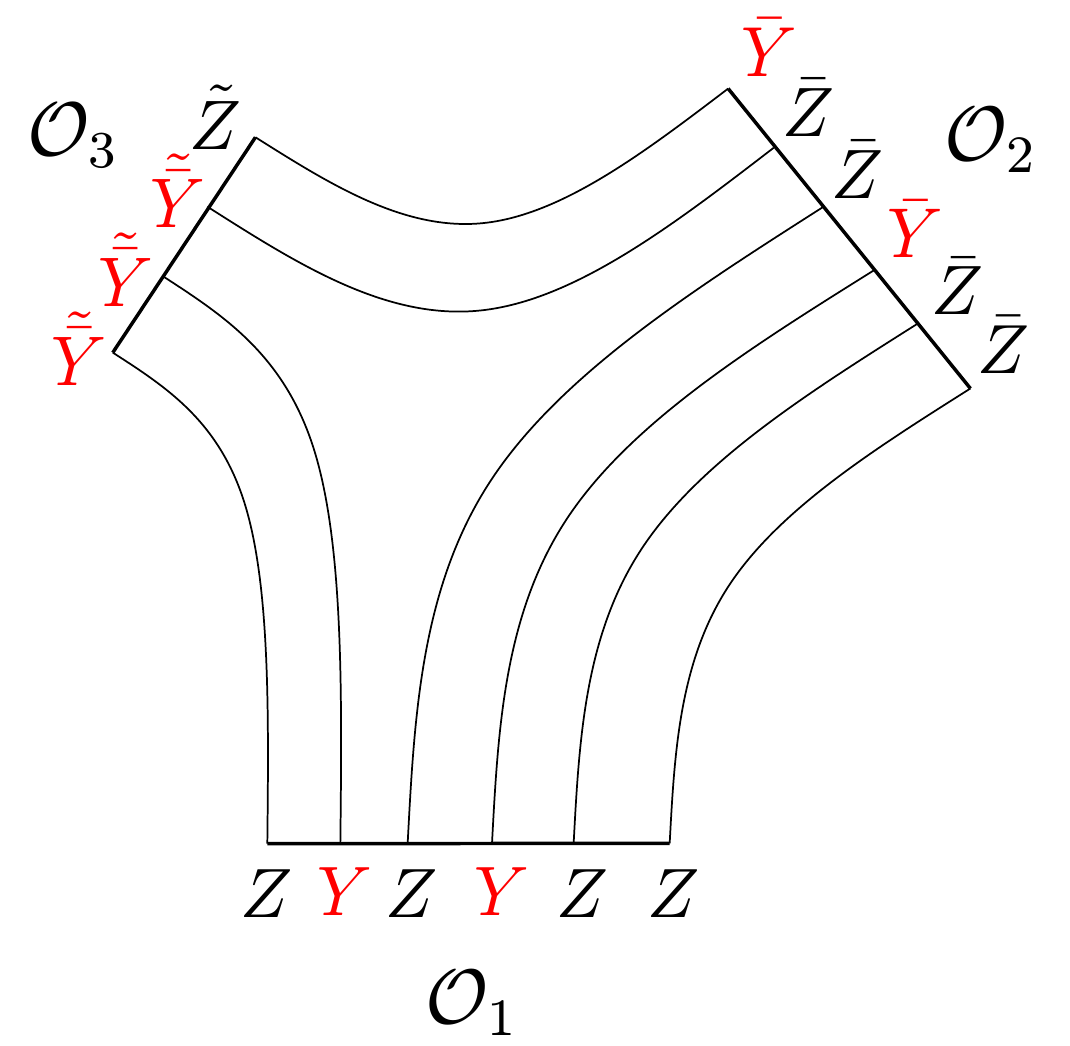}\\
\hspace{-20pt}$(b)$
\end{center}
 \end{minipage}
\caption{Two examples of type I-I-II three-point functions. In both figures, the fields denoted by black letters correspond to the vacuum and the fields denoted by red letters correspond to the magnons. $(a)$ The configuration studied in most of the literature (see e.~g.~\cite{Tailoring1}). It amounts to choosing the polarization vectors as $n_1=n_3=\tilde{n}_1=\tilde{n}_3=(1,0)^{t}$ and $n_2=\tilde{n}_2=(0,1)^t$ $(b)$ The configuration used in \cite{BKV}. $\tilde{Z}$ and $\tilde{\bar{Y}}$ are given by $\tilde{Z}=Z+\bar{Z}+Y-\bar{Y}$ and $\tilde{\bar{Y}}=\bar{Y}-\bar{Z}$ respectively. The polarization vectors in this case are given by $n_1=\tilde{n}_1=(1,0)^{t}$, $n_2=\tilde{n}_2=(0,1)^{t}$ and $n_3=\tilde{n}_3=(1,1)^t$.}
\label{config112}
\end{figure}

The first class of such three-point functions is called {\it type I-I-II}. These are the ones for which two of the operators have magnon excitations  in the same ${\rm SU}(2)$,  whereas the magnons for the third operator are in the other ${\rm SU}(2)$. Examples of such configurations are depicted in figure \ref{config112}. This class of three-point functions were studied extensively in the literature and it was shown in \cite{Foda,KM} that they can be expressed as a product of two Izergin-Korepin determinants \cite{IK,IK2}. From such exact expressions, the semi-classical limit was extracted  in \cite{Tailoring3,Kostov,Kostov2}. In what follows, we shall reproduce it from our result \eqref{forthemoment}.

Let us, for simplicity, consider the case where $\mathcal{O}_1$ and $\mathcal{O}_2$ belong to ${\rm SU}(2)_R$ and $\mathcal{O}_3$ belongs to ${\rm SU}(2)_L$. The structure constant factorizes into the left and the right parts as explained in section \ref{sec:semi-classical} and each part can be expressed in terms of integrals of the type given in \eqref{forthemoment}. To get an explicit expression for $C_{123}$ from \eqref{forthemoment}, we also need to know the BPS three-point functions $C_{123}^{\rm BPS}$. This can be easily computed as they are just a simple product of Wick contractions. The result is 
\beq{
\ln C_{123}^{\rm BPS}= \sum_{\{i,j,k\}\in {\rm cperm}\{1,2,3\}}\frac{L_i+L_j-L_k}{2}\left(\ln \sl{n_i,n_j}+\ln \sl{\tilde{n}_i,\tilde{n}_j}\right) \period
}
Using this expression, we can write down the result for the type I-I-II three-point function as
\beq{\label{finalI-I-II}
\ln C_{123} = \mathcal{K}+\mathcal{L}+\mathcal{R}+\mathcal{N}\comma
}
where each part is given by
\beq{
\mathcal{K}=&\sum_{\{i,j,k\}\in {\rm cperm}\{1,2,3\}}(Q_i+Q_j-Q_k)\ln  \sl{n_i,n_j}+(\tilde{Q}_i+\tilde{Q}_j-\tilde{Q}_k)\ln \sl{\tilde{n}_i,\tilde{n}_j}\comma \\
\mathcal{L}=&\frac{1}{2}\left(\oint_{\Gamma_3}\frac{du}{2\pi}{\rm Li}_2 \left(e^{ip_3+(L_1-L_2)/2u} \right)+\oint_{\Gamma_3}\frac{du}{2\pi}{\rm Li}_2 \left(e^{ip_3+(L_2-L_1)/2u} \right)\right)\label{forL}\comma \\
\mathcal{R}=&\frac{1}{2}\left(\oint_{\Gamma_1\cup \Gamma_2}\frac{du}{2\pi}{\rm Li}_2 \left(e^{ip_1+ip_2-iL_3/2u} \right)+\oint_{\Gamma_1}\frac{du}{2\pi}{\rm Li}_2 \left(e^{ip_1-ip_2+iL_3/2u} \right)\right.\nn\\
&\left. \quad +\oint_{\Gamma_2}\frac{du}{2\pi}{\rm Li}_2 \left(e^{-ip_1+ip_2+iL_3/2u} \right)\right)\label{forR}\comma\\
\mathcal{N}=&-\frac{1}{2}\sum_{k}\oint_{\Gamma_k} \frac{du}{2\pi} {\rm Li}_2 \left( e^{2ip_k}\right) \period 
}
Here $\mathcal{K}$ denotes the contribution determined purely by  kinematics and the ${\rm SU}(2)_{L,R}$ global charges $l_i$ and $r_i$ are given by
\beq{
\begin{aligned}\label{liri}
&\tilde{Q}_1=\frac{L_1}{2}\comma && \tilde{Q}_2=\frac{L_2}{2}\comma&& \tilde{Q}_3 = \frac{L_3}{2}-M_3 \comma\\
&Q_1=\frac{L_1}{2}-M_1\comma && Q_2=\frac{L_2}{2}-M_2\comma&& Q_3 =\frac{L_3}{2}\period
\end{aligned}
}
The second and the third terms $\mathcal{L}$ and $\mathcal{R}$ contain the dynamical information of the three-point functions and come from ${\rm SU}(2)_L$ and ${\rm SU}(2)_R$ respectively.  The last term $\mathcal{N}$ is the part corresponding to the norms of the Bethe states in the exact quantum expression (see for instance \cite{Tailoring1}).
To make a direct connection with the results in \cite{Kostov2}, we now rewrite the second and the third terms in $\mathcal{R}$ by pushing the contour onto the second sheet as we did in figure \ref{contourmanipulation}-$(c)$. Then the two terms read
\beq{\label{rewrite-L}
-\oint_{\Gamma_1}\frac{du}{2\pi}{\rm Li}_2 \left(e^{-(ip_1-ip_2+iL_3/2u)} \right)-\oint_{\Gamma_2}\frac{du}{2\pi}{\rm Li}_2 \left(e^{-(-ip_1+ip_2+iL_3/2u)} \right)\period
}
Now using the dilogarithm identity,
\beq{\label{rewriteLi}
{\rm Li}_{2}\left(\frac{1}{x}\right)= -{\rm Li}_2(x)-\frac{\pi^2}{6}-\frac{1}{2}\ln ^2 (-x)\comma
}
we can show\fn{The terms coming from the second and the third terms in the identity \eqref{rewriteLi} vanish.} that \eqref{rewrite-L} is identical to the first term in \eqref{forR}. By performing similar manipulation, we can also show that the first and the second terms in \eqref{forL} are equivalent. In this way, we can obtain the following alternative expression for $\mathcal{L}+\mathcal{R}$:
\beq{
\mathcal{L}+\mathcal{R}=\oint_{\Gamma_1\cup \Gamma_2}\frac{du}{2\pi}{\rm Li}_2 \left(e^{ip_1+ip_2-iL_3/2u} \right)+\oint_{\Gamma_3}\frac{du}{2\pi}{\rm Li}_2 \left(e^{ip_3+(L_2-L_1)/2u} \right)\period
}
Together with the norm part $\mathcal{N}$, this perfectly agrees with the result in \cite{Kostov2}.
\subsubsection*{Type I-I-I three-point function}
\begin{figure}[tb]
\begin{center}
\picture{clip, height=6cm}{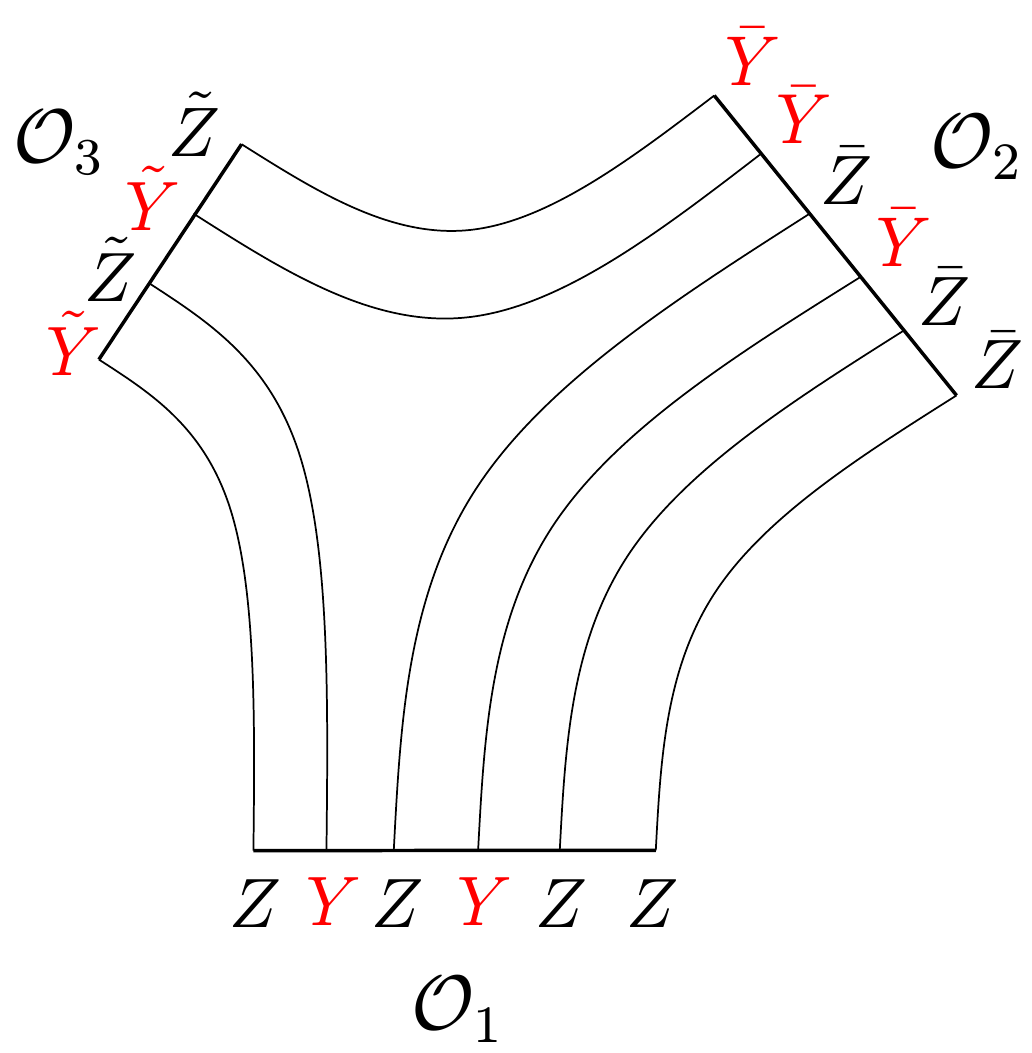}
\end{center}
\vspace{-0.8cm}
\caption{An example of the type I-I-I three-point functions studied in the \cite{BKV}. $\tilde{Y}$ in the figure represents $Y+\bar{Z}$. The polarization vectors are given by $n_1=\tilde{n}_1=(1,0)^{t}$, $n_2=\tilde{n}_2=(0,1)^{t}$ and $n_3=\tilde{n}_3=(1,1)^t$}
\label{config111}\vspace{-0.2cm}
\end{figure}

Let us now turn to the case where all the three operators have magnons in the same ${\rm SU}(2)$-sector. They are called {\it type I-I-I} in \cite{KKN1}. An example of this class of correlators is given in figure \ref{config111}. As compared to the type I-I-II correlators, they have much more complicated structure and the exact results known at weak coupling are given either in terms of the sum of the triple product of determinants \cite{KKN1} or in terms of the multiple-integral expression based on the separation of variables \cite{JKKS}. Both of these expressions are hard to deal with and their semi-classical limit has not been computed. Despite such complications for the 
exact result, the semiclassical result we derive below turned out to take a remarkably simple form.  It would certainly be a challenging future problem to reproduce it from the expressions given in \cite{KKN1} and \cite{JKKS}.

For definiteness, let us consider the case where all the operators belong to ${\rm SU}(2)_R$. In this case, there is no dynamical contribution from ${\rm SU}(2)_L$ and we can write down the full expression using \eqref{forthemoment} as
 \beq{\label{finalI-I-I}
 \ln C_{123}= \mathcal{K}+\mathcal{R}+\mathcal{N}\comma
  }
  with each part given by
\beq{
\mathcal{K}=&\sum_{\{i,j,k\}\in {\rm cperm}\{1,2,3\}}(Q_i+Q_j-Q_k)\ln  \sl{n_i,n_j}+(\tilde{Q}_i+\tilde{Q}_j-\tilde{Q}_k)\ln \sl{\tilde{n}_i,\tilde{n}_j}\comma\\
\mathcal{R}=&\frac{1}{2}\sum_{\{i,j,k\}\in{\rm cperm}\{1,2,3\}}\left(\oint_{\Gamma_i\cup \Gamma_j}\frac{du}{2\pi}{\rm Li}_2 \left(e^{ip_i+ip_j-ip_k} \right)\right)\comma\label{forL2}\\
\mathcal{N}=&-\frac{1}{2}\sum_{k}\oint_{\Gamma_k} \frac{du}{2\pi} {\rm Li}_2 \left( e^{2ip_k}\right)\period
}
Here the definitions of $l_i$ and $r_i$ are modified from \eqref{liri} in the following manner:
\beq{
\begin{aligned}\label{liri2}
&\tilde{Q}_1=\frac{L_1}{2}\comma && \tilde{Q}_2=\frac{L_2}{2}\comma&& \tilde{Q}_3 = \frac{L_3}{2} \comma\\
&Q_1=\frac{L_1}{2}-M_1\comma && Q_2=\frac{L_2}{2}-M_2\comma&& Q_3 =\frac{L_3}{2}-M_3\period
\end{aligned}
}
As advertized,  the expression above for the structure constant  is  as simple as  the one for the I-I-II type. 
\subsubsection*{Remark on the integration contour}
So far, we have been assuming that the cuts in $p_i$ are sufficiently small. In particular, we used this assumption when we derive the analyticity of the Wronskians. Let us briefly explain what we expect when we gradually increase the sizes of the cuts in the integral expression \eqref{forthemoment}.
\begin{figure}[tb]
\begin{center}
\picture{clip, height=4cm}{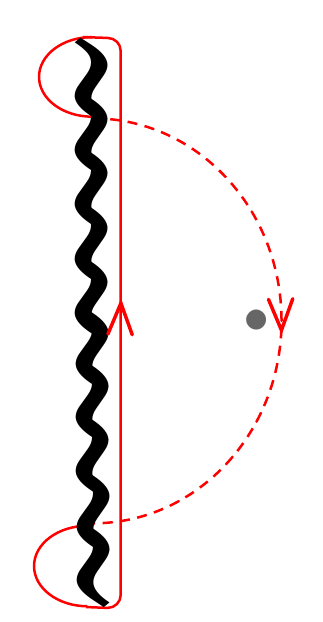}
\end{center}\vspace{-0.5cm}
\caption{The deformation of the contour due to the branch-point singularities. When the branch point (denoted by a black dot in the figure) crosses the cut, the contour around it must be deformed such that it avoids the point.}\label{contourdefpicture}
\end{figure}\vspace{-0.2cm}

Since the dilogarithm ${\rm Li}_2(x)$ has a branch cut emanating from $x=1$, the integrands of \eqref{forthemoment} contain infinitely many branch-point sigularities at $e^{i(p_i+p_j-p_k)}=1$ and $e^{2ip_i}=1$. These correspond to the zeros and the poles of the Wronskians respectively. As we increase the size of the cut, at some point, they start crossing the cut. When this happens, we need to deform the contour as depicted in figure \ref{contourdefpicture} in order to keep the final result continuous with respect to the size of the cut. Thus, if we consider the operators with large cuts, the integration contours will be substantially deformed and will no longer be given by the ones around the cuts. This would explain the observation made in \cite{Kostov2} that one must deform the contours appropriately in order to reproduce the value  obtained by numerics. It would be important to perform detailed numerical computation and confirm the claim we made here.
\section{Application to the strong coupling\label{sec:strong}}
One of the important findings of the present work is that, as far as the 
 semi-classical behaviors  are concerned,  the same structure and 
  the logic  underlie the  three point functions  both at  weak and  strong couplings. In this section we shall 
apply the machineries  developed so far to the computation at strong coupling.
\subsection{Classical integrability of string sigma model on $S^3$\label{subsec:integrabilitystrong}}
Let us first give a brief review\fn{For a more detailed account, see \cite{Vicedo1,Vicedo2,Vicedo3,KK3,KMMZ}.} of the classical integrability of the string sigma model on $S^3$ emphasizing the similarity to and the difference from the Landau-Lifshitz model discussed in section \ref{sec:LL}.

For the string sigma model on $S^3$, we can define two sets of Lax pairs as
\beq{
\left[\del + \frac{j_z}{1-x}\comma \,\,\delbar+\frac{j_{\barz}}{1+x}\right]=0\comma \qquad \left[\del + \frac{x\tilde{j}_z}{1-x}\comma \,\,\delbar-\frac{x\tilde{j}_{\barz}}{1+x}\right]=0\period
}
Here $x$ is the spectral parameter and the currents $j$ and $\tilde{j}$ are defined using the embedding coordinate $Y_i$ $(i=1,\ldots 4)$ as
\beq{
\begin{aligned}
j =G^{-1}dG \comma \qquad \tilde{j}=dGG^{-1}\comma\qquad G\equiv \pmatrix{cc}{Y_1+iY_2&Y_3+iY_4\\-Y_3+iY_4&Y_1-iY_2}\period
\end{aligned}
}
For each Lax pair, we have an auxiliary linear problem and a monodromy matrix:
\beq{
&\left( \del + \frac{j_z}{1-x}\right) \psi =0\comma \quad\!\! \left( \delbar + \frac{j_\zbar}{1+x}\right) \psi=0\comma \quad \Omega(x)\equiv {\rm P} \!\exp \left[ -\oint \left( \frac{j_z dz}{1-x}+ \frac{j_{\barz}d\barz}{1+x}\right)\right] \comma\label{rightset}\\
&\left( \del + \frac{x\tilde{j}_z}{1-x}\right) \tilde{\psi} =0\comma \quad \!\!\left( \delbar + \frac{xj_\zbar}{1+x}\right) \tilde{\psi}=0\comma \quad \tilde{\Omega}(x)\equiv {\rm P} \!\exp \left[ -\oint \left( \frac{x\tilde{j}_z dz}{1-x}- \frac{x\tilde{j}_{\barz}d\barz}{1+x}\right)\right]\period\label{leftset}
}
Note that the two sets of quantities defined above are related with each other by $\tilde{j}=GjG^{-1}$, $\tilde{\psi}=G\psi$ and $\tilde{\Omega}=G\Omega G^{-1}$.
As with the Landau-Lifshitz model, the quasi-momentum $p(x)$ is given by the logarithm of the eigenvalue of the monodromy matrix $\Omega \sim \tilde{\Omega}\sim {\rm diag} (e^{ip},e^{-ip})$. The spectral curve is defined also in a similar way as
\beq{
\det \Big( y  -\Omega(x) \Big)=\det \Big( y  -\tilde{\Omega}(x) \Big) = (y-e^{ip})(y-e^{-ip})=0\period
}

\begin{figure}[tb]
\begin{center}
\picture{clip, height=5.5cm}{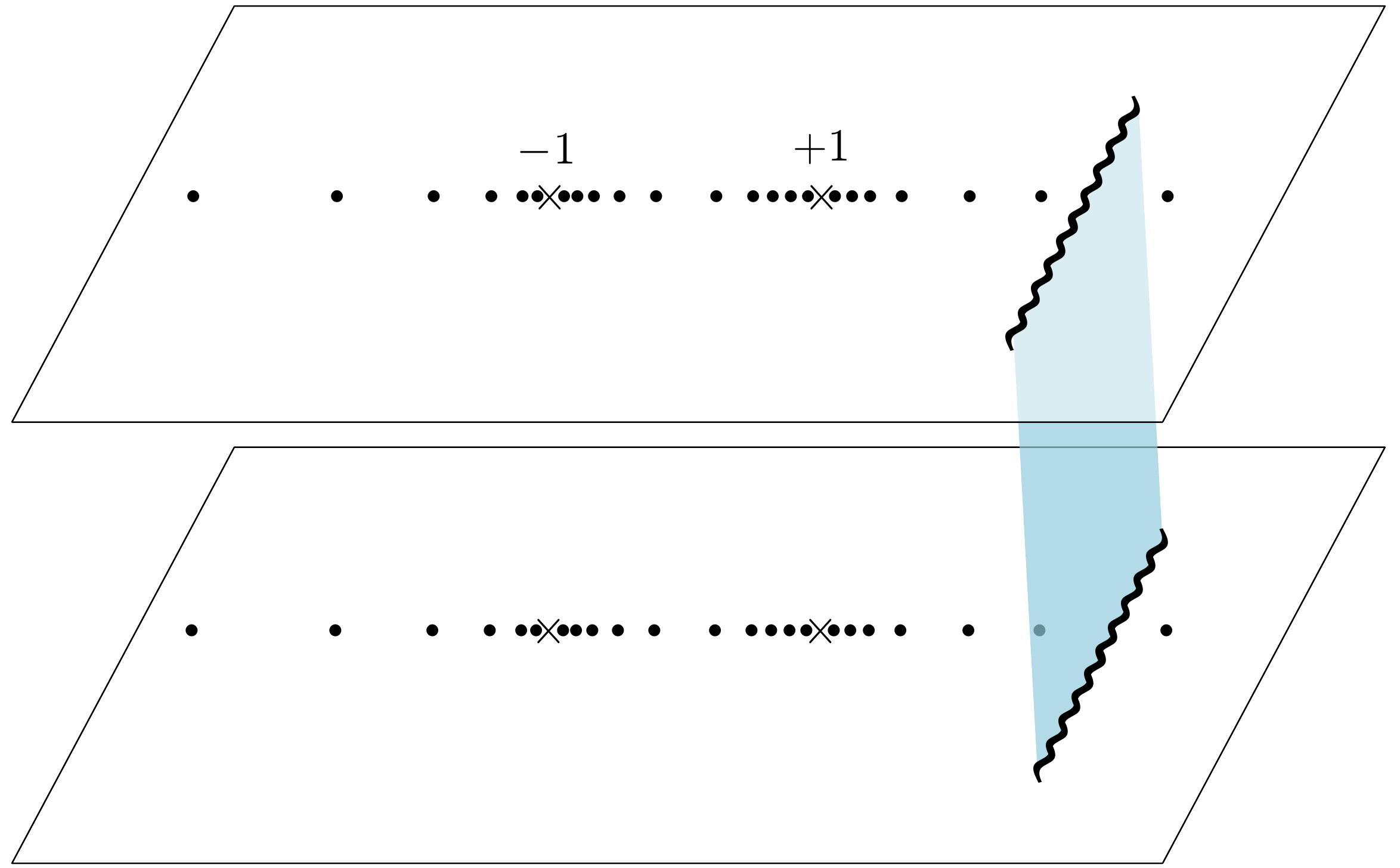}
\end{center}\vspace{-0.5cm}
\caption{The structure of the spectral curve at strong coupling. In addition to the branch cuts, it has infinitely many singular points, denoted by black dots, which accumulate to $x=\pm 1$. Those singular points should be regarded as degenerate branch points.}\label{spectralstrong}\vspace{-0.2cm}
\end{figure}

The asymptotic behavior of the quansi-momentum around $0$ and $\infty$ encodes the information of the global charges\fn{In the most general situation, the quasi-momentum around $x=0$ behaves as $p(x)\sim 2\pi m + x \tilde{Q}/g+\cdots$, where $m$ is an integer called the winding number. Here we are considering the $m=0$ case for simplicity.} as
\beq{
\begin{aligned}\label{asymptpstrong}
p(x)&\sim -\frac{Q}{g }\frac{1}{x} \quad(x\to \infty)\comma\qquad
p(x)\sim \frac{ \tilde{Q}}{g}x \quad(x\to 0)\comma
\end{aligned}
}
where $Q$ and $\tilde{Q}$ are the charges of the ${\rm SU}(2)_R$ and ${\rm SU}(2)_L$ respectively. We should note that, unlike the Landau-Lifshitz model, the quasi-momentum does not have a pole at $x=0$. Instead, it has poles at $x=\pm 1$ with residues given by the worldsheet\fn{$\mathcal{E}$ and $\mathcal{P}$ defined here are the energy and the momentum of the $S^3$ sigma model in the conformal gauge. They therefore do not have definite physical meaning. In particular $\mathcal{E}$ is in general different from the lightcone energy of the string sigma model in $AdS_5\times S^5$.} energy $\mathcal{E}$ and momentum $\mathcal{P}$:
\beq{
&p(x)\sim -\frac{\sqrt{(\mathcal{E}\pm \mathcal{P})/2g}}{x\mp 1} \qquad(x\to \pm 1)\period
}
Owing to this pole structure, the singular points of the spectral curve accumulate to $x=\pm1$ as shown in figure \ref{spectralstrong}.

As in the Landau-Lifshitz sigma model, the filling fractions are given by contour integrals on the spectral curve. However, their explicit forms are slightly modified:
\beq{
\begin{aligned}\label{action-strong}
&S_k \equiv \frac{1}{2\pi i}\oint_{\mathcal{C}_{k}} p(x)du(x)\period
\end{aligned}
}
Here $u(x)$ is the {\it rapidity} variable, given by 
\beq{\label{defofrap}
u(x)=g\left(x+\frac{1}{x} \right)\comma
}
and the integration contour goes around the $k$-th branch cut\fn{As in the Landau Lifshitz sigma model, we should consider the (infinitely many) singular points satisfying $e^{2ip(x)}=1$ also as (degenerate) branch cuts.} $\mathcal{C}_k$ counterclockwise on the first sheet.
\subsection{ ${\rm SU}(2)_L$ and ${\rm SU}(2)_R$ excitations at strong coupling\label{subsec:su2lr}}
One of the conspicuous differences  from the Landau-Lifshitz model is 
 that the filling fractions given by \eqref{action-strong} can be negative at strong coupling, and it turns out that the signs of the filling fractions are tied to whether the state has excitations in the ${\rm SU}(2)_L$ sector or in the ${\rm SU}(2)_R$ sector.

To understand  this point, let us consider the perturbation around the BMN vacuum. It was shown in \cite{GV,GNV} that the quasi-momentum receives the following correction when an infinitesimal cut is inserted at $x=x_\ast$:
\beq{\label{perturbp}
\delta p(x) = n \frac{dx}{du}\frac{1}{x-x_\ast}= n \frac{x^2}{g(x^2-1)}\frac{1}{x-x_\ast}\period
}
Here $n$ is the filling fraction inserted at $x=x_{\ast}$ and the factor $dx/du$ in \eqref{perturbp} is necessary due to the definition of $S_k$ in \eqref{action-strong}. We can also compute the energy shift using the results\fn{The argument roughly goes as follows: As is clear from \eqref{perturbp}, the perturbation modifies the behavior around $x=\pm 1$. Owing to the Virasoro constraint, the AdS quasi-momentum $\hat{p}$ around $x=\pm1$ must also be deformed in the same way. Once we understand how $\hat{p}$ is modified, we can then read off the energy shift from its asymptotic behavior at $x=\infty$.} in \cite{GV,GNV} as
\beq{\label{energy-shift}
\delta \Delta = \frac{2n}{x_{\ast}^2-1}\period
}
In \eqref{energy-shift}, the quantity $1/(x_{\ast}^2-1)$ is positive when $|x_{\ast}|>1$,  while it is negative when $|x_{\ast}|<1$. Since all the physical excitations around the BMN vacuum must have the positive energy shift\fn{In other words, one cannot lower the energy starting from the BMN vacuum.}, this means that $n$ must be positive if $|x_{\ast}|>1$ whereas it must be negative when $|x_{\ast}|<1$. This is in marked contrast with the situation in the Landau-Lifshitz model, where we always needed to take $n$ to be positive to describe the physical states. Physically, this is because the Bethe roots in the region $|x_{\ast}|<1$ correspond to anti-particles: In order to construct a physical state from the anti-particles, we need to insert ``holes'' just as in the Dirac's fermi sea. 

We can show more generally  that the filling fraction defined by \eqref{action-strong} must be positive whenever the cut is outside the unit circle whereas they must be negative whenever the cut is inside the unit circle. Now, to understand the physical meaning of these two types of cuts, let us consider the relation\fn{\eqref{sumgf} follows from the fact that $r$ and $l$ can be expressed as
\beq{
Q= \frac{1}{2\pi i}\oint_{x=\infty} p(x) du(x)\comma \qquad \tilde{Q}=-\frac{1}{2\pi i}\oint_{x=0}p(x)du(x)\period
}} between the global charges and the filling fractions:
\beq{\label{sumgf}
Q-\tilde{Q}+\sum_k S_k=0\period
}
For the BMN vacuum, all the filling fractions are zero and $Q$ and $\tilde{Q}$ are equal. Now, if we insert cuts outside the unit circle, which have positive filling fractions, we must either decrease $Q$ or increase $\tilde{Q}$ in order to satisfy \eqref{sumgf}. However, since the BMN vacuum has the maximal\fn{This is clear in particular at weak coupling. Whenever we excite magnons on the spin chain, the total global charge must always decrease as shown in \eqref{pinfx}.} $Q$ and $\tilde{Q}$, the only way to achieve this is to decrease $Q$. This clearly tells us that those states correspond to the ones  with excitations in ${\rm SU}(2)_R$. By a similar argument, we can show that the states with cuts inside the unit circle correspond to the states with ${\rm SU}(2)_L$ excitations. For a summary, see figure \ref{su2lrpic}. In Appendix \ref{ap:h}, we provide an interpretation of the ${\rm SU}(2)_L$- and ${\rm SU}(2)_R$-sectors from the point of view of the full spectral curve of the $AdS_5\times S^5$ sigma model.
\begin{figure}[tb]
\begin{minipage}{0.5\hsize}
\begin{center}
\picture{clip, height=5cm}{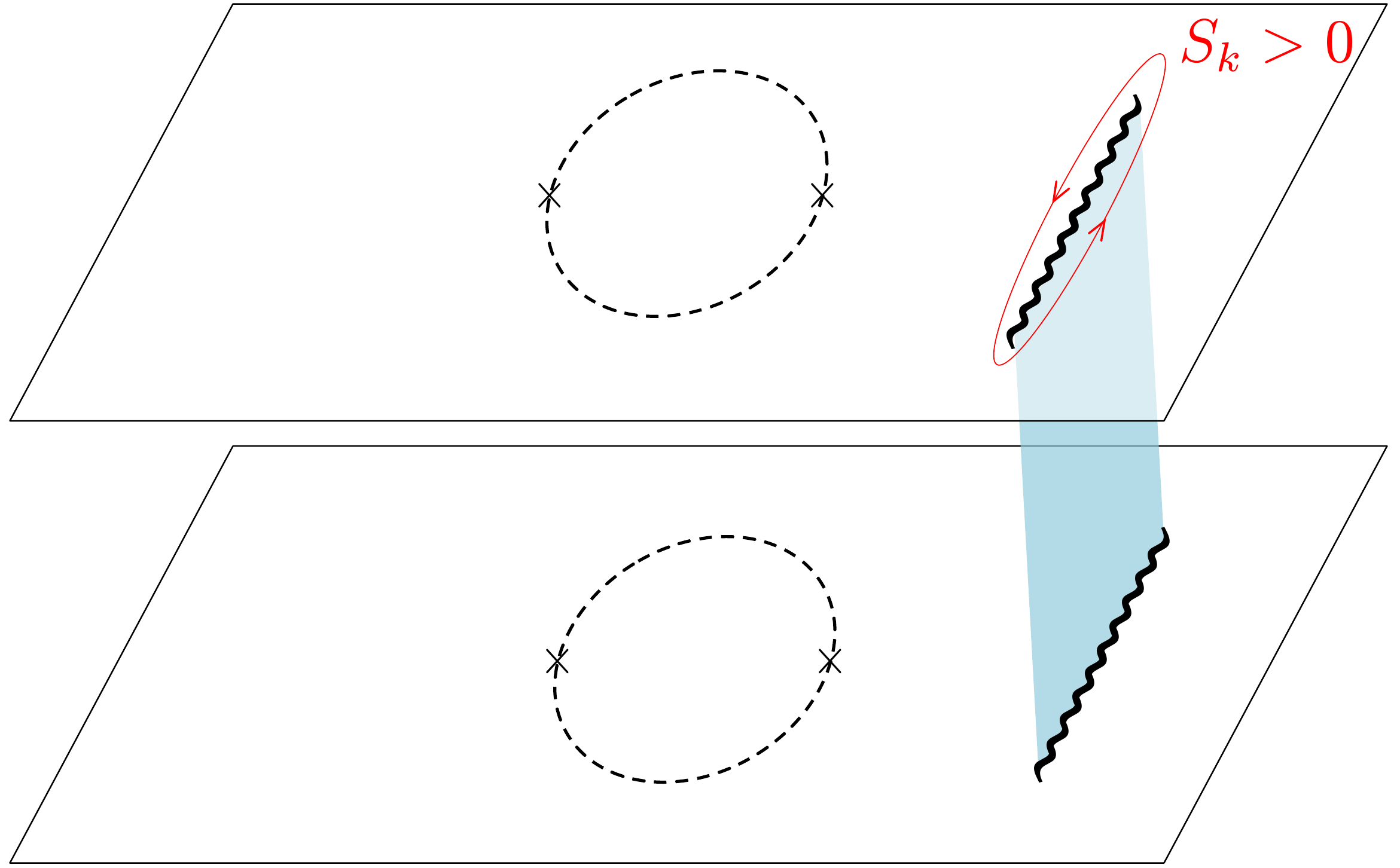}\\
${\rm SU}(2)_R$
\end{center}
\end{minipage}
\begin{minipage}{0.5\hsize}
\begin{center}
\picture{clip, height=5cm}{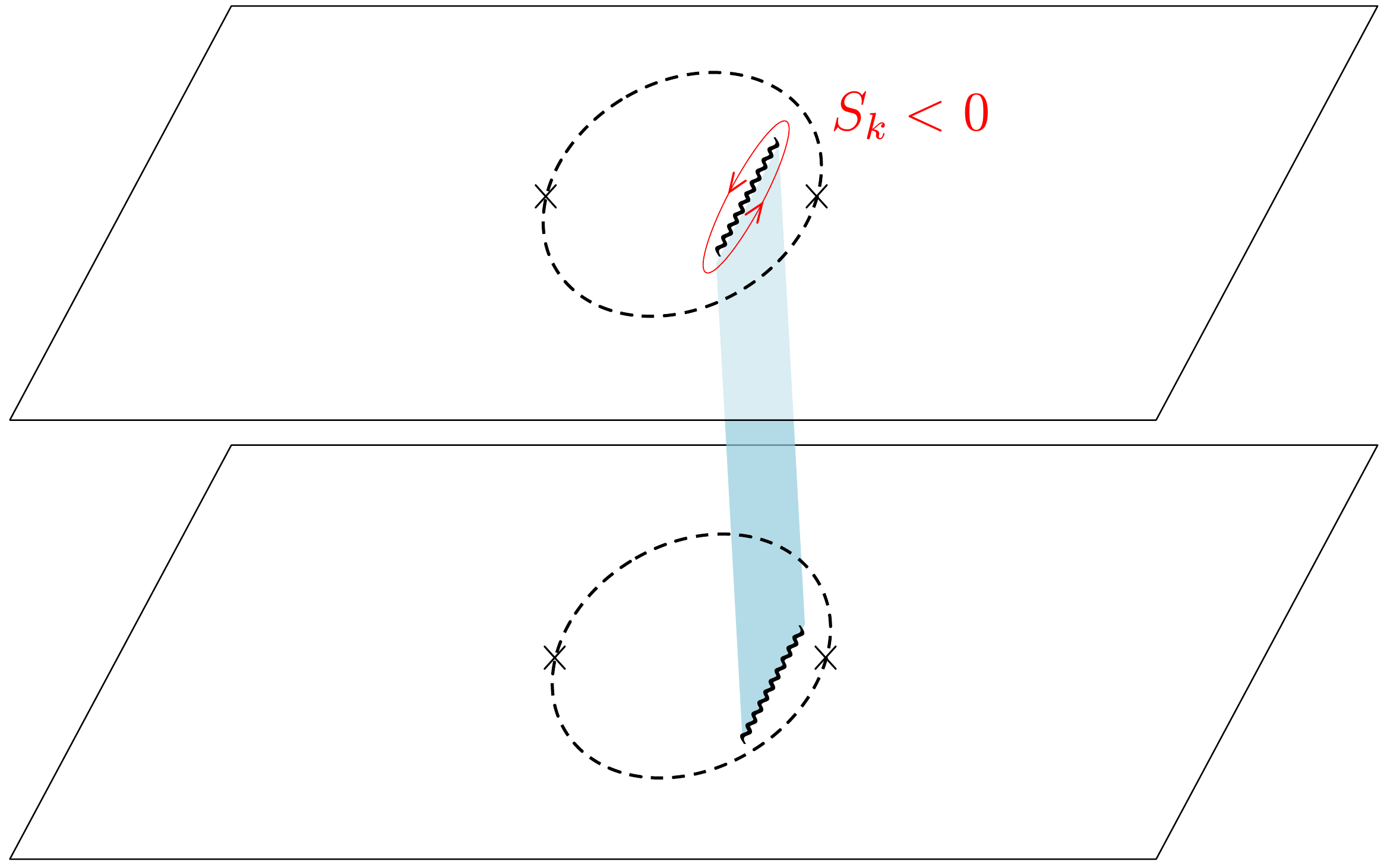}\\
${\rm SU}(2)_L$
\end{center}
\end{minipage}\vspace{-0.2cm}
\caption{The spectral curves for ${\rm SU}(2)_R$- and ${\rm SU}(2)_L$-sectors. The curve for ${\rm SU}(2)_R$ (left figure) contains branch cuts outside the unit circle and the filling fractions are positive. On the other hand, the curve ${\rm SU}(2)_L$ (right figure) has branch cuts inside the unit circle and the filling fractions are negative.\label{su2lrpic}}\vspace{-0.2cm}
\end{figure}

\subsection{Angle variables, $\ln C_{123}$ and Wronskians at strong coupling\label{subsec:anglestrong}}
With this knowledge, we now construct the angle variables which compute the derivative of $\ln C_{123}$, and express them in terms of the Wronskians. Below we shall treat the ${\rm SU}(2)_R$-sector and the ${\rm SU}(2)_L$-sector separately.
\subsubsection*{${\rm SU}(2)_R$-sector}
Let us first discuss the states with ${\rm SU}(2)_R$ excitations. To construct the angle variables, we should study the normalized eigenvectors of the monodromy matrix as in the Landau-Lifshitz model. One important difference in the present situation is that we now have two sets of linear problems and monodromy matrices. For the ${\rm SU}(2)_R$, the appropriate one to use is \eqref{rightset}. This is because \eqref{leftset} is invariant under the ${\rm SU}(2)_R$ transformations and is therefore insensible to the ${\rm SU}(2)_R$ excitations. 

As in section \ref{subsec:action-angle}, the separated variables can be constructed from the poles $\gamma_i$ of the normalized eigenvector $h(x)$, 
\beq{
h(x)\equiv \frac{1}{\sl{n\comma  \psi_{+}}}\psi_+\period
}
Here $\psi_{\pm}$ are the solutions to the auxiliary linear problem \eqref{rightset} satisfying
\beq{
\Omega(x)\psi_{\pm}(x)=e^{\pm ip (x)}\psi_{\pm}(x)\period
}
As shown in \cite{Vicedo2}, a pair of canonically conjugate variables at strong coupling is given not by $(\gamma_i,-ip(\gamma_i))$ but by $(u(\gamma_i),-ip(\gamma_i))$, where $u(x)$ is the rapidity defined by \eqref{defofrap}. This explains the form of the filling fraction given in \eqref{action-strong}.

Now, to construct the angle variables, we need to consider the generating function of the canonical transformation \eqref{generatingfunction} and then differentiate it with respect to $S_k$. As explained in the previous subsection, we should simultaneously decrease the global charge $Q$ when we vary $S_k$. This amounts to adding to $p(x)du(x)$ a one form whose integral does not vanish only for the cycle around $\mathcal{C}_k$ and the cycle at infinity. As a result, we get
\beq{
\phi_k = 2\pi  \sum_j \int^{\gamma_j^{\rm 3pt}}_{\gamma_j^{\rm 2pt}} \omega_k\comma
 }
 where $\omega_k$ is the one form satisfying
 \beq{
 \oint_{\mathcal{C }_j}\omega_k =\delta_{kj}\comma \quad \oint_0 \omega_{k}=0\comma \quad \oint_{\infty}\omega_k=-1\period
 }

Let us next express the derivative of $\ln C_{123}$ in terms of the angle variables. The arguments leading to \eqref{unambiguous} are by and large applicable also to the present case, except for one important point. At strong coupling, in addition to the contribution from the $S^3$ part of the sigma model, we should also include the contribution from the AdS part. In particular, whenever we perturb the filling fraction in the $S^3$ part, we inevitably change the conformal dimension $\Delta_i$, which is a global charge in AdS. This leads to the following modification of \eqref{unambiguous}:
\beq{\label{modifiedder}
\frac{\del \ln C_{123}}{\del S_{k_i}^{(i)}} =i\phi_{k_i}^{(i)}+i \frac{\del \Delta_i}{\del S_{k_i}^{(i)}} \phi_{\Delta}^{(i)}\period
}
Here $\phi_{\Delta}^{(i)}$ is the angle variable conjugate to $\Delta_i$, whose definition is given in Appendix \ref{ap:g}.

Now, following the argument in section \ref{subsec:anglewron}, we can express the angle variable $\phi_{k_i}^{(i)}$ in terms of the Wronskians and the result takes the same form as \eqref{eq-34}. We can perform similar analysis also to the AdS part (see Appendix \ref{ap:g} for details) to get the following expression of the angle variable $\phi_{\Delta}^{(i)}$: 
\beq{\label{AdSanglewron}
\phi_{\Delta}^{(i)}=\frac{i}{2}\ln \left(\frac{|x_i-x_j|^2|x_k-x_i|^2}{|x_j-x_k|^2}\left.\frac{\sl{j_{-}\comma k_{-}}}{\sl{i_{-}\comma j_{-}}\sl{k_{-}\comma i_{-}}}\right|_{x=\infty} \left.\frac{\sl{j_{+}\comma k_{+}}}{\sl{i_{+}\comma j_{+}}\sl{k_{+}\comma i_{+}}}\right|_{x=0} \right) \period
}
Here $x_i$ denotes the position of the operator $\mathcal{O}_i$ and the eigenvectors $i_{+}$'s and $\tilde{i}_{+}$'s are the solutions to the auxiliary linear problems of the $AdS_3$ sigma model.
\subsubsection*{${\rm SU}(2)_L$-sector}
For the ${\rm SU}(2)_L$-sector, the linear problem we should consider is \eqref{leftset},  as  it is the one  that transforms nontrivially under the ${\rm SU}(2)_L$ transformation. 
 
In this case, the separated variables in the ${\rm SU}(2)_L$ sector can be constructed from the poles $\tilde{\gamma}_i$ of the normalized eigenvector $\tilde{h}(x)$,
\beq{
\tilde{h}(x)\equiv \frac{1}{\sl{ \tilde{n}\comma \tilde{\psi}_+}}\tilde{\psi}_{+}\period
}
Here $\tilde{\psi}_{+}$ is the solution to the auxiliary linear problem \eqref{leftset} satisfying
\beq{
\tilde{\Omega}(x)\tilde{\psi}_{\pm} (x)=e^{\pm ip(x)}\tilde{\psi}_{\pm}(x)\period
}
 Then the separated variables can be constructed from the poles at $\tilde{\gamma}_i$ as $(u(\tilde{\gamma}_i),-ip(\tilde{\gamma}_i))$.
 
 From the separated variables, we can construct the angle variables in the same manner as for the ${\rm SU}(2)_L$-sector. The only modification in the present case is that, when we change the filling fraction $S_k$, we need to change $l$ but not $r$ as discussed in section \ref{subsec:su2lr}. This can be achieved by adding to $p(x)du(x)$ a one form whose integral does not vanish only for the cycle around $\mathcal{C }_k$ and the cycle around $x=0$. Then we get the expression,
 \beq{
 \tilde{\phi}_k =2\pi \sum_{j}\int^{\gamma^{\rm 3pt}_j}_{\gamma_j^{\rm 2pt}} {\omega}_k\comma
 }
 where ${\omega}_k$ is the one form satisfying\fn{Here the contour for the second integral goes around $x=0$ on the first sheet counterclockwise.}
 \beq{
 \oint_{\mathcal{C}_j}{\omega}_k =\delta_{jk}\comma \quad \oint_{0}{\omega}_k=-1\comma \quad 
\oint_{\infty}{\omega}_k=0\period
 }
Using these angle variables, we can express the derivative of $\ln C_{123}$ as\fn{We shall not present the derivation here since it closely parallels the one for the ${\rm SU}(2)_R$.}
\beq{\label{deranglel}
\frac{\del \ln C_{123}}{\del S_{k_i}^{(i)}} =i\tilde{\phi}_{k_i}^{(i)}+i \frac{\del \Delta_i}{\del S_{k_i}^{(i)}} \phi_{\Delta}^{(i)}\period
}
Here, as in the previous relation \eqref{modifiedder},   $\phi_{\Delta}^{(i)}$ is the AdS angle variable \eqref{AdSanglewron}.

Let us next express the angle variables in terms of the Wronskians. Although the basic logic in section \ref{subsec:anglewron} applies also to the present case, we have to modify \eqref{eq-33} and \eqref{eq-34} appropriately as follows:
 \beq{\label{anglerepsu2l}
 \begin{aligned}
 \tilde{\phi}_k &=2\pi \sum_j \int^{\gamma_j^{\rm 3pt}}_{\gamma_j^{\rm 2pt}}{\omega}_k=-2\pi \int_{0^{-}}^{0^{+}}\sum_j \tilde{\omega}_{\gamma_j^{\rm 3pt}\gamma_j^{\rm 2pt};k}=i\int^{0^{+}}_{0^{-}}d\ln \frac{\sl{\tilde{n}\comma \tilde{\psi}_{+}^{\rm 3pt}}}{\sl{\tilde{n}\comma \tilde{\psi}_{+}^{\rm 2pt}}}-e_{k}\\
 &=\left.i\ln \left( \frac{\sl{\tilde{n}\comma \tilde{\psi}_{+}^{\rm 3pt}}\sl{\tilde{n}\comma \tilde{\psi}_-^{\rm 2pt}}}{\sl{\tilde{n}\comma \tilde{\psi}_{-}^{\rm 3pt}}\sl{\tilde{n}\comma \tilde{\psi}_+^{\rm 2pt}}}\right)\right|_{x=0}-i\int_{0^{-}}^{0^{+}}e_{k}\period
 \end{aligned}
 }
Here the one forms $\tilde{\omega}_{PQ;k}$ and $e_{k}$ are defined by \eqref{defabelian} and \eqref{eq-32} respectively. 

To express \eqref{anglerepsu2l} in terms of Wronskians,  we use the highest weight condition again. In this case, we should study the behavior of the monodromy matrix $\tilde{\Omega}(x)$ around $x=0$ on the first sheet,
\beq{
\tilde{\Omega}(x)={\bf 1}+ix\pmatrix{cc}{\tilde{S}_3&\tilde{S}_-\\\tilde{S}_{+}&-\tilde{S}_3}+\cdots \period
}
Applying the argument similar to the one in section \ref{subsec:anglewron}, we arrive at the following form of the eigenvectors at $x=0$ (on the first sheet):  
\beq{\label{generalx0l}
\tilde{\psi}_{+}(0)=a \tilde{n}\comma \qquad \tilde{\psi}_{-}(0)=a^{-1}i\sigma_2 \tilde{n}+b\tilde{n}\period
 } 
 
 Using \eqref{anglerepsu2l} and \eqref{generalx0l}, we finally get the expression for the angle variables in terms of the Wronskians:
 \beq{\label{su2lfinalangle}
 \tilde{\phi}_{k_i}^{(i)} = i\ln \left(\frac{\sl{\tilde{n}_i\comma \tilde{n}_j}\sl{\tilde{n}_k\comma \tilde{n}_i}}{\sl{\tilde{n}_j\comma \tilde{n}_k}}\left.\frac{\sl{j_{+}\comma k_{+}}}{\sl{i_{+}\comma j_{+}}\sl{k_{+}\comma i_{+}}}\right|_{x=0} \right)-i\int_{0^{-}}^{0^{+}}e_{k_i}^{(i)}\period
 }
Here the Wronskians are evaluated on the first sheet and 
$\bar{\phi}_{k_i}^{(i)}$ denotes the angle variable of the operator $\mathcal{O}_i$ associated with the $k_i$-th cut whereas $\tilde{n}_i$ is the ${\rm SU}(2)_L$ polarization vector for $\mathcal{O}_i$. To derive \eqref{su2lfinalangle}, we used the fact that the Wronskians among $i_+$'s are equivalent to the Wronskians among $\tilde{i}_{+}$'s, $\sl{i_{+}\comma j_{+}}=\sl{\tilde{i}_+\comma\tilde{j_+}}$. This is because two sets of eigenvectors are related with each other by the similarity transformation, $\tilde{i}_{+}=G \,i_{+}$, and the Wronskians are invariant under such transformation. 
\subsection{Semi-classical orthogonality of on-shell states at strong coupling\label{subsec:orthogonalstrong}}
\begin{figure}[tb]
\begin{center}
\picture{clip,height=5.5cm}{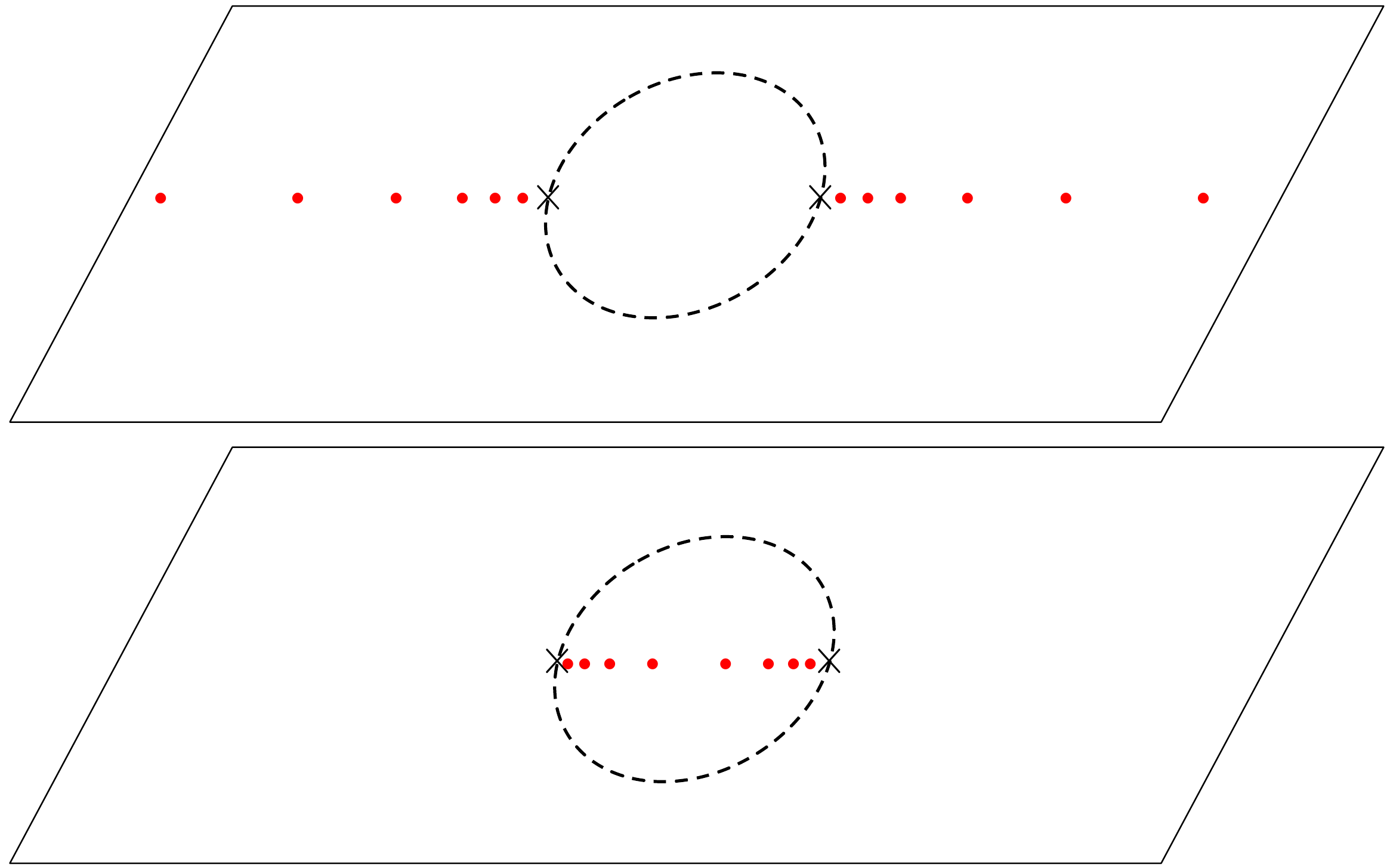}
\end{center}\vspace{-0.5cm}
\caption{The positions of the separated variables for two-point functions at strong coupling. From the orthogonality of on-shell states \eqref{orthogonalstrong}, we conclude that the separated variables are either at the singular points outside the unit circle on the first sheet, or at the singular points inside the unit circle on the second sheet.\label{SoVstrongpic}}\vspace{-0.2cm}
\end{figure}
The key ideas for determining  the analyticity of the Wronskians in the Landau-Lifshitz model were the requirement of the semi-classical orthogonality between two different on-shell states and the assumption of the continuity between the BPS correlators and the non-BPS correlators. Here we apply these two ideas to the analysis at strong coupling.

Just as for weak coupling,  one can construct, at strong coupling,  a different on-shell state by introducing an infinitesimal cut at the position of the singular point. We should, however, be careful about whether the perturbation is physical or not: As explained in section \ref{subsec:su2lr}, in order to obtain a physical state, we should insert a positive filling fraction when the singular point is outside the unit circle, 
 whereas we should insert a negative filling fraction when the singular point is inside the unit circle. With this in mind, we now impose the orthogonality condition
\beq{\label{orthogonalstrong}
\langle \psi |\psi +\delta\psi \rangle=0\period
}
Here $\delta \psi$ must correspond to a physical perturbation in the sense explained above.  Now it is not so hard to verify that the argument in section \ref{subsec:orthogonality} applied to the present case leads to the conclusion that the separated variables are at the singular points outside the unit circle on the first sheet, or at the singular points inside the unit circle on the second sheet (see figure \ref{SoVstrongpic}). Then, repeating the argument\fn{Since the monodromy relation at strong coupling takes exactly the same form as \eqref{semiclassicalmonod}, the equation \eqref{products} holds without modification.}
given in section \ref{subsec:analyticity}, with  the above  modification taken into account, we can determine the poles and the zeros of the Wronskians. The results are summarized in table  \ref{tab-strong2}.

Now, using these analyticity  properties, we can solve the Riemann-Hilbert problem and determine the individual Wronskians as described in section \ref{subsec:RH}. The main difference in the present case is that the Wronskians change the analyticity when they cross $|x|=1$. This leads to extra integration contours around the unit circle. Once the Wronskians are determined, we can compute the angle variable and then determine the structure constants using \eqref{modifiedder} and \eqref{deranglel}. The results will be given explicitly in the next subsection.
\begin{table}[h]
\begin{center}
\begin{tabular}{cc||c|c||c|c}
&&$1/\sin p_i$&$1/\sin p_j$&${\displaystyle \sin \frac{p_i+p_j+p_k}{2}}$&${\displaystyle \sin \frac{p_i+p_j-p_k}{2}}\hspace{10pt}$\\\hline
$|x|>1$&$\sl{i_+,j_+}$&\checkmark&\checkmark&\checkmark&\checkmark\\
&$\sl{i_-,j_-}$&&&&\\\hline
$|x|<1$&$\sl{i_+,j_+}$&&&&\\
&$\sl{i_-,j_-}$&\checkmark&\checkmark&\checkmark&\checkmark
\end{tabular}\\
\vspace{0.5cm}
\begin{tabular}{cc||c|c||c|c}
&&$1/\sin p_i$&$1/\sin p_j$&${\displaystyle \sin \frac{p_i-p_j+p_k}{2}}$&${\displaystyle \sin \frac{-p_i+p_j+p_k}{2}}$\\\hline
$|x|>1$&$\sl{i_+,j_-}$&\checkmark&&\checkmark&\\
&$\sl{i_-,j_+}$&&\checkmark&&\checkmark\\\hline
$|x|<1$&$\sl{i_+,j_-}$&&\checkmark&&\checkmark\\
&$\sl{i_-,j_+}$&\checkmark&&\checkmark&
\end{tabular}
\caption{The analytic properties of $\sl{i_{\pm},j_{\pm}}$ on the $[u,u,u]$-sheet at strong coupling. \label{tab-strong2}}
\end{center}\vspace{-0.2cm}
\end{table}
\subsection{Results and discussions\label{subsec:finalstrong}}
We now write down the results for the three-point functions at strong coupling explicitly and compare them with the results in \cite{KK3}.
\subsubsection*{Type I-I-II three-point functions}
\begin{figure}[t]
   \begin{center}
   \picture{clip,height=3.5cm}{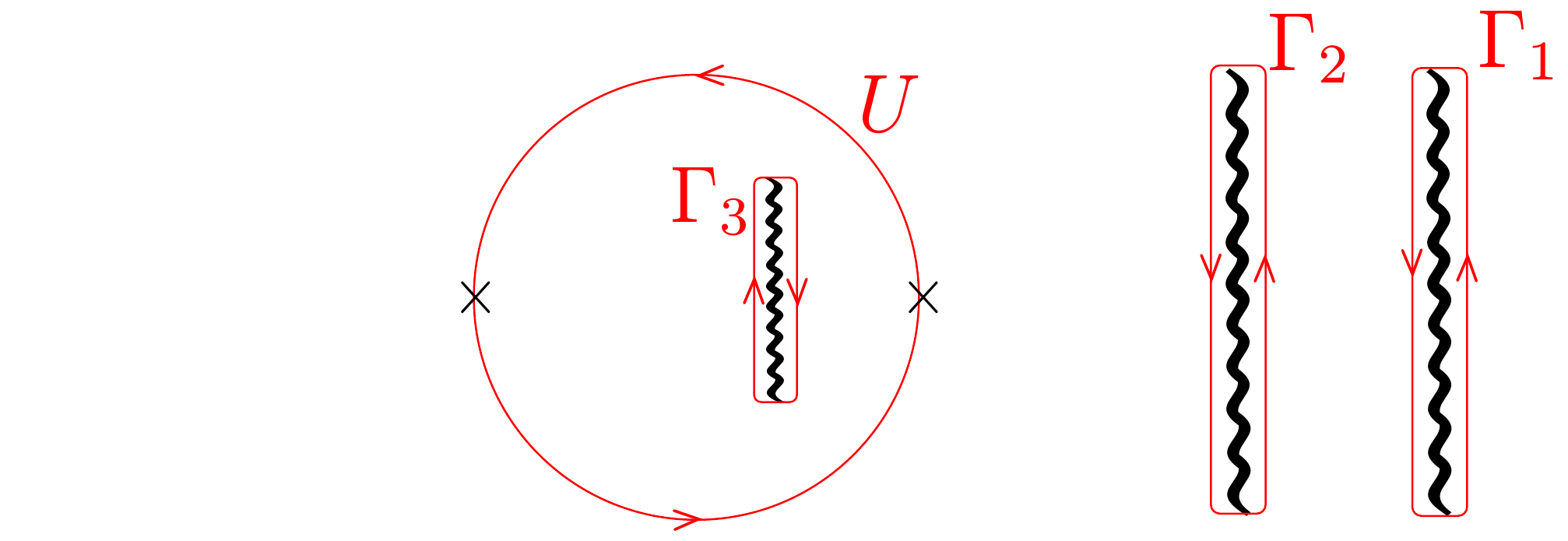}
   \end{center}\vspace{-0.5cm}
   \caption{Integration contours for the type I-I-II three-point functions. $\Gamma_1$ and $\Gamma_2$ encircle counterclockwise the branch cuts of $p_1$ and $p_2$ respectively,  whereas $\Gamma_3$ goes around the branch cuts of $p_3$ clockwise. $U$ is the contour which goes counterclockwise around the unit circle.\label{112contourpic}}\vspace{-0.2cm}
   \end{figure}

Let us first consider the type I-I-II three-point functions. Below we assume that $\mathcal{O}_1$ and $\mathcal{O}_2$ belong to ${\rm SU}(2)_R$  while $\mathcal{O}_3$ belongs to ${\rm SU}(2)_L$.
For such a three-point function, the result has the following structure:
\beq{\label{112strong}
\ln C_{123}= \mathcal{K}+\mathcal{D}_{\rm S}-\mathcal{D}_{\rm AdS}
}
Here $\mathcal{K}$ is the kinematical part given by
\beq{\label{Kstrong112}
\begin{aligned}
\mathcal{K}=\sum_{\{i,j,k\}\in {\rm cperm}\{1,2,3\}}&(Q_i+Q_j-Q_k)\ln  \sl{n_i,n_j}+(\tilde{Q}_i+\tilde{Q}_j-\tilde{Q}_k)\ln \sl{\tilde{n}_i,\tilde{n}_j}\\
&-(\Delta_i+\Delta_j-\Delta_k)\ln |x_i-x_j|\comma
\end{aligned}
}
where $Q_i$ and $\tilde{Q}_i$ are the $S^3$ global charges of the operator $\mathcal{O}_i$, and the term in the second line comes form the AdS part. $\mathcal{D}_{\rm S}$ and $\mathcal{D}_{\rm AdS}$ denote the dynamical parts coming from the $S^3$ part and $AdS_3$ part respectively. Both $\mathcal{D}_{\rm S}$ and $\mathcal{D}_{\rm AdS}$ consist of several factors as
\beq{\label{Dsadsstrong112}
\begin{aligned}
\mathcal{D}_{\rm S} &= \left( \mathcal{L}+\mathcal{R}\right)_{\rm S} + \mathcal{N}_{\rm S}\comma\qquad\mathcal{D}_{\rm AdS} = \left( \mathcal{L}+\mathcal{R}\right)_{\rm AdS} + \mathcal{N}_{\rm AdS}\comma
 \end{aligned}
 } 
and each factor is given as follows: 
\beq{
\begin{aligned}\label{explain2}
\left( \mathcal{L}+\mathcal{R}\right)_{\rm S}=&\frac{1}{2}\left(\oint_{2U}\frac{du}{2\pi}{\rm Li}_2 \left(e^{ip_1+ip_2+ip_3} \right)+\oint_{\Gamma_1\cup\Gamma_2\cup2 U}\frac{du}{2\pi}{\rm Li}_2 \left(e^{ip_1+ip_2-ip_3} \right)\right.\\
& \left.+\oint_{\Gamma_1\cup\Gamma_3\cup 2U}\frac{du}{2\pi}{\rm Li}_2 \left(e^{ip_1-ip_2+ip_3} \right)+\oint_{\Gamma_2\cup\Gamma_3 \cup 2U}\frac{du}{2\pi}{\rm Li}_2 \left(e^{-ip_1+ip_2+ip_3} \right)\right)\comma\\
\mathcal{N}_{\rm S}=&-\frac{1}{2}\sum_{k}\oint_{\Gamma_k\cup 2U} \frac{du}{2\pi} {\rm Li}_2 \left( e^{2ip_k}\right)\comma
\end{aligned}
}
\beq{\label{L+Rstrong112}
\begin{aligned}
\left( \mathcal{L}+\mathcal{R}\right)_{\rm AdS}=&\oint_{U}\frac{du}{2\pi}{\rm Li}_2 \left(e^{i\hat{p}_1+i\hat{p}_2+i\hat{p}_3} \right)+\sum_{\{i,j,k\}\in{\rm cperm}\{1,2,3\}}\oint_{U}\frac{du}{2\pi}{\rm Li}_2 \left(e^{i\hat{p}_i+i\hat{p}_j-i\hat{p}_k} \right)\comma\\
\mathcal{N}_{\rm AdS}=&-\sum_{k}\oint_{U} \frac{du}{2\pi} {\rm Li}_2 \left( e^{2i\hat{p}_k}\right)\period
\end{aligned}
}
The contours of integration are depicted in figure \ref{112contourpic} and $\hat{p}_i$ is the AdS quasi-momentum given by
\beq{
\hat{p}_i = \frac{\Delta_i x}{2g(x^2-1)}\period
  }
    
A few remarks are in order. Firstly, as shown in \eqref{explain2}, the integrals along the unit circle are multiplied by the extra factor of 2 (denoted by $2U$)  as compared to the integrals along the cuts\fn{S.K. would like to thank Y.~Jiang, I.~Kostov and D.~Serban for the correspondence related to this point.}. This factor can be deduced by carefully applying the argument given in section \ref{sec:resultweak} to the strong coupling analysis. Roughly speaking, this is because the integrals along the unit circle exist on every sheet of the eight-sheeted Riemann surface whereas the integrals along the cuts exist only on some (roughly the half) of the sheets (see figure \ref{gamma+++}). Secondly, each integral along $U$ is actually divergent owing to the poles in $p_i$ at $x=\pm 1$. However, such divergences cancel out when we combine 
all the terms in (7.30). 
 To illustrate this point, let us consider the integral
\begin{align}
\int_{U} \frac{du}{2\pi} {\rm Li}_2 \left(e^{i(p_1+p_2+p_3)} \right)\,.
\end{align} 
Since we are interested in the behavior around $x=\pm 1$, where the integrand develops singularities, we can approximate the quasi-momenta by their asymptotic form just as in (7.35), namely $p_i(x) \sim \Delta_i x/(2g(x^2-1))$. 
To see the behavior in the vicinity of $x= \pm 1$ on the unit circle, we parametrize the Zhukowsky variable as $x=e^{i\theta}$ near $x=+1$ and as $x=-e^{-i\theta}$ near $x=-1$ and expand the expression for $p_i(x)$ above with respect to $\theta$. In both cases, the result reads 
\begin{align}
p_i(x) \sim -\frac{i\Delta_i}{4g\theta}+O(\theta)\,.
\end{align}
Plugging this expression into  the dilogarithm, we obtain
\begin{align}
{\rm Li}_2\left( e^{i(p_1+p_2+p_3)}\right)\sim {\rm Li}_2\left(e^{(\Delta_1+\Delta_2 +\Delta_3)/(4g\theta)} \right)\period
\end{align}
Since ${\rm Li}_2 (0)$ is finite, there will be no singularity when $\theta$ approaches zero from below ({\it i.e.}  when the integration variable is on the lower semi-circle). On the other hand, when $\theta$ approaches zero from above, the argument of the dilogarithm diverges and we need to use its  asymptotic expression 
\begin{align}
{\rm Li}_2 (z)\propto -\frac{1}{2} \log^2 (-z) -\frac{\pi^2}{6} +O(z^{-1})\qquad (|z|\to \infty)\,,
\end{align}
to obtain 
\begin{align}
{\rm Li}_2\left( e^{i(p_1+p_2+p_3)}\right) \sim -\frac{1}{2} \left(\frac{\Delta_1+\Delta_2 +\Delta_3}{4g\theta}\pm \pi i\right)^2\period
\end{align}
Here the sign in front of $\pi i$ depends on the choice of the branch of the logarithm. However, as we see below, the final result does not depend on the choice of this sign.
As can be seen clearly, this expression contains a double pole and a single pole with respect to $\theta$. Now  if we combine all the terms contained in  (7.30), we get%
\begin{align}
\begin{split}
-\frac{1}{2} \left[\left(\frac{\Delta_1+\Delta_2 +\Delta_3}{4g\theta}\pm\pi i\right)^2+\left(\frac{\Delta_1+\Delta_2 -\Delta_3}{4g\theta}\pm\pi i\right)^2\right.\\\left.+\left(\frac{\Delta_1-\Delta_2 +\Delta_3}{4g\theta}\pm\pi i\right)^2
+\left(\frac{-\Delta_1+\Delta_2 +\Delta_3}{4g\theta}\pm\pi i\right)^2\right.\\
\left.-\left(\frac{2\Delta_1}{4g\theta}\pm\pi i\right)^2-\left(\frac{2\Delta_2}{4g\theta}\pm\pi i\right)^2-\left(\frac{2\Delta_3}{4g\theta}\pm\pi i\right)^2\right]\,,
\end{split}
\end{align}
which add up  to the finite result $-(\pi^2)/2$. This confirms the absence of the singularities in the full expression (7.30). Thirdly, as in the weak coupling, the integrals along the cuts can be re-expressed by pushing some of the contours onto the second sheet:
  \beq{
 \left.( \mathcal{L}+\mathcal{R}){}_{\text{S}}\right|_{\text{along $\Gamma_i$}}=\oint_{\Gamma_1\cup \Gamma_2}\frac{du}{2\pi}{\rm Li}_2 \left(e^{ip_1+ip_2-ip_3} \right)+\oint_{\Gamma_3}\frac{du}{2\pi}{\rm Li}_2 \left(e^{ip_3+ip_1-ip_2} \right)\period
  }
  Here the first term can be interpreted as the contribution from the ${\rm SU}(2)_R$ whereas the second term can be regarded as coming from the ${\rm SU}(2)_L$. However, such factorization is not complete at strong coupling since the integrals along the unit circles cannot be rewritten in a similar manner.
\subsubsection*{Relation with the hexagon form factor}
Let us make a comment on the relation with the hexagon form factor approach.
As given in \cite{BKV}, the result from the hexagon form factor consists of two parts: One is the {\it asymptotic part}, which is given by the sum over partitions of the physical rapidities, and the other is the {\it wrapping correction}, which is the contribution from the mirror particles. In \cite{BKV}, they showed in simple cases that the integration along the branch cuts arises  from the asymptotic part whereas the integration along the unit circle contains the first leading wrapping correction. More recently, it was demonstrated in \cite{JKKS2} that, by partially resumming the mirror particle contributions, one could get an integral of the dilogarithm along the unit circle and correctly reproduce a part of our results \eqref{112strong}. It would be an very interesting future problem to try to resum all the hexagon form factors at strong coupling and reproduce our full result, which would account for various more complicated processes involving the mirror particles. 
 
\subsubsection*{BPS limit and Frolov-Tseytlin limit}  
  We now study several limits of the result \eqref{112strong} and perform the consistency checks. Let us first consider the three-point functions of the BMN vacuum. As the quasi-momentum for the BMN vacuum does not have any branch cuts, we only have integrals around the unit circle in that case. Furthermore, since the quasi-momenta in $S^3$ and $AdS_3$ coincide for the BMN vacuum, the two dynamical factors become identical, \ie $\mathcal{D}_{\rm S}=\mathcal{D}_{\rm AdS}$,  and cancel out in \eqref{112strong}. Therefore we only have a contribution from the kinematical part in the final answer. This is consistent with the fact that the BPS three-point function does not receive quantum corrections.
  
Let us next study the Frolov-Tseytlin limit \cite{FrolovTseytlin} by taking the charges $r$ and $l$ to be much larger than the coupling constant $g$ while keeping the mode numbers of the cuts to be finite. In terms of the spectral curve, this amounts to pushing the branch cuts far away from the unit circle. More precisely, the branch cuts for $p_1$ and $p_2$ are pushed out into the region $|x|\gg1$ whereas the branch cuts of $p_3$ are confined  to the region $|x|\ll 1$. In such a limit,  we can approximate the quasi-momenta on the unit circle by the quasi-momenta of the BMN vacuum. As explained above, for the BMN vacuum, integrals along the unit circle cancel out between $S^3$ and $AdS_3$. Thus, in the Frolov-Tsyetlin limit, the integrals along the unit circle become negligible. 

To study the remaining contributions, it is convenient to express the result \eqref{112strong} in terms of $\bar{p}_3$ defined by
\beq{\label{lefttoright}
\bar{p}_3(x)\equiv-p_3(1/x)\period
}
  As explained in Appendix \ref{ap:h}, $\bar{p}$ can be interpreted as the quasi-momentum defined on a different sheet in the full eight-sheeted spectral curve and the relation \eqref{lefttoright} is nothing but the $\mathbb{Z}_4$ automorphism of the string sigma model in $AdS_5\times S^5$. It is $\bar{p}_3$ that is connected to the quasi-momentum for the ${\rm SU}(2)_L$-sector at weak coupling. Now, to write down the expression in the Frolov-Tseytlin limit, we need to know the limiting forms of the quasi-momenta and the rapidity variable. In the region $|x|\gg 1$, $p_1(x)$ and $p_2(x)$ become the quasi-momenta in the Landau-Lifshitz model, whereas if $|x|\ll 1$ they approach their asymptotic forms around $x=0$,
  \beq{
  p_{1,2}\sim \frac{\tilde{Q}_{1,2}}{g}x\period
  }
Similarly, $\bar{p}_3(x)$ becomes the quasi-momentum of the Landau-Lifshitz model if $|x|\gg 1$ whereas it approaches
    \beq{
    \bar{p}_3 \sim \frac{Q_3}{g}x\comma
    }
    in the region $|x|\ll 1$. As for the rapidity variable, it takes the following asymptotic form:
    \beq{
    u(x) \sim \begin{cases}g x & |x|\gg 1\\g/x&|x|\ll 1\end{cases}\period
    }
    Using these asymptotic forms and replacing the global charges $Q_i$ and $\tilde{Q}_i$ with the spin-chain variables as given in \eqref{liri}, we can verify that \eqref{112strong} in the Frolov-Tseytlin limit coincides with the result at weak coupling \eqref{finalI-I-II}.
    
    One can also study the next-leading order correction to the Frolov-Tseytlin limit and compare it with the results in \cite{JKLS}. In \cite{JKLS},  based on the previous results at strong coupling \cite{KK3}, they concluded that the next-leading order correction to the Frolov-Tseytlin limit agrees with the one-loop structure constant at weak coupling except for integration contours. Since we now have contours\fn{The relation between the results in this paper and the results in \cite{KK3} will be briefly discussed later.} which coincide with the weak coupling ones in the Frolov-Tseytlin limit, the results match also at this order. For details, see section 6 of \cite{JKLS}. 
  
\subsubsection*{Type I-I-I three-point functions}
\begin{figure}[tb]
\begin{center}
\picture{clip,height=3.5cm}{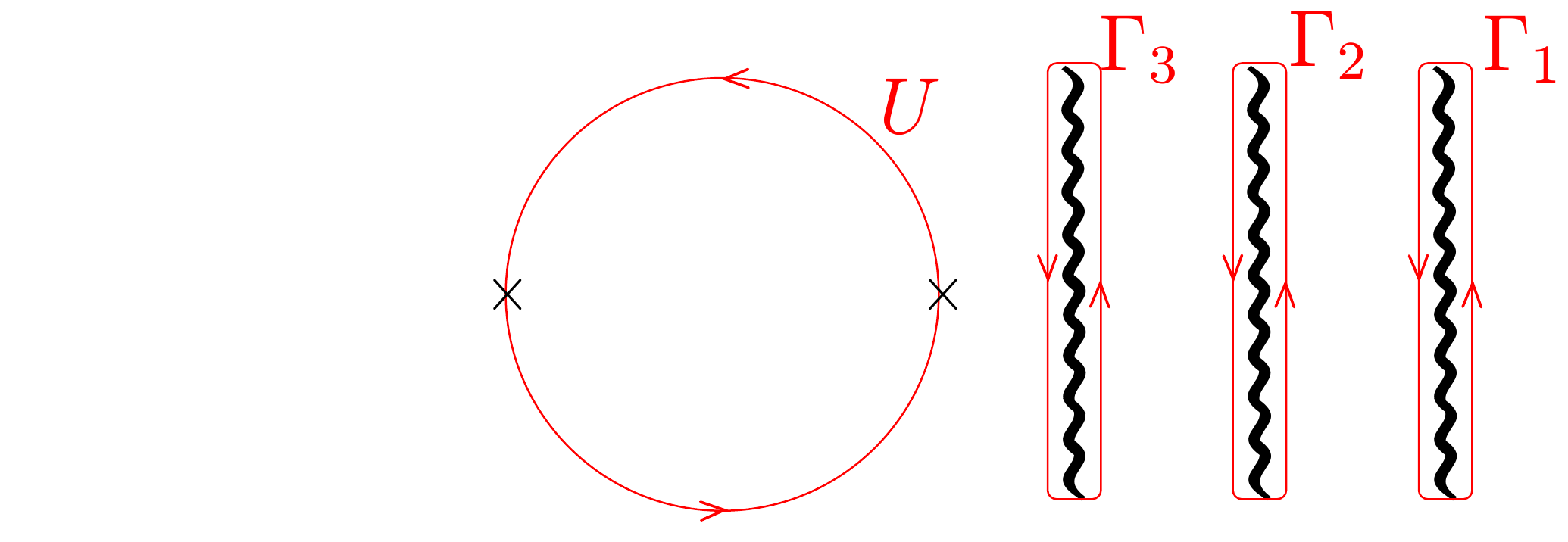}
\end{center}\vspace{-0.5cm}
\caption{Integration contours for the type I-I-I correlators. $\Gamma_i$ encircles the branch cuts of $p_i$ counterclockwise. Here again $U$ is the contour
that goes counterclockwise  around the unit circle.\label{111contourpic}}
\vspace{-0.2cm}\end{figure}
Next we consider the Type I-I-I three-point functions. As in section \ref{subsec:finalweak}, we consider the case where all the operators belong to ${\rm SU}(2)_R$.
Also in this case, the result can be expressed as
\beq{
\ln C_{123} =\mathcal{K} +\mathcal{D}_{\rm S} -\mathcal{D}_{\rm AdS}\period
}
Here $\mathcal{K}$ and $\mathcal{D}_{\rm AdS}$ are given by the same expressions as before, namely \eqref{Kstrong112}, \eqref{Dsadsstrong112} and \eqref{L+Rstrong112}. On the other hand, $\mathcal{D}_{\rm S}$ for the Type I-I-I three-point function is given by
\beq{
\mathcal{D}_{\rm S}= \left(\mathcal{L}+\mathcal{R} \right)_{\rm S} +\mathcal{N}_{\rm S} \comma
  }  
 with 
 \beq{
\begin{aligned}\label{L+R111strong}
\left( \mathcal{L}+\mathcal{R}\right)_{\rm S}=&\oint_{U}\frac{du}{2\pi}{\rm Li}_2 \left(e^{ip_1+ip_2+ip_3} \right)+\frac{1}{2}\sum_{\{i,j,k\}\in{\rm cperm}\{1,2,3\}}\left(\oint_{\Gamma_i\cup \Gamma_j\cup 2U}\frac{du}{2\pi}{\rm Li}_2 \left(e^{ip_i+ip_j-ip_k} \right)\right)\comma\\
\mathcal{N}_{\rm S}=&-\frac{1}{2}\sum_{k}\oint_{\Gamma_k\cup2 U} \frac{du}{2\pi} {\rm Li}_2 \left( e^{2ip_k}\right)\period
\end{aligned}
}
The integration contours in \eqref{L+R111strong} are depicted in figure \ref{111contourpic}.

We can study the Frolov-Tseytlin limit also in this case and the result again matches with the result at weak coupling \eqref{finalI-I-I}. 
\subsubsection*{Comparison with the result in \cite{KK3}}
Before ending this section, let us comment on the relation with the previous results for the three-point functions at strong coupling \cite{KK3}.

 In \cite{KK3}, we determined the analyticity of the Wronskians assuming that the saddle-point configuration of the worldsheet is smooth except at the positions of the vertex operators. The integration contours obtained under this assumption are more complicated than what we have found  in this paper and the result in the Frolov-Tseytlin limit did not  quite agree with the result at weak coupling. This implies that the assumption of smoothness is not quite correct and the saddle-point configuration has extra singularities. Although counterintuitive it may seem at first thought, such extra singularities are not so uncommon as already pointed out in \cite{KK3}. For instance, consider the finite gap solution for the two-point function whose spectral curve contains more than one cuts. Such a solution is given in terms of the ratio of the theta functions defined on the higher-genus Riemann surface. Although those ratios are free of singularities on the Lorentzian worldsheet, they have infinitely many poles\fn{Such poles do not correspond to the insertion of vertex operators and do not affect the monodromy relation.} on the Euclidean worldsheet, which is more appropriate for studying the correlation functions. Such extra poles, if present, can affect the argument in \cite{KK3} and change the integration contours. By contrast, the logic presented in this paper is based on the orthogonality of the on-shell states, which is the {\it exact} quantum property of the system, and therefore would be more universal and reliable. 

\section{Conclusion and prospects\label{sec:conclusion}}
In this paper, we studied the semi-classical three-point function in the ${\rm SU}(2)$-sector of $\mathcal{N}=4$ super Yang-Mills theory in four dimensions at weak coupling. The key idea was to express it as a saddle-point value of the coherent-state path integral and utilize the classical integrability of the Landau-Lifshitz model. This revealed the nature of the semi-classical structure constant as a generating function of the angle variables. For the computation of the angle variables, many of the machineries developed for the strong coupling analysis could be transplanted, the most important among which are the expression of the angle variables in terms of the Wronskians and the functional equation for the Wronskians. To solve the functional equation, we developed a new logic to determine the analyticity, which is based on the orthogonality of two different on-shell states. The final results agree with the results in the literature and also make predictions  for as-yet-unknown semi-classical structure constants for certain types of three-point functions. 

We then re-examined the strong coupling analysis based on our new logic. It led to a modification of the integration contours of the result obtained in \cite{KK3} and rendered the result in the Frolov-Tseytlin limit to be in agreement with the weak coupling one. In addition, the new result is consistent with the recent hexagon form factor approach \cite{BKV}.

 As for the prospects, of paramount importance is to further explore the implication of the cognate integrability structure at weak and strong coupling, which we discussed in this paper. Given the importance of the monodromy relation at the tree level and at strong coupling, a natural next step is to study it at higher loops. This may lead to a first-principle derivation of the integrable structure for three-point functions at finite coupling. Another important structure worth mentioning in this regard is the striking similarity between our functional equations \eqref{products}, which are the direct consequence of the monodromy relation, and the relations\fn{See (5.26) and (5.27) in \cite{BJ}.} constraining the lightcone string vertex in the pp-wave background \cite{BJ}. It would be interesting to figure out the reason for this similarity. More generally, clarifying the integrable structure threading gauge and string theories would be a cornerstone for deeper understanding of the AdS/CFT correspondence. It may also yield  practical merit  if it leads to a new formulation of integrability for structure constants, which is more powerful than the existing approaches.

Apart  from such  challenging  and  far-reaching questions, there are numerous future directions that could be explored  with the results and the techniques developed in this paper. Below we briefly address some of them.
\vspace{-10pt}
\subsubsection*{\small Semi-classical limit of type I-I-I three-point functions}
\vspace{-10pt}
In this paper, we made predictions for the semi-classical limit of type I-I-I three-point functions at weak coupling \eqref{finalI-I-I}. It would be an interesting  problem to reproduce it from the exact quantum expression given in \cite{KKN1,JKKS}. Since the result in \cite{KKN1,JKKS} has a more complicated structure than the type I-I-II three-point function, we probably need to develop new tools for studying it.  
\vspace{-10pt}
\subsubsection*{\small Resummation of the hexagon form factor at strong coupling}
\vspace{-10pt}
Another interesting direction of research is to analyze the strong-coupling semiclassical limit using the hexagon form factor formalism. It was shown in \cite{BKV,JKKS2} that a part of our result can be reproduced from the resummation of the hexagon form factor at strong coupling. It is important to further push this line of research and try to obtain the full strong coupling result from the hexagon form factor. This would be a litmus test for the hexagon form factor approach. 
\vspace{-10pt}
\subsubsection*{\small One-loop corrections at strong coupling}
\vspace{-10pt}
In the spectral problem, the power of the classical integrability and the associated spectral curve was not limited to the leading strong coupling limit. It also provided an efficient framework to study one-loop corrections around the classical solution \cite{GV,GNV}. The main idea there was to describe fluctuations as infinitesimal cuts inserted in the classical spectral curve. In this paper, we employed a very similar idea to determine the analytic properties of the Wronskians. It would then be extremely interesting, 
 by extending our argument,  to try to include the one-loop corrections . As a first step, it may be simpler to first analyze the weak coupling limit since the next-leading correction in the semi-classical limit was already computed by other means \cite{Bettelheim}. 
\vspace{-10pt}
\subsubsection*{\small Application to other quantities}
\vspace{-10pt}
It would also be interesting to apply the method discussed here to other quantities in $\mathcal{N}=4$ SYM. Of particular interest among them is the four-point function. The four-point function at the tree-level was studied in the paper \cite{EC} using integrability. However, even at that level, the resultant expression is rather involved owing to the complicated combinatorics of Wick contractions. In order to uncover a hidden structure, it might be helpful to study their semi-classical limit using our formalism. Such a structure is already known at strong coupling where it was shown that the four-point function of semi-classical operators can be described by the functional equation called $\chi$-system \cite{CT}. It would be interesting to try to construct the weak-coupling counterpart of the $\chi$-system. In addition, it might also be possible to use our framework to study non-planar observables such as the non-planar dilatation operator.
\vspace{-10pt}
\subsubsection*{\small Entanglement entropy in integrable spin chains and field theories} 
\vspace{-10pt}
Another interesting possibility is to apply the ideas and the techniques of this paper to the computation of entanglement entropy in general integrable spin chains and field theories.
To compute the entanglement entropy, one must first construct a reduced density matrix. In the case of spin chains, this can be achieved by preparing two identical states, cutting them into two halves and gluing the left (or the right) halves. This procedure is similar to the tailoring method for the three-point function \cite{Tailoring1}. Thus, it may be possible to study the entanglement entropy of the semi-classical state, which contains a large number of long wave-length excitations, using the formalism developed in this paper. This would be of particular interest since the entanglement entropy for such a highly excited state is difficult to compute by other methods.

We hope to revisit some of these questions in the future.
\par\bigskip\noindent
{\large\bf Acknowledgment}\par\smallskip\noindent
 The research of  Y.K. is supported in part by the 
 Grant-in-Aid for Scientific Research (B) 
No.~25287049, while that of T.N. is supported in part 
 by JSPS Research Fellowship for Young Scientists, from the Japan 
 Ministry of Education, Culture, Sports,  Science and Technology. The research of S.K. is supported by the Perimeter Institute for Theoretical Physics. Research at Perimeter Institute is supported by the Government of Canada through Industry Canada and by the Province of Ontario through the Ministry of Economic Development and Innovation. 
\appendix
\section{From Heisenberg spin chain to the Landau-Lifshitz model \label{ap:a}} 
In this appendix, we shall give a brief description of how to obtain 
 the Landau-Lifshitz model  from the Heisenberg spin chain in the 
 semi-classical limit. 
\subsection*{Coherent state representation of SU(2) }
\label{sec:cohstate}
To pave our way,  we shall quickly  review  the coherent state representation  of a Heisenberg spin chain (see  \cite{Fradkin} for the description relevant  to the present context)  and comment on its physical meaning. 
As mentioned in the main text, it is a 
 representation of SU(2) on the functions on the coset space SU(2)$/$U(1), which is isomorphic to a unit sphere. 

In this subsection we shall  focus on a single spin $1/2$ state. 
Let $\upket$ be the eigenstate of 
 $S_3$ with the eigenvalue $1/2$. Then, a U(1) operator  $h=e^{i\al  S_3}$ around this direction only produces a phase and an arbitrary SU(2) element $g$ can be decomposed  as $g =\Omega h$, where $\Omega$ belongs to the coset SU(2)$/$U(1). Thus, $g \upket =\Omega \upket e^{i\al/2} $. On general grounds,  $\Omega$ can be parametrized using the remaining generators  $S_\pm= S_1 \pm iS_2$ as $\Omega(\eta) = \exp \left( \eta S_+ -\etabar S_-\right)$, where $\eta$ is a complex parameter. For the Landau-Lifshitz model, one usually adopts the  representation where  the target space is easily seen to be a unit sphere. 
This is achieved  by the choice of the parametrization\footnote{The minus sign 
 in front in $\eta$  is a convention to conform to the one in \cite{WuYang1}.}  $\eta =- (\theta/2)e^{-i\phi}$. Then 
\begin{align}
\Omega(\eta) \upket = \exp \left(- i\theta (S_2\cos\phi -S_1\sin\phi)
 \right) \upket \label{Omegaeta} \period
\end{align}
Now let ${\tt n}_0=(0,0,1)$ be a unit vector in the $z$ direction and ${\tt n} 
 = (\sin\theta \cos\phi, \sin\theta \sin\phi, \cos\theta)$ be a unit vector in 
 a general direction. Then, it is easy to see that $|{\tt n}_0 \times {\tt n}| =
\sin\theta$ and 
\begin{align}
{{\tt n}_0 \times {\tt n} \over |{\tt n}_0 \times {\tt n}|} = (-\sin\phi, \cos\phi, 0)
\period
\end{align}
Comparing with (\ref{Omegaeta}) we find 
\begin{align}
\ket{{\tt n}} &\equiv \Omega(\eta)\upket = 
\exp \left(- i\theta {{\tt n}_0 \times {\tt n} \over  |{\tt n}_0 \times {\tt n}|} \cdot \vec{S}
\right) \upket = \cos{\theta \over 2} \upket +e^{i\phi} \sin{\theta\over 2} \ket{\!\downarrow}  \period \label{ketn}
\end{align}

At this point  an alert reader may have noticed that the pair of coefficients 
  $(\cos(\theta/2), e^{i\phi} \sin(\theta/2))$ coincide with the components 
 of the so-called monopole harmonics, introduced in  \cite{WuYang1, WuYang2}
as $Y_{q,l,m}$ defined on a unit sphere,  in the case where $q=eg =\half$,  with  $e$ and $g$,  respectively,  being the electric charge of a particle 
 on the sphere and the magnetic charge of a monopole situated  at the origin. 
Actually, as it is a section of a non-trivial U(1) bundle and  it has to be  defined in two overlapping open sets, like those around the northern 
 and the southern hemispheres, separately in such a way that in the overlap 
 its expressions are connected by a non-trivial gauge transformation. 
What is happening is that in order to produce a spin $1/2$ representation 
 out of a vector ${\tt n}$, which obviously carries spin 1, it must be 
combined  with an extra spin of magnitude $1/2$, which can be interpreted as 
provided by a ``minimum"  charge-monopole system. 

The monopole harmonics associated with the vector ${\tt n}$ 
 as above corresponds to  $Y_{\half,\half,m}({\tt n})$, $(m=\pm \half)$. 
As described in \cite{WuYang1}, an important property of the monopole harmonics is that under the action of a rotation matrix $D({\tt n}^{\prime})_{m' m}$ around the direction  ${\tt n}^{\prime}$, the monopole harmonics does not simply rotate into a 
linear combination of monopole harmonics. This is because,  under the rotation, 
while  the open sets with respect to which the monopole harmonics is defined 
get rotated into different regions, the gauge connection $A_\mu(x)$ is not changed. Therefore in order  to recover the same relative configuration of the open sets and the form of the  connection one must make a suitable gauge transformation of $A_\mu(x)$. This produces an extra U(1) phase factor of the form $\exp(i\Phi({\tt n}, {\tt n}')q)$, where $\Phi({\tt n}, {\tt n}')$ is the area of  the triangle on the unit sphere  
the vertices of which are defined by ${\tt n},{\tt n}'$ and the vector ${\tt n}_0$. It is clear from the preceding discussions that this phase, to be 
 called {\it the Wess-Zumino phase}, is an essential ingredient in  realizing the spin $\half$ representation in terms of  the coherent states  $\ket{{\tt n}}$. 

An important quantity in which  this phase appears is the inner product of the 
coherent states. One can show by direct calculation that 
\beq{
\langle{\tt n}^{\prime}|{\tt n}\rangle &=\cos \frac{\theta}{2}\cos \frac{\theta^{\prime}}{2} +e^{i(\phi-\phi^{\prime})} \sin\frac{\theta}{2}\sin\frac{\theta^{\prime}}{2}\nn\\
&=\exp \left(i\frac{\Phi ({\tt n}^{\prime}\comma{\tt n})}{2}\right) \sqrt{1-\frac{({\tt n}-{\tt n}^{\prime})^2}{4}}\comma\label{ape-2}
}
where 
\begin{align}
\tan \frac{\Phi ({\tt n}^{\prime}\comma{\tt n})}{2}
&=\frac{ \left( {\tt n}^{\prime}\times{\tt n}\right)\cdot{\tt n}_{0}}{1+{\tt n}_0\cdot{\tt n}+{\tt n}_0\cdot{\tt n}^{\prime}+{\tt n}\cdot{\tt n}^{\prime}} \period \label{ape-2-2}
\end{align}
More intuitive expression of the Wess-Zumino phase will also be given shortly. 

Before leaving  this subsection, let us record two basic relations we will use. 
One is the (over)completeness relation which reads 
\beq{
{\bf 1} =\frac{1}{2\pi}\int d^3 {\tt n} \,\delta ({\tt n}^2-1)\ket{{\tt n}}\bra{{\tt n}}\period \label{ape-3}
}
This can be readily  verified by substituting the explicit form of $\ket{{\tt n}}$ 
given in \eqref{ketn} and performing the integration. One then obtains 
that the RHS is indeed equal to $\ket{\!\uparrow}\bra{\uparrow\!}+\ket{\!\downarrow}\bra{\downarrow\!}=1$. Another basic relation of use is 
$\bra{{\tt n}}\vec{S}\ket{{\tt n}}=\frac{1}{2}{\tt n} $, which can also 
 be checked with ease. 
\subsection*{Brief derivation of the Landau-Lifshitz model}
\label{sec:derLLmodel}
Making use of the coherent state representation of the SU(2)
 spin $1/2$ state explained  above,  we now briefly describe how the 
Landau-Lifshitz model arises  from  the Heisenberg spin chain  in the  semiclassical limit. 

Let us denote by $\ket{\vec{{\tt n}}}=\ket{{\tt n}_1} \otimes \cdots \otimes \ket{{\tt n}_L}$ a coherent state of the spin chain and consider the 
 transition amplitude 
$\bra{\vec{{\tt n}}_{\rm final}}e^{-iHt}\ket{\vec{{\tt n}}_{\rm initial}}$ 
from the initial state to the final state, through the Hamiltonian of the Heisenberg  spin chain given (up to a convenient constant) by
\beq{
H=4g^2 \sum_{i=1}^{L} \left( \frac{1}{4}-\vec{S}_i\vec{S}_{i+1}\right) \period
}
By the standard procedure, namely by performing  the time evolution in  
infinitesimal steps with the insertions of  the completeness relation \eqref{ape-3}
at each step, one obtains the coherent state path-integral representation 
\beq{
\bra{\vec{{\tt n}}_{\rm final}}e^{-iHt}\ket{\vec{{\tt n}}_{\rm initial}}=\int \mathcal{D}\vec{{\tt n}}(t) e^{iS} \comma
}
with  the action $S$  given by 
\beq{
S= \sum_{i=1}^{\ell}\int dt  \left[ \frac{\left( {\tt n}_i \times\del_{t}{\tt n}_i\right) \cdot {\tt n}_0}{2(1+{\tt n}_i\cdot {\tt n}_0)}-\frac{g^2}{2} ({\tt n}_i-{\tt n}_{i-1})^2\right] \period\label{aeq-2}
} 
By taking the continuum limit of this expression, we obtain the well-known action of the Landau-Lifshitz model. 
\beq{
S=\int dt\int_{0}^{\ell} d\sigma  \left[ \frac{\left( {\tt n} \times\del_{t}{\tt n}\right) \cdot {\tt n}_0}{2(1+{\tt n}\cdot {\tt n}_0)}-\frac{g^2}{2} \del_{\sigma}{\tt n}\cdot\del_{\sigma}{\tt n}\right] \period\label{aeq-3}
}
The first term on the RHS represents the Wess-Zumino phase  produced through the inner product as given  in \eqref{ape-2} and \eqref{ape-2-2}. 

Just as in the Wess-Zumino-Novikov-Witten model, for example, such a Wess-Zumino term has a representation in terms of an integral one dimension higher 
 (in this case as a three dimensional integral) of the form 
\beq{
\frac{1}{2} \int _0^{1} ds\, \int dt\int_{0}^{L} d\sigma 
\,  {\tt n} \cdot \left(\del_t {\tt n} \times \del_s {\tt n}\right)\comma \label{eq:wesszumino}
}
where $s$-dependence of ${\tt n}$ is defined such that ${\tt n}(s=1)=(0,0,1)$ and ${\tt n}(s=0)={\tt n}$. The expression \eqref{eq:wesszumino} has a rather intuitive meaning. Since ${\tt n}$ is a unit vector,  $\del_t {\tt n}$ and $\del_s {\tt n}$ are perpendicular to ${\tt n}$. Therefore 
 the exterior product $\del_t {\tt n} \times \del_s {\tt n}$ is 
  in the direction of ${\tt n}$ and  ${\tt n} \cdot \left(\del_t {\tt n} \times \del_s {\tt n}\right)dt ds$ is nothing but the infinitesimal area element. 
Hence the integration gives the area and together with 
 the factor of $1/2$, which is the value of  $q=eg$ discussed in the previous subsection, we get the exponent of the Wess-Zumino phase factor. 
%
\section{Poisson brackets and the r-matrix for the Landau-Lifshitz model}
\label{ap:b}
As described in section \ref{subsec:action-angle}, the classical r-matrix for the Landau-Lifshitz model 
 can be obtained quickly as the classical limit of the well-known form 
 of the quantum R-matrix of the Heisenberg spin chain. 

However, it would be of interest to supply the first principle derivation of the r-matrix from the computation of the Poisson brackets  among the coherent state variables $n_i(\sig,\tau)$. Below we give a sketch of such a derivation. 

\subsection*{Poisson brackets}
First we derive the Poisson (Dirac) bracket structure of the Landau-Lifshitz model. The most straightforward way is to start from the action \eqref{ceq-1}, regard $\vec{\tt n}$ as the fundamental variable and derive the Dirac brackets, taking into account the constraints $\vec{\tt n}^2 =1$. However, in practice 
it turned out to be much easier to first parametrize the 2-sphere by $\theta$ and $\phi$ and then compute the Dirac brackets. In terms of these angle 
 variables, the action of the Landau-Lifshitz sigma model takes the form 
\beq{
S=-\int d\tau d\sigma \left[ \frac{1}{4}\left( \cos \theta \del_{\tau} \phi +  \phi\sin \theta \del_{\tau}\theta\right) + \frac{g^2}{2} \left( \del_{\sigma} \theta \del_{\sigma} \theta +\sin ^2 \theta \del_{\sigma} \phi\del_{\sigma} \phi\right)\right]  \period \label{aeq-7}
}
From this action, the conjugate momenta can be determined as
\beq{
\Pi_{\phi}=-\frac{1}{4}\cos \theta \comma \qquad \Pi _{\theta} =-\frac{1}{4}\phi \sin \theta\period
}
Evidently, these two equations should be regarded as the constraints. The commutation relation of these two constraints is given by
\beq{
\{\Pi_{\phi}+ \frac{1}{4}\cos \theta\big| _{\sigma},\Pi _{\theta} +\frac{1}{4}\phi \sin \theta\big| _{\sigma^{\prime}} \}=-\frac{\sin \theta }{2}\delta (\sigma -\sigma^{\prime} )\period
}
Thus, the Dirac bracket for any dynamical variables $A$ and $B$ for this 
 system is given by 
\beq{
&\{ A,B\}_{D} = \{ A ,B\}\nn\\
&\,\,+\int d\sigma \frac{2}{\sin \theta }\left( \{A,\Pi_{\phi} + \frac{1}{4}\cos \theta\}\{ \Pi _{\theta} +\frac{1}{4}\phi \sin \theta ,B\}- \{A,\Pi _{\theta} +\frac{1}{4}\phi \sin \theta \} \{ \Pi_{\phi} + \frac{1}{4}\cos \theta ,B\} \right)\period\label{aeq-8}
}
Applying this formula to the variables ${\tt n}_i(\sig)$ and ${\tt n}_j(\sig)$ at equal time 
yields 
\beq{
\{ {\tt n}_i(\sigma) \comma {\tt n}_j(\sigma^{\prime})\}_D =2\epsilon_{ijk}{\tt n}_k \delta (\sigma-\sigma^{\prime})\comma
}
which is nothing but the classical  commutation relations for the spin variables. (In what follows, we omit writing the subscript $D$.)
\subsection*{Classical r-matrix}
Having derived the commutation relations for the variables $\vec{n}$, 
we can now derive the Poisson bracket between the Lax matrices and determine the classical r-matrix. The Poisson bracket between $J_{\sigma}$ given in 
 \eqref{lax-2} can be calculated as
\beq{
\{ J_{\sigma} (\sigma |u )\,\, \overset{\otimes}{\comma}\,\, J_{\sigma}(\sigma^{\prime}| v)\} &=-\frac{1}{16\pi ^2 uv}\{{\tt n}(\sigma)\cdot \vec{\sigma}  \,\,\overset{\otimes}{\comma}\,\,{\tt n}(\sigma^{\prime})\cdot\vec{\sigma}\} \nn\\
&=-\delta(\sigma-\sigma^{\prime})\frac{1}{8\pi ^2 uv} \epsilon_{ijk}{\tt n}_k (\sigma) \sigma_i \otimes \sigma_j\period \label{aeq-9}
}
One can  simplify this  expression by using the Fiertz identity
\beq{
(\sigma _a )_{ij}(\sigma _b)_{kl} &=\sum_{c,d} \frac{{\rm tr} \left( \sigma_c \sigma_a \sigma _d\sigma_b\right)}{4}(\sigma_c )_{il} (\sigma _d)_{kj}\comma
}
where the indices $c$ and $d$ run from $0$ to $3$ and  $\sigma_0$ is 
 defined to be equal to  ${\bf 1}$.
Applying this identity, the factor $\epsilon_{ijk} \sigma _i \otimes \sigma _j$ can be re-expressed as
\beq{
\epsilon_{ijk}(\sigma_i)_{\alpha \beta}(\sigma_j)_{\gamma \delta}= \frac{i}{2} \left( (\sigma _k )_{\alpha \delta} \delta_{\beta \gamma} -(\sigma _k )_{\beta \gamma} \delta_{\alpha \delta}\right)\period
}
Utilizing such formulas, we can arrive at the following expression\fn{To arrive at the expression \eqref{aeq-10}, we use $(uv)^{-1}=\left( v^{-1}-u^{-1}\right) (u-v)^{-1}$.} for the Poisson bracket:
\beq{
\{ J_{\sigma} (\sigma |u )\,\, \overset{\otimes}{\comma}\,\, J_{\sigma}(\sigma^{\prime}| v)\} =\delta (\sigma-\sigma^{\prime})\left[ {\bf r}(u-v)\comma\, - \left( J_{\sigma}(u)\otimes {\bf 1}+{\bf 1}\otimes J_{\sigma}(v) \right) \right]\period\label{aeq-10}
}
In this expression,  ${\bf r}(u)$ is the so-called classical r-matrix,  which in this case is given by 
\beq{
{\bf r}(u)=\frac{\mathbb{P}}{ u}\period
}
The symbol  $\mathbb{P}$  denotes the operator which permutes the two spaces in the tensor product: $V_1\otimes V_2 \longmapsto V_2\otimes V_1$. It is well-known\fn{A proof of \eqref{aeq-11} below can be found in page 106-107 of \cite{Inverse}.  } that when the Poisson bracket between the Lax matrices can be expressed 
in terms of  the classical r-matrix as in \eqref{aeq-10}, the Poisson bracket between the monodromy matrices can also be expressed by the classical r-matrix as
\beq{
\{\Omega (u)\overset{\otimes}{\comma}\, \Omega (v)\} = \left[\Omega (u)\otimes \Omega (v)\comma\,  {\bf r}(u-v)\right]\period\label{aeq-11}
}
\section{Highest weight condition on the semi-classical wave function}\label{ap:c}
Here we study constraints from the highest weight condition on the semi-classical wave function and show that the constant vector $n$ appearing in the normalization condition \eqref{normvec} must be  equal to the polarization vector.

As explained in section \ref{subsec:on-shellBethe}, the states constructed on the rotated vacuum characterized by the polarization vector $n=(n^1,n^2)^t$ satisfy the highest weight condition
\beq{\label{hwsemi}
S^{\prime}_{+}\ket{\Psi}=0\comma
}
with $S^{\prime}_{+}$ given in \eqref{defSprime}. To understand the consequence of this condition in the semi-classical limit, let us recall the form of the semi-classical wave function (in the action-angle basis), 
\beq{
\Psi = \exp \left(i \sum_k S_k \phi_k \right)\period
}
As explained in section \ref{subsec:action-angle},  the angle variables $\phi_k$ can be constructed from the poles of the factor $\langle n^{\prime} ,\psi_{+}\rangle$, where $n^{\prime}$ is a constant vector which defines the normalization condition\footnote{Thus in literature this vector is usually referred to as the normalization vector.}. Thus, in order to gurantee the highest weight property of the semi-classical wave function, we need to choose $n^{\prime}$ such that $\langle n^{\prime} ,\psi_{+}\rangle$ is invariant under the transformation generated by $S_{+}^{\prime}$.

For this purpose, let us first go back to the Heisenberg spin chain. In the Heisenberg spin chain, the Lax operator is given by
\beq{
L (u)= \pmatrix{cc}{1+iS_3/u&iS_{-}/u\\iS_{+}/u&1-iS_3/u}\period
}
By the straightforward computation, one can show that it transforms under $e^{a S_{+}^{\prime}}$ as
\beq{
\begin{aligned}\label{heisentransf}
e^{a S_{+}^{\prime}}L(u)e^{-a S_{+}^{\prime}}=&A L(u)A^{-1}\comma
\end{aligned}
}
where the matrix $A$ is given by
\beq{
A=N \pmatrix{cc}{1&-a\\0&1}N^{-1} \period
}
Now, since the Landau-Lifshitz sigma model is obtained by taking the continuum limit of the Heisenberg spin chain, \eqref{heisentransf} implies the following transformation rule of the Lax matrix of the Landau-Lifshitz sigma model:
\beq{\label{Jtransform}
e^{a S_{+}^{\prime}}\Big( J_{\sigma}\Big)= A J_{\sigma}A^{-1}\period
}
This means that a solution $\psi_+$ to the auxiliary linear problem transforms as $\psi_+ \to A \psi_+$ in order to compensate for the transformation \eqref{Jtransform}. Thus the Wronskian $\langle n^{\prime} ,\psi_{+}\rangle$ gets transformed as 
\beq{
\langle n^{\prime} ,\psi_{+}\rangle\mapsto
\langle n^{\prime} ,A\psi_{+}\rangle=\langle A^{-1}n^{\prime} ,\psi_{+}\rangle\comma
}
where the equality follows from the SL(2) invariance of the skew-symmetric product. It is then easy to see that the invariance under the transformation requires $n^{\prime}= A^{-1}n^{\prime} $ and this leads to the identification $n^{\prime}=n$.
\section{Construction of the separated variables}
\label{ap:d}
In this appendix, we will describe how the separated variables are obtained 
for the Landau-Lifshitz model. 
\subsection*{Expressions of the Poisson brackets obtained from the r-matrix} 
First, let us give a list of Poisson bracket relations between the components of 
the monodromy matrix written as 
\beq{
\Omega (u)\equiv \pmatrix{cc}{\mathcal{A}(u)&\mathcal{B}(u)\\\mathcal{C}(u)&\mathcal{D}(u)}\period
}
With the form of the r-matrix given in \eqref{rmatrix} and the basic Poisson bracket formula \eqref{defrmatrix} involving the r-matrix, 
the Poisson bracket relations between the components of $\Omega(u)$  can be 
easily computed as
\beq{
&\{\mathcal{A}(u)\comma \mathcal{B}(v)\}= \frac{-1}{u-v}\left(\mathcal{A}(u)\mathcal{B}(v)-\mathcal{A}(v)\mathcal{B}(u) \right)\comma 
\quad\{\mathcal{A}(u)\comma \mathcal{C}(v)\}=\frac{1}{u-v}\left(\mathcal{A}(u)\mathcal{C}(v)-\mathcal{A}(v)\mathcal{C}(u) \right)\comma\nn\\
&\{\mathcal{A}(u)\comma \mathcal{D}(v)\}=\frac{1}{u-v}\left(\mathcal{B}(u)\mathcal{C}(v)-\mathcal{B}(v)\mathcal{C}(u) \right)\comma
\quad\{\mathcal{B}(u)\comma \mathcal{C}(v)\}=\frac{1}{u-v}\left(\mathcal{A}(u)\mathcal{D}(v)-\mathcal{A}(v)\mathcal{D}(u) \right)\comma\nn\\
&\{\mathcal{B}(u)\comma \mathcal{D}(v)\}=\frac{1}{u-v}\left(\mathcal{B}(u)\mathcal{D}(v)-\mathcal{B}(v)\mathcal{D}(u) \right)\comma
\quad\{\mathcal{C}(u)\comma \mathcal{D}(v)\}= \frac{-1}{u-v}\left(\mathcal{C}(u)\mathcal{D}(v)-\mathcal{C}(v)\mathcal{D}(u) \right)\comma\nn\\
&\{\mathcal{A}(u)\comma\mathcal{A}(v)\}=\{\mathcal{B}(u)\comma\mathcal{B}(v)\}=\{\mathcal{C}(u)\comma\mathcal{C}(v)\}=\{\mathcal{D}(u)\comma\mathcal{D}(v)\} =0\period\label{poisson7}
}
These basic relaitions will be utilized in what follows. 
\subsection*{Separated variables \`a la Sklyanin}
Having displayed  the explicit expression for the Poisson brackets, we now construct the separated canonically conjugate variables by the so-called Sklyanin's magic recipe\cite{Sklyanin}. In this method, such variables are obtained  as associated 
 to  the poles of the normalized  eigenvector $h$ of the monodromy matrix, defined in the following way\fn{In Sklyanin's original formulation, the normalization condition is expressed in terms of the ordinary inner product as $n^{\prime}\cdot h=1$. Here we are instead using the skew-symmetric inner product in order to make connection with the Wronskian. It is equivalent to the original formulation under the identification of $n^{\prime}$ with $i\sigma_2 n$.}:
\beq{
\Omega (u) h (u)=e^{ip(u)}h (u)\comma \quad \langle n\comma  h\rangle=1\period
}
Here $n=(n^1,n^2)^t$ is the polarization vector. To simplify the construction 
 it turns out to be  convenient to first transform the  monodromy matrix $\tilde{\Omega}(x)$ by a similarity transformation into the form 
\beq{
\tilde{\Omega}(x) \equiv \pmatrix{cc}{n^2&n^1\\ -n^1&n^2}\Omega(x) \pmatrix{cc}{n^2& -n^1 \\n^1&n^2}\equiv \pmatrix{cc}{\tilde{\mathcal{A}}(x)&\tilde{\mathcal{B}}(x)\\\tilde{\mathcal{C}}(x)&\tilde{\mathcal{D}}(x)}\period
}
As the Lax pair equations are invariant under such a transformation, the components of $\tilde{\Omega}$ satisfy the same Poisson-bracket relation as those  of 
the components of $\Omega$   displayed  in \eqref{poisson7}. 

Let us denote the poles of $h$ by $\gamma_i$. Then the components of $\tilde{\Omega}$ satisfy the following relation\fn{To see this, it is helpful to consider the relation between the {\it normalized} eigenvector $h$ and the {\it unnormalized} eigenvector $\psi_+$. The normalized eigenvector can be constructed from the unnormalized eigenvector by $h = \psi_+/\langle n\comma \psi_+\rangle$. Therefore the poles of the normalized eigenvector arise when the unnormalized eigenvector satisfy $\langle n\comma \psi_+\rangle=0$. Thus, at the poles of the normalized eigenvector, the vector parallel to $n$ becomes the eigenvector of the monodromy matrix. Then, it is easy to see that \eqref{eq-1} follows.}.
\beq{
\tilde{B}(\gamma_i)=0\comma\quad \tilde{\mathcal{D}}(\gamma_i)=\tilde{\mathcal{A}}(\gamma_i)^{-1}=e^{ip(\gamma _i)}\period\label{eq-1}
}
In what follows, we make use of  these relations to derive the commutation  relations  between $\gamma_i$'s and $p(\gamma_i)$'s. 

We start from the analysis of $\{ \tilde{\mathcal{B}}(u)\comma \tilde{\mathcal{B}}(v)\}=0$. Since $\tilde{B}$ has zeros at $\gamma_i$ and $\gamma_j$ $(i\neq j)$, it can be expressed in the form  $\tilde{\mathcal{B}}(u)=(u-\gamma_i)\mathcal{B}^{\prime}(u)$ or $\tilde{\mathcal{B}}(u)=(u-\gamma_j)\mathcal{B}^{\prime\prime}(u)$. The functions $\mathcal{B}^{\prime}(u)$ and $\mathcal{B}^{\prime\prime}(u)$ are not known but what is important is that they have the properties $\mathcal{B}^{\prime}(\gamma_i)\neq 0$ and $\mathcal{B}^{\prime\prime}(\gamma_j)\neq 0$. Then the commutation relation between $\tilde{\mathcal{B}}(u)$ and $\tilde{\mathcal{B}}(v)$ can be rewritten as
\beq{
&(u-\gamma_i )(v-\gamma_j )\{ \mathcal{B}^{\prime}(u)\comma \mathcal{B}^{\prime\prime}(v)\} -(v-\gamma_j )\mathcal{B}^{\prime}(u)\{ \gamma_i \comma \mathcal{B}^{\prime\prime}(v)\}\nn\\
&-(u-\gamma_i )\mathcal{B}^{\prime\prime}(v)\{\mathcal{B}^{\prime}(u) \comma \gamma_j\} +\mathcal{B}^{\prime}(u)\mathcal{B}^{\prime\prime}(v) \{ \gamma_i \comma \gamma_j\}=0\period
}
Now at this stage, we can safely take the limit $u\to \gamma_i$ and $v\to \gamma_j$. Then the first three terms vanish   the last term gives 
 the relation 
\beq{
\{\gamma_i \comma \gamma_j\}=0\period\label{eq-2}
}

Next consider the commutation relation between $\tilde{\mathcal{A}}(u)$ and $\tilde{\mathcal{B}}(v)$. Here again, we should substitute  
the expansion $\tilde{\mathcal{A}}(u) =\tilde{\mathcal{A}}(\gamma_i) + (u-\gamma_i)\mathcal{A}^{\prime}(u)$ as well as the ones for $\mathcal{B}^{\prime}$ and $ \mathcal{B}^{\prime\prime}$. Then similarly to the previous case, 
the limit $u\to \gamma_i$ and $v\to \gamma_j$ can be easily taken 
and, making use of the relation \eqref{eq-2}, we can deduce 
the important relation 
\beq{
\{ \tilde{\mathcal{A}}(\gamma _i)\comma \gamma _j\} = \tilde{\mathcal{A}} (\gamma_i ) \delta _{ij}\period\label{eq-3}
}

As the last step, a similar calculation for $\{ \tilde{\mathcal{A}}(x)\comma \tilde{\mathcal{A}}(x^{\prime})\}=0$ leads to 
\beq{
\{\tilde{\mathcal{A}}(\gamma_i )\comma \tilde{\mathcal{A}}(\gamma_j )\} =0\period\label{eq-4}
}
Using the expression of $\tilde{\calA}(\ga_i)$ and $p(\ga_i)$ given in \eqref{eq-1} and the equations \eqref{eq-2}-\eqref{eq-4}, we can obtain the commutation relations among   $\gamma_i$'s and $p(\gamma_j)$'s as
\beq{
\{\gamma_i \comma \gamma_j \}=\{p(\gamma_i)\comma p(\gamma_j)\}=0\comma\quad -i\{\gamma_i \comma p(\gamma_j)\}=\delta_{ij}\period 
}
This shows  that $\left(\gamma_i\comma - i p(\gamma_i) \right)$'s are the separated canonical pairs of variables associated to the poles of the normalized 
 eigenvector. 
 \section{Baker-Akhiezer vectors for the two-point functions}\label{ap:e}
 In the case of two-point functions, the explicit solutions can be constructed by the finite gap method \cite{Vicedo1}. For the general spectral curve with genus $g$, the solutions to the auxiliary linear problem evaluated at $(\tau,\sigma)=(0,0)$ reads\fn{See (4.13) in \cite{Vicedo1}.}
 \beq{\label{Eeq1}
 \psi^0_{+}(u)=\pmatrix{c}{k_{-}(u)\\k_{+}(u)}\comma \quad\psi^0_{-}(u)=\pmatrix{c}{k_{-}(\hat{\sigma}u)\\k_{+}(\hat{\sigma} u)} \period
 }
 where $\hat{\sigma}$ is the holomorphic involution and the functions $k_{-}(u)$ and $k_{+}(u)$ are characterized uniquely by their divisors and the normalization at infinity: 
 \beq{
 \begin{aligned}\label{Edivisor}
&\left(k_{+}\right)=\infty^{+}+\sum_{i=1}^{g} \gamma_i^{\prime}-\sum_{j=1}^{g+1}\hat{\gamma}_j \comma \qquad k_{+}(\infty^{-})=1 \comma\\
&\left(k_{-}\right)=\infty^{-}+\sum_{i=1}^{g} \gamma_i-\sum_{j=1}^{g+1}\hat{\gamma}_j \comma \qquad k_{-}(\infty^{+})=1\period
  \end{aligned}
 }
 Here $\gamma^{\prime}_i$ are the initial values of the separated variables parametrizing the moduli for two-point functions, and $\gamma_i$ and $\hat{\gamma}_i$ are the divisors satisfying\fn{The symbol $a\sim b$ means that there is a single-valued function on the Riemann surface which has poles at $a$ and zeros at $b$.}
 \beq{\label{Eequi}
 \{\hat{\gamma}_j \}\sim\{\infty^{-},\gamma_i\}\sim \{\infty^{+},\gamma^{\prime}_i\}\period
 }
 As noted in \cite{Vicedo1}, the solutions \eqref{Eeq1} describe the highest weight eigenstate of $S_3$. This means that the corresponding polarization vector is $n=(1,0)^t$. The solutions for more general rotated vacua can be obtained by the global rotation.
 
  The solutions \eqref{Eeq1} do not satisfy the normalization conditions $\langle \psi_{+}^0\comma \psi_{-}^0\rangle=1$. To normalize the solutions, we need to divide them by $\sqrt{\langle \psi_{+}^0 \comma \psi_-^{0}\rangle}$ as in \eqref{rescalingk}. After the division, we obtain
 \beq{
 \psi_{+}(u) = C(u)\pmatrix{c}{k_{-}(u)\\k_{+}(u)}\comma \quad\psi_{-}(u)=C(u)\pmatrix{c}{k_{-}(\hat{\sigma}u)\\k_{+}(\hat{\sigma} u)}\comma
 }
 with $C(u)$ given by
 \beq{
 C(u)=\frac{1}{\sqrt{\langle \psi_{+}^0 \comma \psi_-^{0}\rangle}}=\frac{1}{\sqrt{k_-(u)k_{+}(\hat{\sigma}u)-k_+(u)k_{-}(\hat{\sigma}u)}}\period
}
Now, owing to \eqref{Edivisor}, $C(u)$ contains $2(g+1)$ square-root zeros at $\hat{\gamma}_i$ and $\hat{\sigma}\hat{\gamma}_i$. In addition, as argued in section \ref{subsec:normalization}, it must contain the square-root singularity at the positions of the branch points $b_k$. Thus the divisor of $C(u)$ is given by
\beq{
\left(C\right)=\frac{1}{2}\sum_{j=1}^{g+1}\hat{\gamma}_j+\frac{1}{2}\sum_{j=1}^{g+1}\hat{\sigma}\hat{\gamma}_j-\frac{1}{2}\sum_{k=1}^{2(g+1)}b_k\period
}
Combined with \eqref{Edivisor}, this determines the divisor of the factor $\langle n\comma \psi_{+} \rangle$ to be
\beq{
\left(\langle n \comma \psi_{+} \rangle\right) = \infty^{+}+\sum_{i=1}^{g} \gamma_i^{\prime} +\frac{1}{2}\sum_{j=1}^{g+1}\left(\hat{\sigma}\hat{\gamma}_j-\hat{\gamma}_j\right)-\frac{1}{2}\sum_{k=1}^{2(g+1)}b_k\period
}
This shows that $\langle n \comma \psi_{+} \rangle$ has spurious zeros and poles at $\hat{\gamma}_j$ and $\hat{\sigma}\hat{\gamma}_j$ unless we choose $\hat{\gamma}_j$ to be invariant under the holomorphic involution.

For the genus $0$ solutions including the ones corresponding to the BPS operators, we can confirm that it is always possible to choose $\hat{\gamma}_j$ to be invariant under $\hat{\sigma}$ by analyzing the explicit form of the solution. On the other hand, the situation for the higher genus solutions is less obvious since it is in general not clear if we can choose $\hat{\gamma}_j$ to be invarint under $\hat{\sigma}$ without violating the relation \eqref{Eequi}. However, when the cuts are sufficiently small, the solution would be very close to the BPS one, and, therefore from the continuity argument similar to the one used in section \ref{sec:ev}, we expect that it is possible to choose $\hat{\gamma}_j$ to be invariant under the involution (at least for some  appropriate choices\fn{Different choices of $\gamma^{\prime}_j$ in the moduli of two-point functions only change the overall phase of the structure constant. Thus, for the computation of the three-point functions, we can choose a convenient one.} of $\gamma^{\prime}_j$.)
\section{Quasi-momentum in the full spectral curve}\label{ap:h}
In this appednix, we shall clarify the relation \eqref{lefttoright}. 

For this purpose, consider the monodromy matrix for the full $AdS_5 \times S^5$ is a $(4|4)\times (4|4)$ matrix given by
\beq{
\Omega_{AdS_5\times S^5} (x) \sim \diag \left(e^{i\tilde{p}_1}\comma e^{i\tilde{p}_2}\comma e^{i\tilde{p}_3}\comma e^{i\tilde{p}_4}|e^{i\hat{p}_1}\comma e^{i\hat{p}_2}\comma e^{i\hat{p}_3}\comma e^{i\hat{p}_4}\right)\period
}
Here $\tilde{p}_i$ and $\hat{p}_i$ denote the quasi-momenta for the $S^5$ part and for the $AdS_5$ part respectively. 
\begin{figure}[tb]
\begin{center}
\picture{clip,height=6cm}{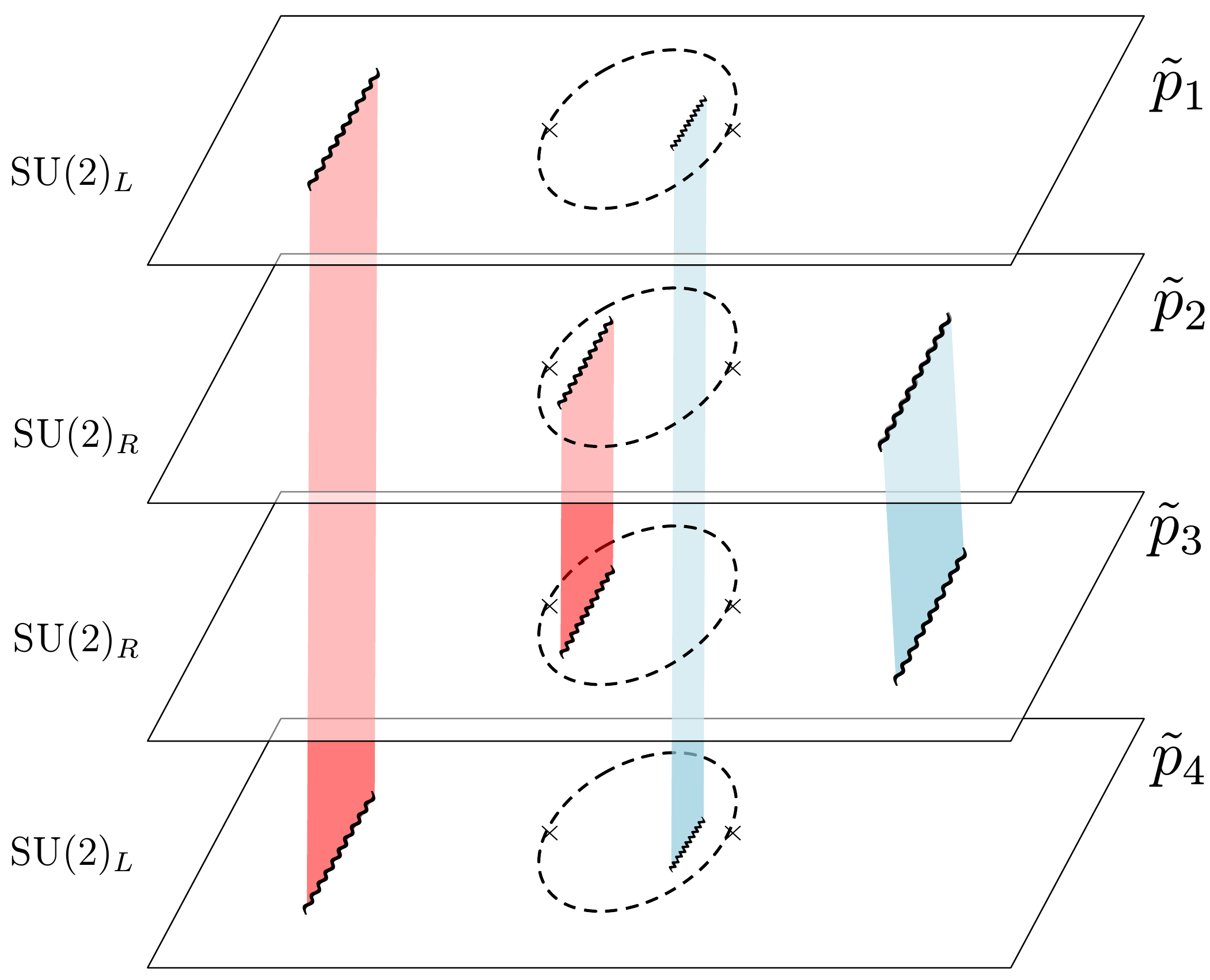}
\label{fullcurve}
\end{center}
\vspace{-0.5cm}
\caption{The $S^5$ part of the full eight-sheeted spectral curve. The cuts denoted in the same color are related with each other by the $\mathbb{Z}_2$ automorphism. The ${\rm SU}(2)_L$ and ${\rm SU}(2)_R$ sectors discussed in this paper correspond to the first and the fourth, and the second and the third sheets respectively.\label{fullcurve}}\vspace{-0.2cm}
\end{figure}

The quasi-momenta in the SU$(2)_L\times $SU$(2)_R$ sector, which we studied in this paper, are identified with those in the full $AdS_5 \times S^5$ as follows (see figure \ref{fullcurve} above):
\beq{
\left. p (x)\right|_{\text{SU}(2)_R} = \tilde{p}_2 -\tilde{p}_3\comma \qquad\left. p (x)\right|_{\text{SU}(2)_L} = \tilde{p}_1 -\tilde{p}_4\period
}
As explained in \cite{BKSZ}, owing to the $\mathbb{Z}_2$ automorphism of the coset, the quasi-momenta obey the following involution relation,
\beq{
\begin{aligned}
\tilde{p}_{1,2} (1/x)=-\tilde{p}_{2,1} (x)\comma \qquad\tilde{p}_{3,4} (1/x)=-\tilde{p}_{4,3} (x) \period
\end{aligned}
}
In terms of the SU$(2)_L\times $SU$(2)_R$ quasi-momenta, this reads
\beq{
\begin{aligned}
\left. p (1/x)\right|_{\text{SU}(2)_R}= -\left. p (x)\right|_{\text{SU}(2)_L}\comma \qquad \left. p (1/x)\right|_{\text{SU}(2)_L}= -\left. p (x)\right|_{\text{SU}(2)_R} \period
\end{aligned}
}

The quasi-momentum $p(x)$ used in the strong-coupling analysis in section \ref{sec:strong} is the SU$(2)_R$ quasi-momentum. On the other hand, at weak coupling, the result factorizes into the SU$(2)_R$ and the  SU$(2)_L$ sectors and the contribution from the SU$(2)_R$ (SU$(2)_L$) sector is expressed purely in terms of SU$(2)_R$ (SU$(2)_L$) quasi-momenta. Thus in order to make direct comparison between the weak-coupling and the strong-coupling results in the Frolov-Tseytlin limit, we need to rewrite a part of the strong-coupling result in terms of the SU$(2)_L$ quasi-momentum. This is precisely what we did in \eqref{lefttoright} and $\bar{p}$ defined there corresponds to the SU$(2)_L$ quasi-momentum.

\section{Zeros of $\sl{i_+,j_-}$}\label{ap:f}
Here we explain how to determine the zeros of the Wronskian for eigenfunctions with opposite sign eigenvalues, namely  $\sl{i_+,j_-}$,  by applying the argument given in \cite{KK3}.

As shown in \eqref{products}, the product of $\sl{i_+,j_-}$ and $\sl{i_-,j_+}$ contains zeros at $\sin (p_i-p_j+p_k)/2=0$ and $\sin (-p_i+p_j+p_k)/2=0$. For definiteness, we focus on zeros at $\sin (p_i-p_j+p_k)/2=0$ in what follows since the generalization to the zeros associated with $\sin (-p_i+p_j+p_k)/2=0$ is straightforward. When $\sin (p_i-p_j+p_k)/2=0$, all possible  products of Wronskians which vanish  are
\beq{\label{setofw}
\sl{i_+,j_-}\sl{i_-,j_+}\comma \quad \sl{j_-,k_+}\sl{j_+,k_-} \comma \quad \sl{i_+,k_+}\sl{i_-,k_-}\period
}
An important feature of \eqref{setofw} is that all the Wronskians that appear are the ones between the eigenstates in the same group, 
 $\mathcal{S}_1=\{i_+,j_-,k_+\}$ or $\mathcal{S}_2=\{i_-,j_+,k_-\}$. 
 Now, let us first note that the following lemma holds:
\begin{itemize}
 \item[] {\small \bf Lemma}:
In each product of two Wronskians in \eqref{setofw}, only one of the Wronskians can vanish.
  \end{itemize}
 This is because, if both of them vanish simultaneously, the product will have a double zero, and contradicts  the fact that $\sin (p_i-p_k+p_k)/2$ only has simple zeros. Now, using this lemma, we will prove the following main theorem:
 \begin{itemize}
 \item[] {\small \bf Theorem}:
 There are only two distinct possibilities concerning the zeros of the Wronskians in \eqref{setofw}: Either (a) all the Wronnskians among  the members of $\mathcal{S}_1$ are zero and those among $\mathcal{S}_2$ are nonzero, or (b) all the Wronskians among $\mathcal{S}_2$ are zero and those among $\mathcal{S}_1$ are nonzero.
 \end{itemize}
 A proof goes as follows. As stated  in the Lemma, there are three distinct Wronskians which vanish at $\sin (p_i-p_j+p_k)/2$. This means that at least two of such Wronskians will be between the members of the same set, which can be $\mathcal{S}_1$ or $\mathcal{S}_2$. When the Wronskians vanish, the two eigenvectors in the Wronskian become parallel to each other. Since each set contains only three vectors, if two different Wronskians among the same set vanish, all three eigenvectors in that set become parallel simultaneously. Then, the third Wronskian in that set must also vanish. This argument shows that all the Wronskians among one of two sets,  $\mathcal{S}_1$ or $\mathcal{S}_2$,  vanish  simultaneously. Now, using the Lemma, we can conclude that the Wronskians among the other set must not vanish. This proves the theorem.
 
Since we already know the analyticity of the Wronskians of  the same sign type, \ie  $\sl{i_+,k_+}$ and $\sl{i_-,k_-}$, it is now straightforward to determine the zeros of the Wronskians with opposite signs using the Theorem above. This leads to the rule given in section \ref{subsec:analyticity}.
\section{Angle variable for the AdS part}\label{ap:g}
In this appendix,  we sketch the construction and the evaluation of the angle variable for the AdS part given in \eqref{AdSanglewron}(see also section 6.2 of \cite{KK3}).

Since we are studying the solutions with no nontrivial motion in AdS, the quasi-momentum for the AdS part  does not have any cut:
\beq{\label{adsquasiap}
\hat{p}_i = \frac{\Delta_i x}{2g(x^2-1)}\period
}
However, for the analysis of the angle variables, it turns out to be convenient to first consider the one-cut solution and then shrink the cut to get the result for \eqref{adsquasiap}. For one-cut solutions, there are two independent action variables, 
\beq{
S_{\infty}= \frac{1}{2\pi i}\int_{\infty} p (x)du(x)\comma \quad S_0  =\frac{1}{2\pi i}\int_{0} p (x)du(x)\period
}
Since the conformal dimension $\Delta$ is given by $S_0-S_{\infty}$, the angle variable conjugate to $\Delta$ is given by $(\phi_0-\phi_{\infty})/2$, where $\phi_{0}$ and $\phi_{\infty}$ are the variable conjugate to $S_{0}$ and $S_{\infty}$ respectively.

Each angle variable $\phi_0$ and $\phi_{\infty}$ can be constructed and evaluated in the similar manner as for the $S^3$ part. As a result, we obtain
\beq{
\begin{aligned}\label{ads0inf}
\phi_0^{(i)} &= i \ln \left(\left.\frac{\langle\tilde{n}_i\comma \tilde{n}_j \rangle\langle\tilde{n}_k\comma \tilde{n}_i \rangle}{\langle\tilde{n}_j\comma \tilde{n}_k \rangle}\frac{\langle j_{+}\comma k_{+}\rangle}{\langle i_{+}\comma j_{+}\rangle\langle k_{+}\comma i_{+}\rangle} \right|_{x=0^{+}}\right)\comma\\
\phi_{\infty}^{(i)} &= i \ln \left(\left.\frac{\langle n_i\comma n_j \rangle\langle n_k\comma n_i \rangle}{\langle n_j\comma n_k \rangle}\frac{\langle j_{-}\comma k_{-}\rangle}{\langle i_{-}\comma j_{-}\rangle\langle k_{-}\comma i_{-}\rangle} \right|_{x=\infty^{+}}\right)\period
\end{aligned}
}
As discussed in section 6.2 of \cite{KK3}, the polarization vectors in the AdS part are identified with the insertion points of the operators as
\beq{\label{adsnnn}
n_i = \pmatrix{c}{1\\x_i} \comma \qquad \tilde{n}_i = \pmatrix{c}{1\\\bar{x}_i}\period
}
Substituting \eqref{adsnnn} to \eqref{ads0inf} and computing $\phi_{\Delta}=(\phi_0-\phi_{\infty})/2$, we arrive at the expression \eqref{AdSanglewron}.

\end{document}